# A REVIEW ON THE PROGRESS OF POLYMER NANOSTRUCTURES WITH MODULATED MORPHOLOGIES AND PROPERTIES, USING NANOPOROUS AAO TEMPLATES.

Carmen Mijangos[1,2,3*], Rebeca Hernández[1] and Jaime Martin[4]

[1] Instituto de Ciencia y Tecnología de Polímeros, CSIC, c/Juan de la Cierva 3, 28006 Madrid, Spain

[2]IKERBASQUE, Basque Foundation for Science, 48011, Bilbao, Spain

[3]POLYMAT, University of the Basque Country EHU-UPV, Edificio Joxe Mari Korta, Avda. Tolosa 72, 20018 Donostia-San Sebastian (Guipúzcoa)

[4]Centre for Plastic Electronics and Department of Materials, Imperial College London, Exhibition Road, London, SW7 2AZ

***To whom correspondence should be addressed: cmijangos@ictp.csic.es**

## ABSTRACT

Polymers with the same chemical composition can provide different properties by reducing the dimension or simply by altering their nanostructure. Recent literature works report hundreds of examples of advances methods in the fabrication of polymer nanostructures accomplished following different approaches, soft lithography, self-assembly routes, template assisted methods, etc. Polymer nanostructures with modulated morphologies and properties can be easily achieved from anodized aluminum oxide (AAO) templates assisted methods. In the last decade, fabrication of polymer nanostructures in the nanocavities of AAO has raised a great interest since allows the control and tailoring of dimension of a huge number of polymer and polymer-based composites materials. The fact that polymer dimension can be adjusted allow the study of size-dependency properties. Moreover, modulated polymer nanostructures can be designed for specific applications from AAO templates methods. Taking into account the last considerations, this review present an overview of recent and new insights in the fabrication methods of polymer nanostructures from hard porous Anodic Aluminum Oxide (AAO) templates with emphasis on the study of polymer structure/property relationship at nanometric scale and stressing the potential interest in particular applications. It includes i) a description of the anodization methods and strategies to obtain AAO templates with adjusted dimensions; ii) a summary of different infiltration methods, starting with the infiltration of a polymeric fluid (melt or solution) into the nanocavities of AAO template, to conform a great number of the polymer nanostructures with different morphologies, compiled on a table. It includes the very last approach to obtain directly polymer nanostructures by in-situ polymerization of a monomer within AAO nanocavities and how the polymerization kinetics is affected by confinement in AAO nanoreactor; iii) an overview of how the effects of confinement alter the structural aspects, dynamical processes and mechanical, thermal and rheological properties of the polymer; iv) some examples of polymer nanostructures as precursor of applications bio-, adhesion, optical and electrical related; and finally v) a summary of conclusions and suggested challenges.

TABLE OF CONTENTS

# 1. Introduction to polymer nanostructuring and nanopatterning

One-dimensional polymeric nanostructures (1DPN) present novel and intriguing properties suitable for the fabrication of nanoscale devices employed in a wide range of applications such as energy storage or chemical sensors [1], optoelectronic devices[2], biomedical applications [3] or superhydrophobic surfaces [4]. In addition, it is accepted that 1D nanostructures provide a good system to investigate the dependence of electrical, thermal and mechanical properties on dimensionality and size reduction[5]. One-dimensional polymeric nanostructures have attracted plenty of attention from both fundamental and applied point of view as demonstrated by the increasing number of publications since 2000 shown in figure 1.

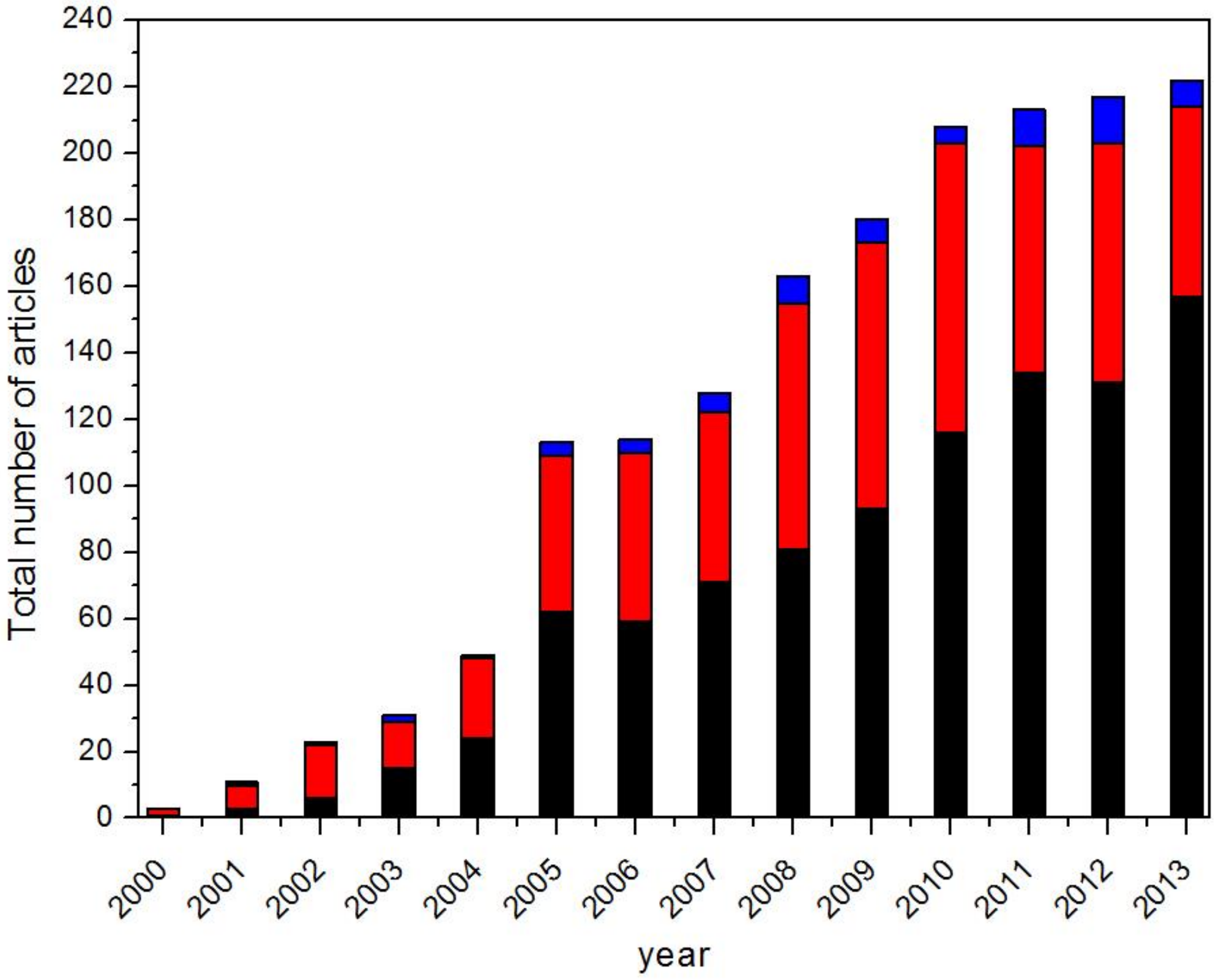

**Figure 1.** Total number of articles and reviews published on the topic of 1DPN since 2000. Polymer nanofibers (black bars), polymer nanotubes (red bars) and polymer nanorods (blue bars)

The SCOPUS search is done on the basis of a broad classification of 1DPN as a function of their geometry, so that publications on polymer nanofibers, polymer nanotubes and polymer nanorods are taken into account. Polymer nanorods present lengths in the range of several hundred nanometers.and therefore, a lower aspect ratio than polymer nanofibers (length/diameter). Due to this, they are normally supported on surfaces giving rise to more or less regular arrays, which results in nanostructured surfaces. Polymer nanotubes have been increasingly studied recently because of the structural versatility and functionality of polymers as well as their hollow 1D morphology.

The focus of this review is the fabrication, study of polymer properties and potential applications of 1DPN obtained by means of template-assisted fabrication employing nanoporous aluminium oxide templates. In section 1 we provide the reader with a snapshot of the current state of the research activity regarding fabrication of 1DPN in which, apart from the employment of template-assisted fabrication, we summarize different routes for the synthesis of 1DPN such as electrospinning, template-assisted fabrication, soft lithography or self-assembly routes. In section 2 we describe different anodization processes and conditions to obtain AAO templates with varying architectures and pore sizes. Section 3 is devoted to different preparation methods of 1DPN starting from infiltration processes of a polymer (nanomoulding) to the direct synthesis of the polymer from the corresponding monomer (nanoreactor). The study of the polymer properties under confinement will be reported on section 4 and polymer nanostructures for applications on section 5.

### 1.1. Electrospinning

Electrospinning has been shown to be a simple but powerful technique for the preparation of 1D nanostructures (mainly nanofibers; other types of 1D nanostructure, such as nanorods, can be cut from nanofibers by special methods [6, 7]). In fact, more than 80% of the total number of papers shown in figure 1 employs this method for the preparation of polymer nanofibers. This is an interesting fabrication process in terms of material diversity, cost, throughput, and the potential for high-volume production. Polymer nanofibers are produced by using an electrostatically driven jet of polymer solution (or polymer melt)[8]. In an ordinary electrospinning process, the jet is simultaneously pulled, stretched, elongated, and bent by the electric forces. Solvent evaporates rapidly, causing the shell jet to solidify, thus producing nano/micro fibers [9, 10].

Besides the preparation of polymer nanofibers, electrospinning has also been employed to prepare polymer nanotubes by using a coaxial, dual-capillary spinneret Co-electrospinning requires a polymer solution in the shell and either a polymer solution or a nonpolymeric Newtonian liquid or even a powder to fill the inner core[10-12].The mechanism by which the core/shell structure is transformed into hollow fibers is based on the evaporation of the core solution[13]. However, there are still some problems in obtaining high-quality nanotubes for coaxial electrospinning, such as the selection of the inner solvent and the control of electrospinning parameters, and so forth[14]. The technique presents some difficulties such as precise control over the fiber diameter and to achieve diameters below 100 nm. Furthermore, the nanofiber mat, which is usually obtained, presents a low degree of organization.

## 1.2. Nanofabrication techniques

Nanofabrication involves processes and methods of constructing engineered nanostructures and devices having minimum dimensions lower than 100 nm. Generally, the nanofabrication methods are divided into two major categories: “bottom–up” and “top–down” methods according to the processes involved in creating nanoscale structures[15]. Bottom up approaches rely on the spontaneous organization or self-assembly of molecules or objects into stable, well defined structures by noncovalent forces including atomic and molecular nanostructures at a surface [16-18].

### *1.2.1. Self-assembly*

Self-assembly has emerged as a powerful bottom-up approach for the synthesis of 1DPN because it provides a high level of control over both the functionality and architecture of the final material. Two different approaches are currently employed, the first one is the self-assembly of a rod-coil block copolymer with designed structure parameters in a selective solvent so that the coil block forms polymer nanotubes directly [19]. A second approach involves the bulk morphology of the microphase separated block copolymer and it is more robust, especially for cylindrical nano-objects [20-22]. This is due to the fact that the block copolymers show cylindrical morphology in a relatively wide range of composition window and no extra parameters such as solvent interaction needs to be considered. Polymer self-assembly can also be driven through hydrogen bonding to give rise to polymer nanorods [23]

A drawback of “bottom-up” self-assembly techniques are that these processes are energetically driven so that they are sensitive to the chemical composition of the initial polymer. Therefore, the spectrum of materials which can integrate 1DPN is limited.

### *1.2.2. Soft litography*

Top–down approaches employ nanofabrication tools that are controlled by external experimental parameters to create nanoscaled structures/functional devices with the desired shapes and characteristics starting from larger dimensions and reducing them to the required values[15]. Nanolitographic techniques such as electron-beam (e-beam) or focused-ion-beam (FIB) writing, proximal probe patterning and X-ray or extreme-UV lithography are referred as `conventional´ top-down methods of nanofabrication. These techniques are limited to the direct patterning of resist materials on planar, rigid substrates and are often incompatible with emerging technologies that involve unconventional and soft materials, i.e. polymers[24]. Soft lithography techniques developed over the past 15 years constitutes a low-cost alternative to conventional lithography and are suitable for the fabrication of 1D polymer nanostructures. There are four general methods of replicating patterns of nanoscale features using soft lithography: printing, molding, phase-shifting edge lithography and nanoskiving [25]. The most common material employed for soft lithographic stamps is an elastomeric polymer, polydimethylsiloxane (PDMS) [26, 27]. PDMS has been widely used to form polymeric arrays of micrometer sized posts for a variety of applications, including control of cellular adhesion and wettability [28-30]. The pattern transfer is simple and thus it constitutes an effective nanofabrication tool for fabricating ultra-small features. On the other hand, a large-scale production of densely packed nanostructures is difficult, it is also dependent on other lithography techniques to generate the template, and usually not cost-effective[15]. In addition, due to the low level of stiffness of PDMS (Young's modulus on the order of MPa), one of the main concerns in their design is the stability of the patterned structures, in fact, irreversible collapse has been shown to occur in high-aspect-ratio arrays[31].

### *1.2.3. Template-assisted fabrication*

Template-based approach facilitates the fabrication of nanoobjects with very defined high aspect ratio. Other advantages of this method of fabrication in comparison to lithography-based fabrication methods include cost effectiveness, wide accessibility and the capability of top-bottom fabrication with nanoscale precision[32, 33]. Template-based approach consists of the nanomolding the polymer into the pores of a template, either a soft-template or a hard-template used as a mold. It is also possible to employ the template as a support for the in situ synthesis of 1DPN. For example, $MnO_2$ has been employed as a reactive template for the synthesis of polypyrrole (PPy). The $MnO_2$ acts not only as chemical oxidative seeds to initiate pyrrole polymerization but also as rigid backbones for the subsequent growth of PPy nanostructures[34].

Soft template approach involves the use of structural directing molecules such as organic dopant anions[35], surfactants[36-38], porous diblock copolymers[39] or supramolecular gels that comprise 3D networks of self-assembled fibers[40, 41]. Vertically oriented zinc oxide (ZnO) nanowire arrays can also been employed as soft templates because of the easily scalable deposition techniques[42, 43], the large variety of nanostructures that can be used as templates, and its facile removal through etching [44, 45]. Hard template approach includes the use of track-etched polycarbonate membranes[46] or anodic aluminum oxide (AAO) and silica membranes prepared by electrochemical etching (or anodization) of the base metals aluminum or silicon to produce ordered nanoporosity on the surface of these materials[47-50].

Among hard templates, porous anodic aluminum oxide (AOO) is by far the most used. It provides large versatility with respect to the diameter and length of the pores, it is of relatively low cost and has a long range ordered arrangement of homogeneous pores. The AAO membrane templates provides aligned straight channels whose pore size and shape are tunable, which have already been extensively used to synthesize polymer nanofibers, nanotubes and their arrays with varied compositions [51, 52]. AAO templates are thermally stable in the temperature range relevant to the processing of polymers. Therefore, mesoscopic structure formation processes such as crystallization, mesophase formation and phase separation, which affect the mechanical properties of the polymer nanorods can be controlled by application of specific temperature profiles.

The wetting of porous templates with polymer melts and solutions or polymer-containing mixtures is a simple and versatile method for the preparation of tubular structures with diameters ranging from a few tens of nanometers to micrometers[53], as it will be described in section 3. AAO templates can be employed for the fabrication of polymer nanorod arrays with uniform structural parameters (diameter, height, and center-to-center spacing) over a large area. A critical drawback to this method is that once AAO is removed, most fabricated nanorods might collapse [54]. This is overcome by placing an AAO membrane on a conducting flexible polymer substrate surface modified to increase the adhesion with the AAO membrane. In this way, ultrahigh-density array of freestanding and vertically aligned poly (3-hexylthiophene) P3HT nanorods on a gold-coated flexible polymer substrate have been fabricated. The highly ordered array of nanorods could be used for optoelectronics, high-density data storage materials, sensors, and photovoltaic cells with high conversion efficiency[55].

### 1.3. Combination of nanofabrication techniques

Combination of “bottom-up” and “top-down” approaches have been efficiently used in producing functionally important nanoscale materials[16]. Specifically, for the case of 1DPN, block copolymer lithography involves a combination of bottom–up self-assembly and top–down lithographic processes that result in domains with high periodicity (10 nm within a template or highly elaborate patterns)[56]. Self-assembly strategies can also be combined with template approaches to achieve polymeric nanotubes. That is the case of the self-assembly of block copolymers BCPs in AAO membranes which give rise to BCP nanorods and nanotubes that, with a weak base or acid, can easily be released from the template [57]. For example, nanorods of polystyrene-*b*-polybutadiene (PS-*b*-PB) and polystyrene-*b*-poly(methylmethacrylate) (PS-*b*-PMMA) have been successfully prepared within AAO membranes and morphological changes under cylindrical nanoconfinement have been studied[58-60]. In addition, nanotubes of polystyrene-*b*-polyacrylonitrile (PS-*b*-PAN) and polystyrene-*b*-poly(4-vinylpyridine) (PS-*b*-P4VP) were prepared within AAO membranes and used to generate nanoporous carbon nanotubes[61, 62].

The layer-by-layer (LbL) assembly method can be regarded as a versatile bottom-up nanofabrication technique[63]. In the LbL process, different charged species-polymeric, colloidal, nano-and microscale-are assembled on a substrate in sequences of alternating charges[64] or other complementary interactions such as hydrogen bonding or host-guest interactions[65]. The fabrication of polymer nanotubes via layer-by-layer (LbL) assembly of polyelectrolyte solutions employing AAO membranes gives rise to hierarchical nanostructures previously challenging to obtain via other means (e.g., solution-casting, melt-wetting)[66]. There exists an optimum set of conditions (ionic strength, molecular weight) that ultimately control chain dimension relative to geometric dimension The tube wall thickness was found to be thinner or thicker than those assembled on planer surfaces depending on pore size [67, 68]. A broad range of polymer materials have been LbL-assembled within the cylindrical nanopores of AAO membranes to yield polymer nanotubes[69]. For example, aqueous dispersions of polyallylamine hydrochloride (PAH) as the anionic and sodium polystryrenesulfonic (PSS) as the cationic component were deposited into an AAO membrane with pore diameter of about 200 nm. After etching the AAO membrane with aqueous NaOH solution, flexible $(PAH/PSS)_3$ complex nanotubes were obtained[70].

# 2. Nanoporous aluminium oxide templates

## 2.1. General Features

Anodic aluminium oxide (AAO) templates consist of a dense array of cylindrical nanopores, which lay mutually parallel and perpendicular to the underlying Al substrate (Figure 2). Pores in AAO templates are arranged in well-ordered domains of close-packed hexagonal symmetry over 10-20 interpore distances, while on a larger scale, domains are randomly oriented. The pores grow originally blind (with closed ends), because a non-porous, constant thickness aluminium oxide layer, named barrier layer, separates the empty space of pores and the underlying Al substrate. However, barrier layer as well as aluminium substrate can be easily etched so that through-hole nanomebranes are achieved.

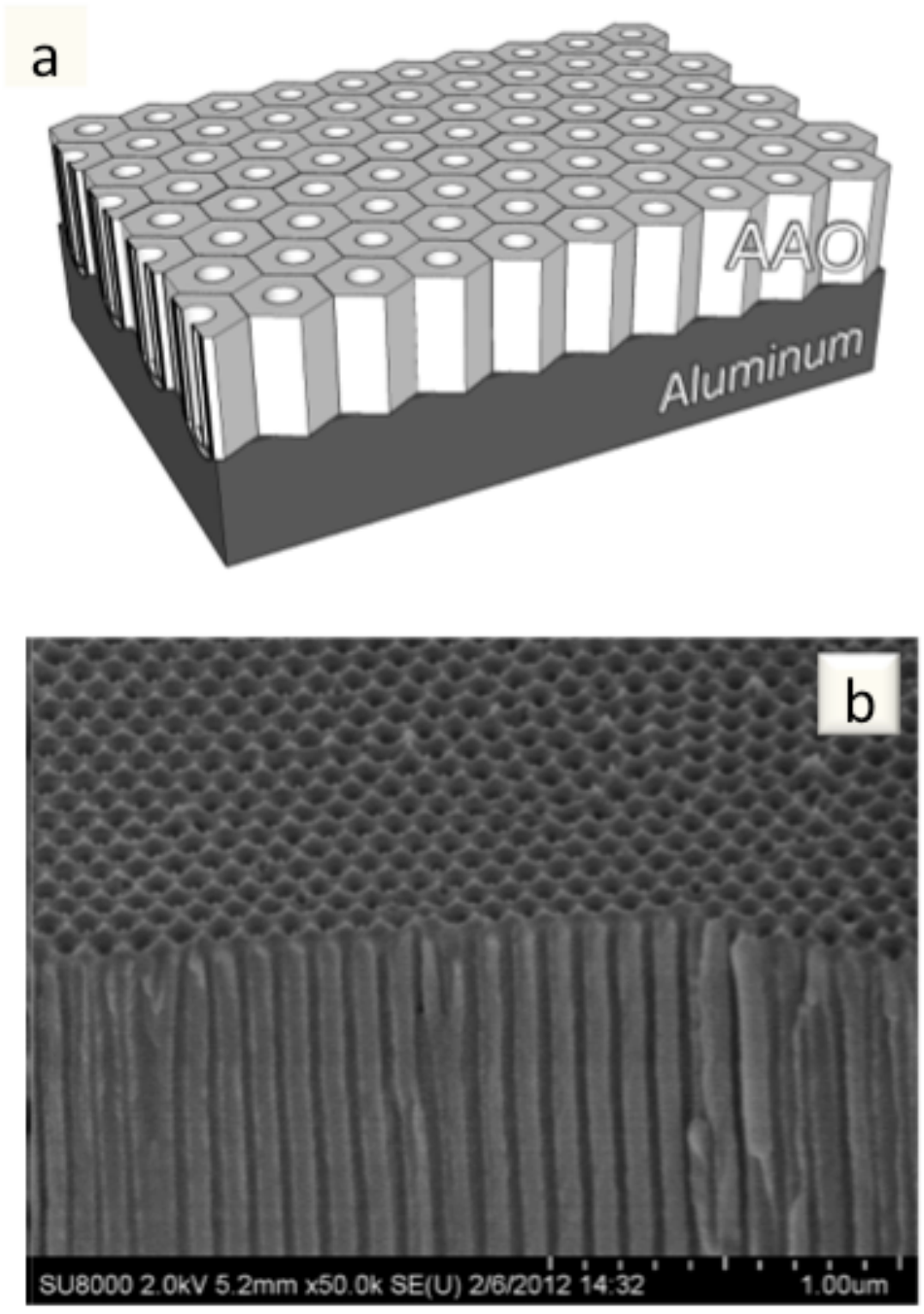

**Figure 2.** (a) Idealized structure of anodic porous alumina and a cross-sectional view of the anodized layer. (b) SEM micrograph of a AAO showing surface and cross section of the template. Reprinted with permission of [71]. Copyright (2013) American Chemical Society.

The formation of the AAO porous honeycombs have been extensively investigated and reported by various groups along the 20th century, starting by the pioneering works by Keller [72], Thomson [73, 74] and others. Nevertheless, AAO drawn the attention from scientist in the nanotechnology arena in 1995, when Masuda and co-workers reported the reaction conditions leading to self-ordering of pores [47] and the subsequent development of a method for achieving ordered AAO templates, i.e. the so-called two-step anodization process [75]. This essentially consists of two consecutive anodization reactions (electrochemical oxidations) over the surface of an aluminum foil. Briefly, a high purity aluminum foil is first cleaned, degreased and polished. Then, a first anodization reaction is performed in an acidic electrolyte under dc constant voltage, which provokes the formation of a first porous AAO layer. The pores in this AAO layer are not ordered at the top surface but are hexagonally arranged at the regions close to the oxide-aluminum interface if the anodization is carried out under specific conditions. On their growth during first anodization, the pores sculpt hemispherical concaves on the aluminum substrate, which thus tend to organize together with the pores. Subsequently, the first AAO layer is dissolved and a second anodization reaction is performed. In the second anodization, the concaves sculpted in the aluminium surface act as nucleation points for the growth of the new pores. Thereby, the hexagonal pattern developed in the first anodization is transferred to the new porous oxide creating the self-ordered nanoporous AAO template.

Since the development of the two-step anodization method by Masuda [75], a number of significant achievements have been realized by groups such as Gösele's [76-79] and Masuda's himself [80-82], to name some. These achievements include the discovery of new self-ordering conditions enabling different structural sizes [47, 76, 80, 83, 84], the fabrication of monodomain pore arrays [85], templates with square and triangular lattices[82], fast fabrication procedures [78, 86], ultra-small pores diameters [84, 87, 88], AAOs with complex pore architectures [78, 79, 89] and multidimensional porous network [90]. However, a detailed description of porous AAO is out of the scopes of this report, which merely aims to offer a general overview of AAO as templates for polymer nanostructure manufacturing. Thus, for detailed information on porous AAO, the reading of a set of review articles recently published by Lee [91], Sulka [92] and Jani [32] is heartily recommended.

Nanoporous self-ordered AAO has been typically used in the polymer field for two purposes, either for polymer nanostructure fabrication by replication of the AAO nanocavities, or as a confining medium for the study of low dimensionality effects on the physicochemical properties of polymers.

As a ***porous template***, AAO present several features that set it apart from other templates when it comes to fabricate polymer nanostructures. Most important are:

(i) AAO presents high flexibility regarding achievable pore diameters, pore lengths and interpore distances, which enables polymer nanomaterials with tailored dimensions.

(ii) Nanopores are arranged in orderly fashion, which allows for the patterning of ordered polymer nanostructures.

(iii) Pores in AAO are highly monodisperse in size and shape, which lead to the production of polymer nanostructures of high morphological quality.

(iv) The template-based approach allow for the patterning of any polymer capable of being molten or in solution [93].

(v) AAO can be easily dissolved in acidic or alkaline without affecting the polymer nanostructures embedded into the pores, so that free polymer nanostructures can be achieved.

However, in the last years, AAO is emerged as one of the main tools ***to study confinement effects*** on the physical behavior of polymers. AAO nanopores impose a high-quality two-dimensional spatial confinement over the polymer contained in them. Hence, the employment of AAO to that end is becoming more and more frequent. Rationales for this interest are:

(i) The degree of confinement experienced by the polymer material can be easily modulated by adjusting the pore diameter.

(ii) The monodisperse, cylindrical pores together with the rigid pore walls lead to a completely well-defined geometry of confinement.

(i) AAO templates present high pore densities ($10^9$-$10^{11}$ pore/cm$^2$) and high-aspect ratio pores (>$10^4$) can be obtained. These enable having relatively high amounts of polymer confined into the pores (tens of mg), which is crucial for many characterization techniques.

(ii) The high surface energy of its hydroxilated pore walls enables the simple infiltration of polymers even inside ultra-high aspect ratio pores.

(iii) The AAO is inert and stable in the temperature ranges where most of physical processes of polymers take place (e.g. below 500 °C), which allow for the study of a number of thermally activated processes, such as crystallization, melting, molecular and collective dynamics, order-disorder transitions and the like.

## 2.2 Types of self-ordered AAO templates

Since main uses of AAO in polymer science are directly connected with the tunability of its nanopores (when it is used both as a negative template and as confining medium), the dimensions of pores and the shape of pores are selected as criteria to present the different AAO templates. Note that the following subsections are focused on providing a concrete presentation of the variety of AAO templates available for polymer uses, rather than on offering a deep study on the formation of each AAO structure.

*2.2.1 Ordered AAO templates with different pore sizes*

The dimensions of the structural features in AAO will determine the dimensions of the replicated polymer nanostructure when used as a template. Therefore, dimensions of the original template impact directly on the properties and functionality of the polymer nanostructure. Likewise, pore dimensions define the degree of spatial confinement experienced by the polymer material when it used as a confining medium. Hence, the great importance of size control in nanoporous AAO templates. The characteristic dimensions of pore structure in a self-ordered AAO template can be defined in terms of pore diameter, pore length and interpore distance (lattice constant of the hexagonal symmetry). These can be easily tuned by applying the appropriate set of reaction parameters during the anodizations, namely constant output voltage, anodization length, temperature, stirring and nature and concentration of the electrolyte. Note that in this report, we focus our attention on self-ordered AAO templates produced by mild anodization (MA) aproaches -which employ low current densities, as their use is more widespread and their dimensions can be readily tailored.

***Pore length***. The longitudinal growth rate of pores in AAO templates synthesized under MA conditions is typically low (1-10 μm/h) and quite constant. Therefore, second anodization time allows precise control over the length of the nanopores from tens of nanometers up to hundreds of microns.

***Interpore distance.*** It is well known that the interpore distance is linearly proportional to anodizing potential [72, 94]. So far, 5 self-ordering regimes have been identified under MA conditions, which lead to 5 different lattice constants. Interpore distances between 450-500 nm are obtained when anodizing at applied voltages of 195-205 V using phosphoric acid as electrolyte [83, 95, 96]. Anodization at voltages around 40 V using oxalic acid solutions yield a lattice constant of 100 nm [47]. Applying an anodization voltage of 25 V and using sulfuric acid solutions as electrolyte lead to an interpore distance of 65 nm [80]. Potential values of 18-19 V in sulfuric acid electrolytes yield a lattice a constant of 50 nm [87, 97, 98]. Furthermore, Masuda et al reported on AAO templates with a interpore distance of 25-30 nm by anodizing at 10-12 V in concentrated sulfuric acid electrolytes [99]. It is also worth noting that Masuda et al. managed to produce ordered AAO with a 13-nm interpore distance by combination of colloidal lithography and a subsequent anodization process [88]. Finally, Lee at al. recently discovered that a routinely employed industrial process using high current densities, the so-called hard anodization (HA), enabled AAO templates with interpore distance between 220 and 300 nm [78], a range not covered by MA self-ordering regimes.

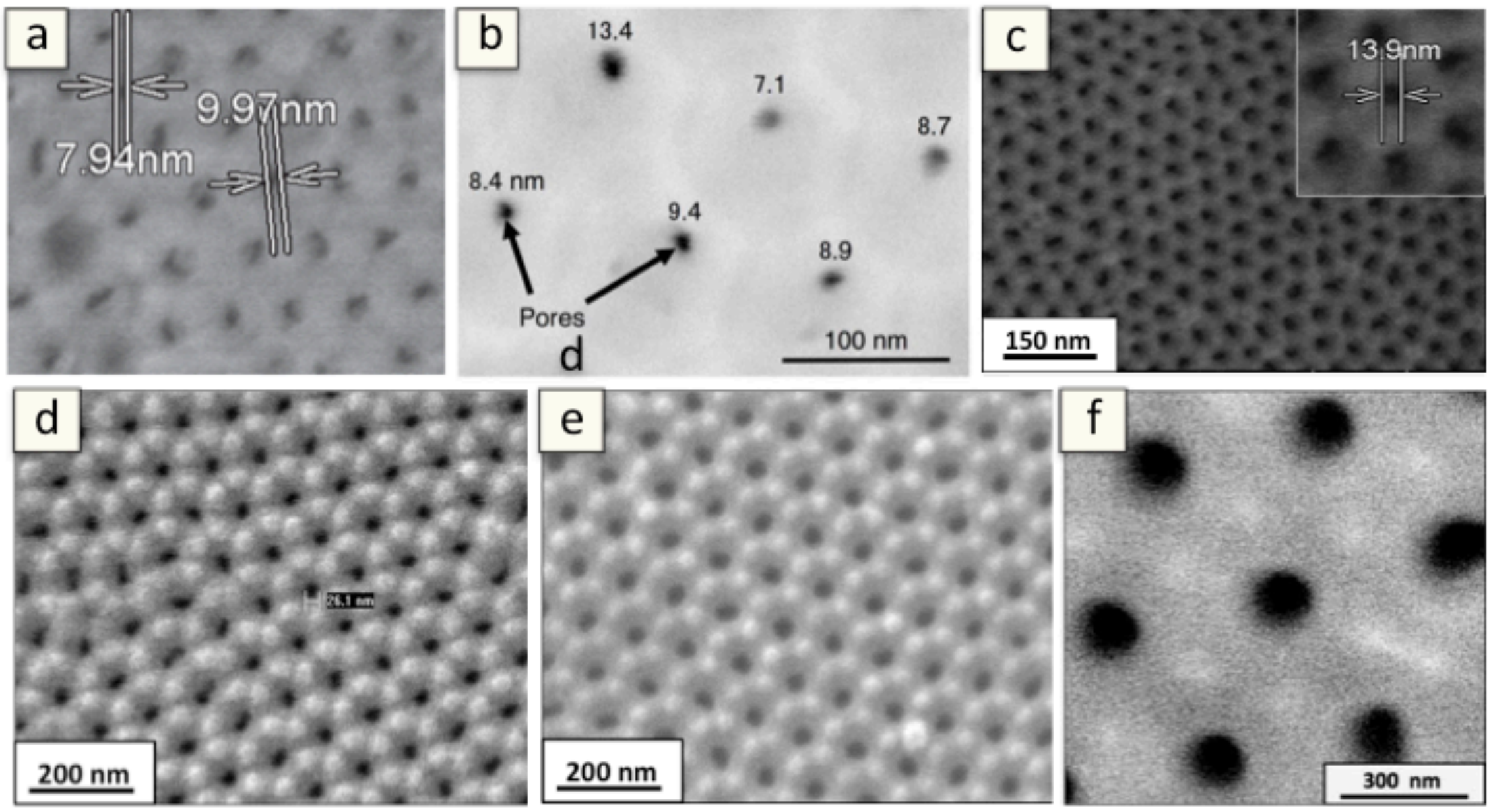

**Figure 3.** Surface views of AAO templates having different pore diameters. (a) and (b) sub-10 nm pores, (c) 14nm pores, (d) 25 nm pores, (e) 35 nm pores, and (f) 130 nm pores. Reproduced with permission [87], Copyright 2014; [84], Copyright 2013; [97] Copyright 2013 American Chemical Society; [100] Copyright 2013; [101] Copyright 2012.

***Pore diameter.*** Pore diameter is also linearly proportional to applied potential [91, 92]. However, whereas interpore distance is established during the first anodization –as it is related pore ordering- and it is fixed for each value of applied voltage, the diameter of pores can be further adjusted in a third step of the fabrication process. In this third step, originally obtained pores are enlarged by the controlled dissolution of the pore walls using acidic solutions (typically $H_3PO_4$ solutions). Thereby, through the selection of the appropriate anodization parameters and subsequent pore widening process, one can cover almost the entire range of diameters from 10 to 400 nm. The AAO template having the smallest diameter pores so far has been obtained by Nishinaga and coworkers [84] (Figure 3a) and Manzano and coworkers [87] (Figure 3b). Both groups managed to fabricate AAOs possessing pores in the 10-nm-range, although through different approaches. Nishinaga et al. reported the discovery of a new self-ordering regime using selenic acid ($H_2SeO_4$) as electrolyte, which yield the formation of 10 nm pores [84]. Manzano et al., in contrast, produced 8-12 nm pores[87] reducing the rate of pore-wall dissolution during the second anodization in already known self-ordering conditions [97]. Masuda et al. have reported pores well below 20 nm by using concentrated sulfuric acid solutions [99, 102]. However, since conditions used for second anodizations were far from those of self-ordering, high aspect-ratio pores are hard to be obtained by these methods. Conversely, 15 nm in diameter pores without aspect-ratio limitations were achieved by Martín et al. by anodizing in ethylene glycol containing sulfuric acid electrolyte at 19 V [97] (Figure 3c). The diameter range between 25 and 50 nm can be well covered by AAO templates anodized in 0.3 M sulfuric acid at 25 V and subsequent controlled widening of pores [80] (Figure 3d). Anodization in 0.3 M oxalic acid electrolyte at 40 V yield 35 nm in diameter pores [47, 75] (Figure 3e), which can be afterwards enlarged up to 80 nm. Pores between 130 and 400 nm in diameter can be produced anodizing in phosphoric acid solutions at voltages values of 195-205 and subsequent widening of these [83, 95, 103] (Figure 3f).

*2.2.2 Ordered AAO templates with different pore architectures*

The development of AAO templates with complex pore architectures has meant a step further in template-based nanofabrication, as cavities different from the typically isolated cylindrical nanopores can be exploited for creating complex, hierarchical nanostructures with novel properties. So far, researchers have been able to produce pores with a number of different morphologies. Most important are pores with diameter modulations, branched pores and multidimensional (3D) porous structures.

***Pores with diameter modulations***. Subtle variations of pore diameter through the thickness of the AAO templates can be achieved simply by changing the electrolyte, as shown by Krishnan et al [104]. Lee et al. managed to realize diameter modulations of higher amplitude by combining conventional MA and HA in a process that each modulation required the exchange of the electrolyte solution [78] (Figure 4a). More recently, Lee at al. developed the so-called "pulse anodization", in which MA steps and HA pulses are alternated without the necessity of electrolyte change [79]. Lee and coworkers [79, 105, 106] as well as Sulka and coworkers [107, 108] have demonstrated that pulse anodization enable a high control of pore diameter modulations in terms of shape and position. Losic et al. reported the "cyclic anodization", in which current profiles in the form of waves were applied to produce the pore modulations [89, 109]. Furthermore, spontaneous current oscillations taking place during HA process in oxalic and sulfuric acid electrolytes induce also pore diameter modulations, as reported by Lee at al. [110] and Schwirn et al. [86] (Figure 4b).

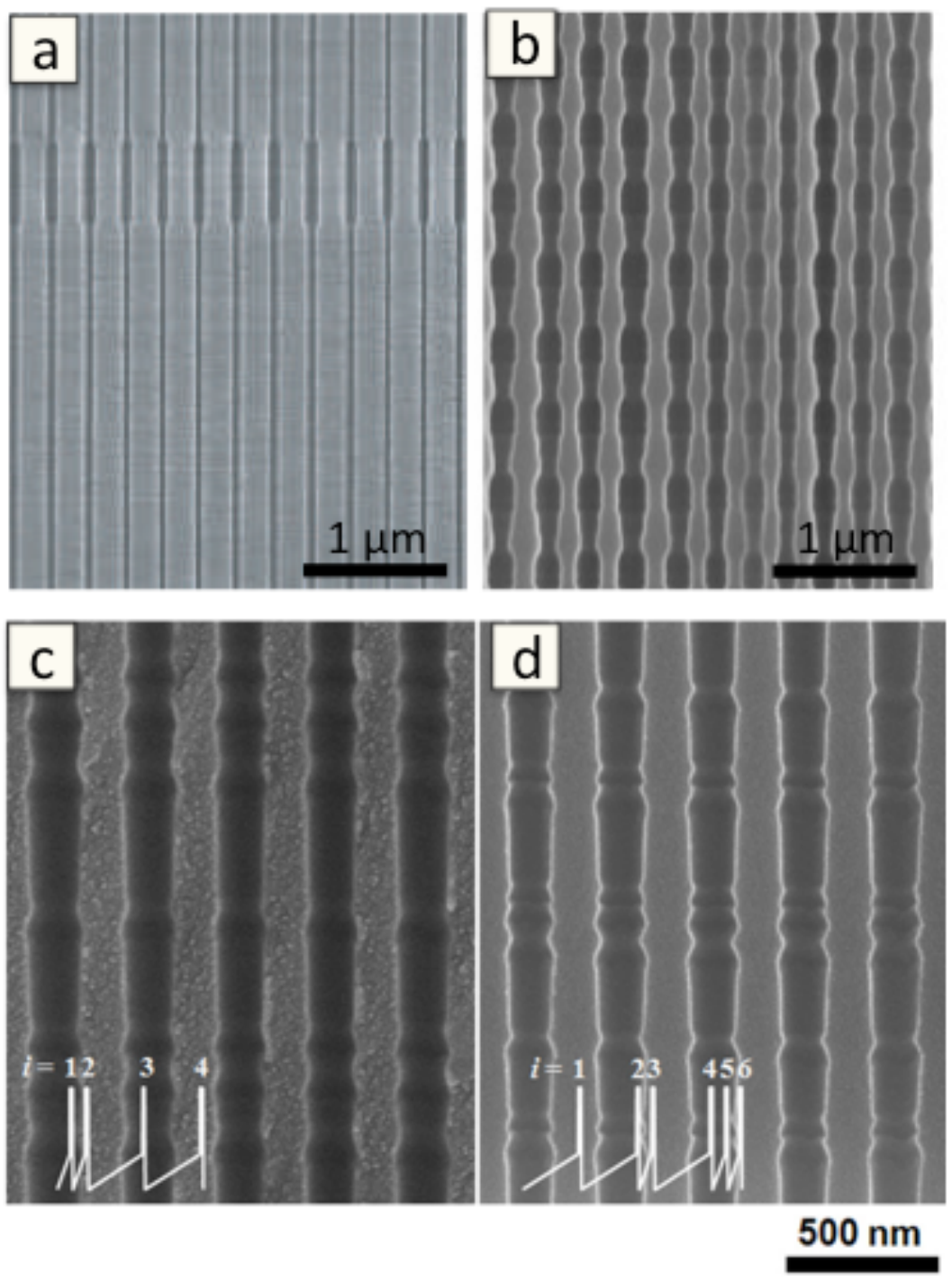

**Figure 4.** (a) Cross sectional SEM images of AAO templates with pore diameter modulations obtained by alternation of mild anodization and hard anodization processes [78] and (b) spontaneous current oscillations . Reproduced with permission [78], Copyright 2006; [110], Copyright 2010.

***Branched pores***. Ordered AAO templates with Y shaped pores can be produced by a step reduction of the applied voltage (by a factor of 1/√2), as reported by Li and coworkers [111]. The method was further extended to achieve multitetered AAO templates consisting of pores branching into smaller pores in succeeding tiers [112]. Meng and coworkers managed to produce a number of different multibranched pore structures adjusting the voltage step reduction [113, 114]. Moreover, exponential reductions of the applied voltage also yield branched pore structures, as has been showed by Chen et al [115].

***Three-dimensional porous structures***. Three-dimensional (3D) ordered nanostructures are currently gaining increasing attention as they promise the low-dimensionality effects associated to nanoscopic elements, while being a macroscale-size pieces, hence, easy to be manipulated, analyzed and employed. However, their fabrication turns to be challenging, as the macroscopic extension of 3D nanostructures restricts their fabrication approaches to bottom-up strategies, i.e those based on self-organization. Thus, a promising method for achieving 3D nanostructures are template approaches using self-organized porous templates having 3D porous structures. Motivated by this idea, a number of groups have ventured into the development of AAO templates with ordered three-dimensionally interconnected nanopores. Lee et al. showed that pore modulations originated from self-induced current oscillations may be an starting point to create a well-defined 3D porous structure at a local scale [110]. Likewise, Losic et al. reported on AAOs with interconnected pores from cyclic anodization [109], but the non-perfect organization of longitudinal pores in the AAOs together with lengthy etching treatments lead to severe structural damages in the material. In this context, is worth noting a recent report by Martín and coworkers, in which they presented an 3D AAO template having a well-defined, ordered, tunable and homogeneous 3D nanotubular network in the sub 100 nm range (Figure 5) [90].

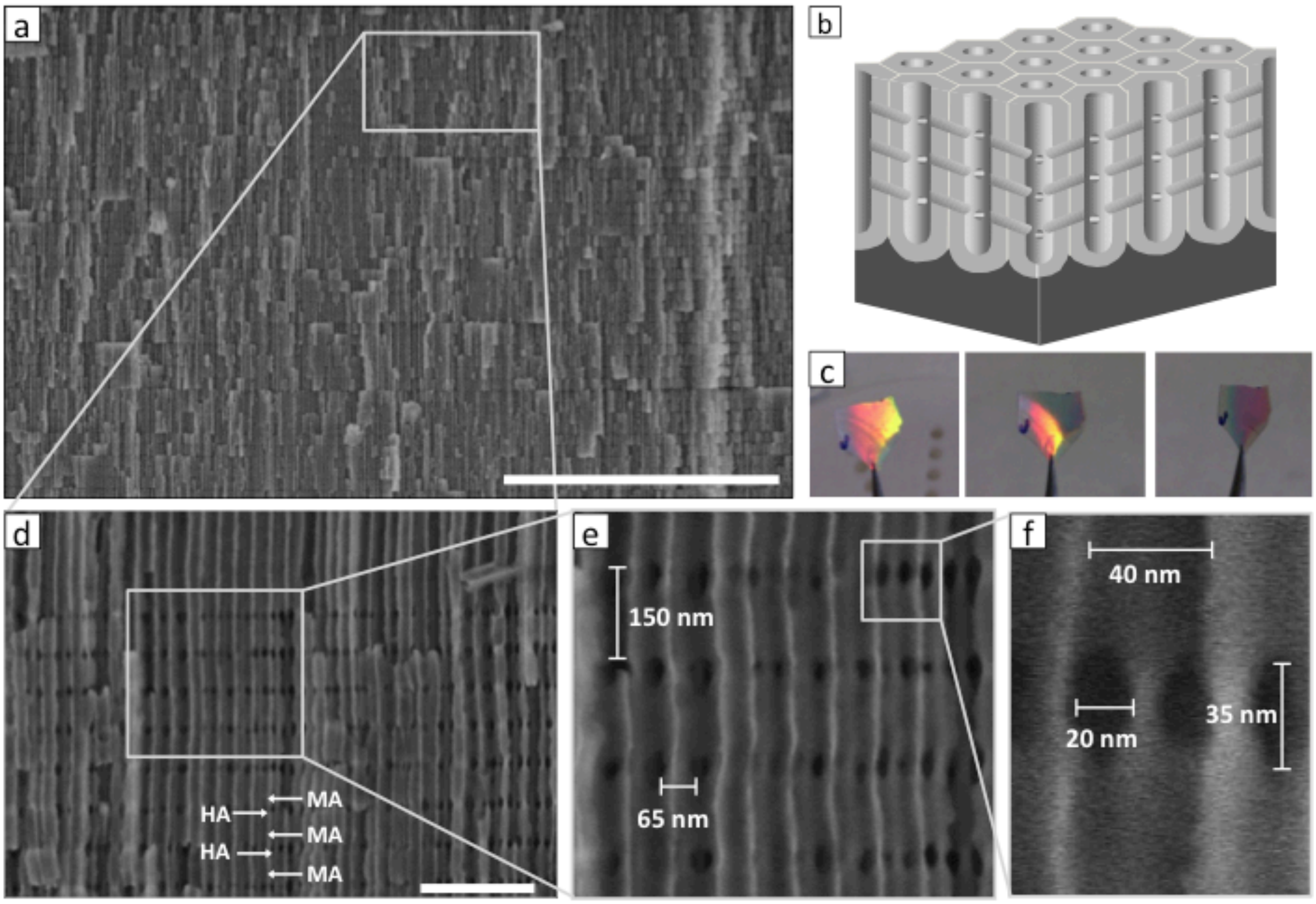

**Figure 5.** **(a)** Cross-section SEM image of the 3D-AAO template fabricated by pulse anodization ($V_{MA}$=25 V, $t_{MA}$=100 s; $V_{HA,N}$ =32 V, $t_{HA}$= 2 s) and subsequent partial chemical etching of the HA regions with $H_3PO_4$ 5 wt %, at 30 °C, for 18 min. **(b)** Schematic cartoon of the interconnected nanochannel network structure of the 3D-AAO templates. **(c)** Optical photographs of a 3D-AAO acquired from different viewing angles (the sample is suspended with tweezers). **(d)** Cross section micrograph of a 3D-AAO template, in which the nanostructure of the template can be appreciated. **(e)** Enlarged view of the area marked in (d) with a rectangle, in which both the longitudinal pores (in the vertical direction) and the transversal pores (in the horizontal and in the out-of-plane direction of the image) can be distinguished. The distance between longitudinal nanopores ($D_{int,long}$) is 65 nm, while the separation of the planes containing the transversal nanopores ($D_{int,trans}$) is 150 nm for this specific 3D-AAO **(f)** Detailed view of the nanochannel structure from which the dimensions of different nanopores can be measured. The diameter of longitudinal nanopores is 40 nm. The section of the transversal nanochannels is rather ellipsoid, presenting a short diameter of 20 nm and a large diameter of approximately 35 nm. Reproduced with permission [90], Copyright 201

# 3. Methods of preparation of polymer nanostructures using AAO templates

Several strategies have been developed in order to prepare polymer nanostructures, among them, polymer nanostructures with modulated morphologies and topographies have been produced with the aid of ordered anodized aluminum oxide (AAO) templates used as shape-defining nanomolds or nanoreactors. A great variety of tailored polymer nanotubes, nanofibers, nanospheres, nanocapsules or more complex nanostructures, such as 3D architectures, can be fabricated by infiltration of a polymer melt or solution in a suitable AAO template, under appropriate conditions or by “in-situ” polymerization of a monomer. The method is high flexible regarding achievable dimensions of the nanostructures, meaning diameter of the nanorods/tubes (ranging from 15 to 400 nm), length (ranging from hundreds of nanometers up to several hundreds of micrometers).

The concept of “Template Synthesis”, first introduced by C. R. Martin et al., was focused on the electrochemical or chemical template synthesis[116]. Later, the polymer template synthesis was widely studied and developed by many different groups, such as U Gosele’s [60, 117, 118]; J. H. Wendorff’s [119-121]; Steinhart’s [118, 122-130]; T. Russell’s [58, 131-136]; J. T. Chen’s [45, 137, 138]; Z. Jin´s [139-142]; J. Lee´s [143]; J. Wang´s [144]; Marsal´s [145]; and more recently, by our group [71, 100, 146-156] among other groups that will be mentioned along this section.

The “Template synthesis” method is a nanomolding process that starts with the infiltration of a polymeric fluid (melt or solution) into a mold, the AAO in this case, and the polymer solidification within the cavity. If necessary, the molded polymeric material can be removed from the cavity in a next step. With the few exceptions of the pioneering works, most of the reports in the literature to prepare polymer nanostructures by means of AAO template assisted methods, are dated on the last ten years and entail the use of laboratory made porous AAO templates. .

In this section, we summarize the preparation methods by infiltration of a polymer (nanomoulding) in section 3.1 and by direct synthesis of the polymer from the corresponding monomer (nanoreactor) in section 3.2. A recent revision by Jani et al. [32] shows the advances in surface engineering and emerging applications of nanopores anodic aluminum oxide templates. Even though, polymer modifications on the AAO templates are also reported by the authors, the review is mostly focused on the fine control of the surface chemistry, functionality, density and thickness of the AAO templates and not to the fabrication of polymer nanostructures using AAO templates. Therefore, the present revision based on the polymer nanostructures, could be considered complementary to that.

### 3.1. Template synthesis methods

Steinhart et al [118] were the first to develop a simple method for the fabrication of polymer nanotubes of polystyrene and polytetrafluoroethylene (PTFE) with a monodisperse size distribution, from the polymer melt or solutions, using ordered porous alumina templates. This method was based on the wetting phenomena that occur if a polymer solution or melt is placed in contact with a substrate of high surface energy, like the porous of AAO templates. As a consequence, the polymer will spread to form a thin film. This method is known as precursor film. In this work, authors also reported the conditions to process the polymer melts, i.e. polymers must be placed on the top of the AAO pore array at a temperature well above its glass transition temperature, in the case of amorphous polymers, or its melting point, in the case of partially crystalline polymers (Figure 6).

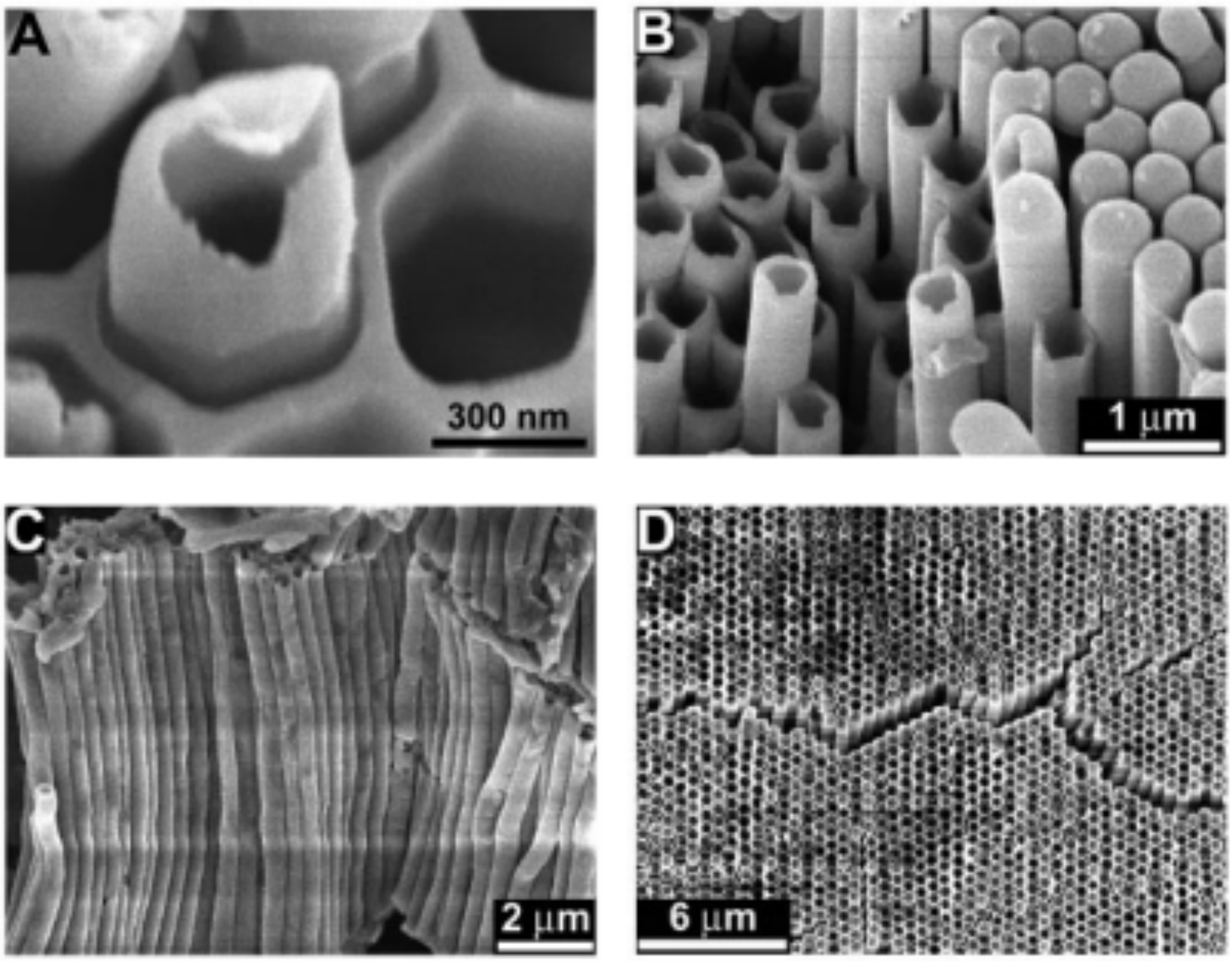

**Figure 6.** Scanning electron micrographs of nanotubes obtained by melt-wetting. (**A**) Damaged tip of a PS nanotube (*Mn* ; 850,000 g/mol) protruding from a porous alumina membrane. The substrate, on which the pore array was located, has been removed to uncover the tube tips. (**B**) Ordered array of tubes from the same PS sample after complete removal of the template. (**C**) Array of aligned PTFE tubes. (**D**) PMMA tubeswith long-range hexagonal order obtained by wetting of a macroporous silicon pore array after complete removal of the template. Reproduced with permission [118], Copyright 2002: Science

Owing to the versatility of this technique, this approach has been largely employed in many different laboratories as a route toward the preparation of polymer and polymer-based composites nanotubes from many different chemical structures and compositions. A particular case of the melt precursor film infiltration method is when the thickness of this precursor film is larger than the pore radius of the AAO template [147, 149]. In this case, solid polymer nanofibers are obtained instead of hollow nanotubes as can be observed in Figure 7. This procedure has been largely applied to systematically prepare polymer nanofibers of PEO, PVDF, PI, PVDF-TrFE, P3HT, etc with dimensions ranging from 15 to 65 nm, in studies of correlation between polymer properties and degree of confinement, as it will described in section 4.

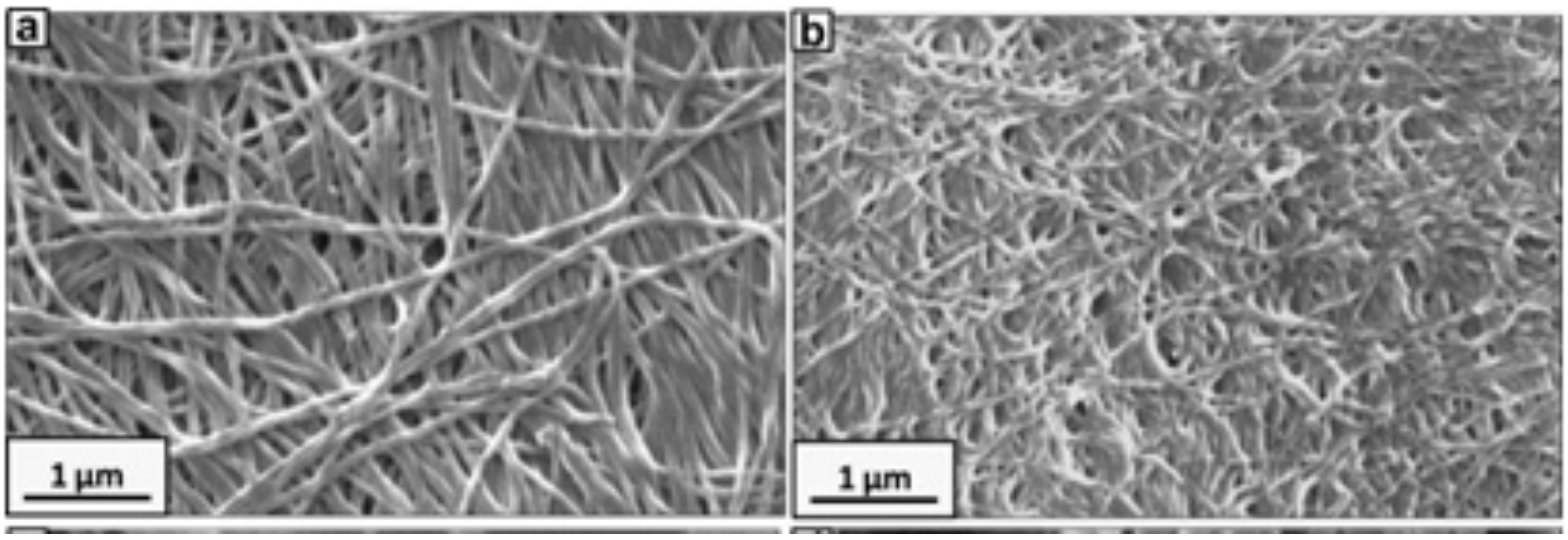

**Figure-7** SEM micrograph of polymer-based solid one-dimensional nanostructures. (a, b) PVDF nanofibers of 35 and 20 nm in diameter. Adapted with permission[100] Copyright 2012: Elsevier

Based also on wetting phenomena, if the polymer infiltration in AAO templates is carried out under partial wetting regime conditions, the melt infiltrates into the pores by capillarity. **Capillary process** of polymer liquids into templates is also a spontaneous process that has been studied by Russell et al,. [131-133], McCarthy et al,.[157] and previously by Whitesides et al for other patterns [158]. This is the case, when during infiltration polymer chains are in a low mobility regime within the melt (temperature slightly above the glass transition temperature) and are unable to form a melt precursor film.  However, the polymer melt is still a low surface energy liquid that tends to wet the pore walls. In this way, the polymer melt infiltrates into the pore, not only by advancing over the alumina pore walls, but also through the whole pore section. In this case, solid polymer solid nanorods or nanofibers are obtained. Generally, capillary infiltration is orders of magnitude slower than precursor wetting infiltration, and it is characterized by a meniscus in the advance extreme of the melt (Figure 8).

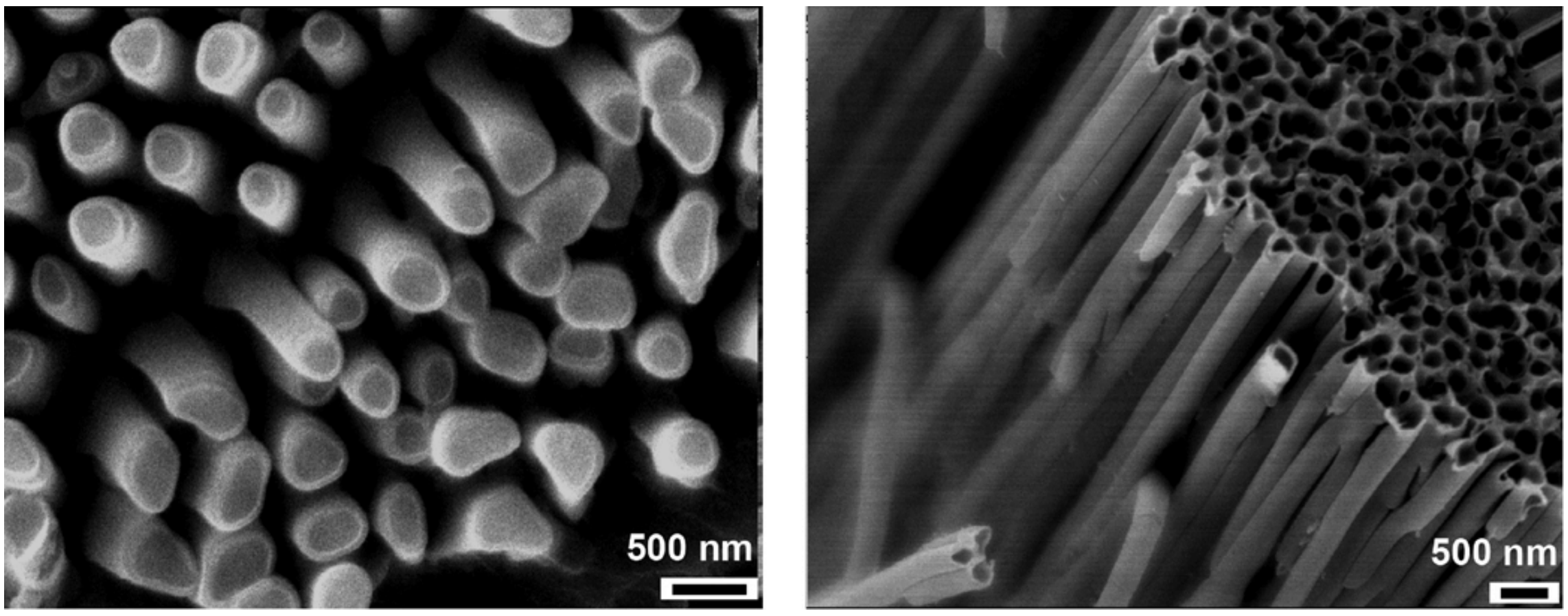

**Figure 8.** SEM images of PS-2 (*M*n ) 83.3 kg/mol) nanostructures generated within alumina membranes: (a) nanorods obtained after annealing at 125 C for 2 h (scale bar 500 nm); (b) nanotubes obtained after annealing at 192 C for 2 h (scale bar 500 nm). Reproduced with permission [131]Copyright 2006: American Chemical Society

A systematic study of the wetting of cylindrical alumina nanopores with polystyrene melt within cylindrical alumina nanopores, has been reported by M Zang et al [131]. In this work they observed that, as the annealing temperature was increased, a transition from partial to complete wetting, resulting in the formation of very different one-dimensional polymeric nanostructures (nanorods and nanotubes). They also found that the wetting transition temperature was dependent on the polymer molecular weight. Moreover, the large difference in the wetting rate between partial and complete wetting observed was used to fractionate polymers with different molecular weights.

Besides the study with homopolymers, T. Russell et al [133] investigated the morphologic changes experienced when lamella-, cylinder-, and sphere-forming block copolymers of polystyrene-*b*-polybutadiene were introduced into the pores of anodized aluminum oxide (AAO) membranes in the melt, by capillary forces, after a thermal annealing process. In order to investigate the effect of confinement on the microphase separation of the BCPs, they also varied the size of the pores in the AAO and the molecular weight of the BCPs. They found concentric cylinders for the lamella-forming

BCP and torus-like morphologies as the degree of confinement increased. For the bulk cylinder-forming BCPs, they observed a rich variety of morphologies, not seen in the bulk, that included stacked torus-like morphologies and single-, double-, and triple-helical morphologies. Finally they found that the specific morphology depended on the rapport between the AAO pore diameter and the period of the block copolymer BCP in the bulk (Figure 9)

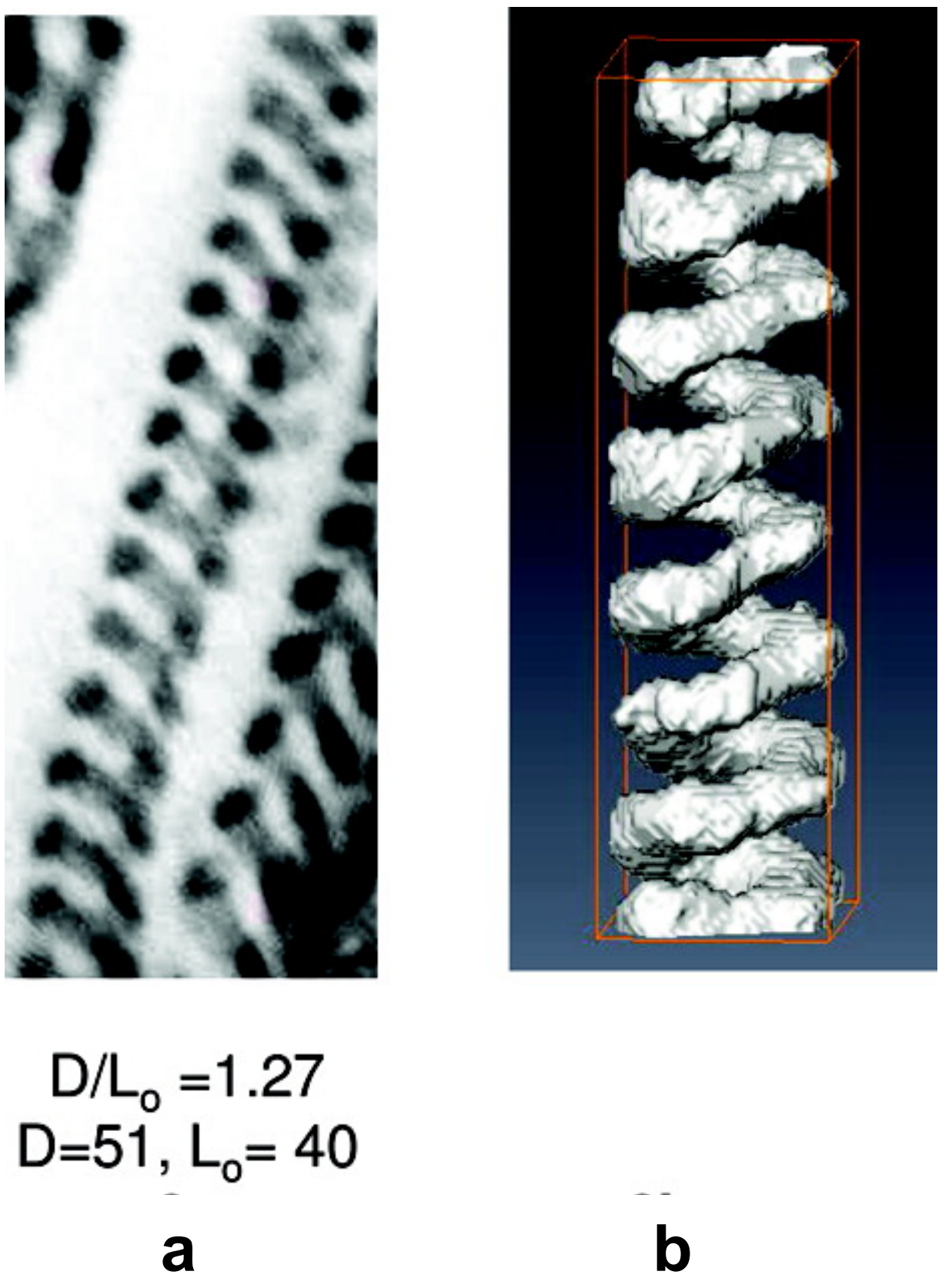

**Figure 9-** TEM images of bulk cylinder-forming PS-b-PB BCP inside AAO. View along the AAO pore of a nanorod with (a) PBD single helice b). Corresponding TEMT image. Scale bar: 50 nm. Reproduced with permission [133] Copyright 2009: American Chemical Society

**Solution wetting.** Polymer infiltration by immersion of templates into a polymer solution ("solution wetting infiltration") is probably the most traditional methods used for the preparation of polymer nanostructures. Polymer solutions usually infiltrate into pores of AAO filling their complete volume; however, depending on the concentration of the solution and preparation conditions, nanotubes, nanofibers, nanospheres or, nanocapsules, etc can be obtained. In particular, polymer nanotubes can be also obtained by infiltration of a dilute solution in the AAO template and subsequent drying of the system. Therefore, it can be considered an alternative for the preparation of polymeric nanotubes when polymers do not melt at the temperatures at which the precursor wetting infiltration takes place. However, it should be noted that the nanofibers often present lack of material, which means that incomplete nanofibers are obtained.

Wendorff el al have demonstrated that wetting from polymer solution, in addition to polymer melt, is also an effective way toward the preparation of polymer nanotubes, with particular emphasis on the influence of the molecular weight of the polymer used for wetting [120]. They observed that in the preparation of stable cylindrical structures, both, the quality of the solvent used and polymer concentration are important parameters. Moreover, they found a correlation between the variation of the molecular weight of the polymer used and the morphology of the obtained sample, i.e. nanowires, nanotubes, or arranged cylindrical structure with a regular arrangement of voids.

X Feng et al [139] reported the fabrication of polymer nanospheres, nanocapsules, and hemispherically capped nanorods simply through the wetting of anodic aluminum oxide (AAO) membranes with polymer solutions. They demonstrated that the formation of nanorods is dependent upon the solvents used (e.g., tetrahydrofuran and methyl ethyl ketone) which have strongly adsorbent nature toward alumina surfaces and not on the molecular weight of the polymer, as had been previously reported [120]. They also observed a coarsening process from spheres to capsules to rods in AAO cylindrical nanopores during solvent evaporation. The coarsening process served as the formation mechanism for nanorods which had the similar diameter as the AAO nanopores as observed in Figure 10.

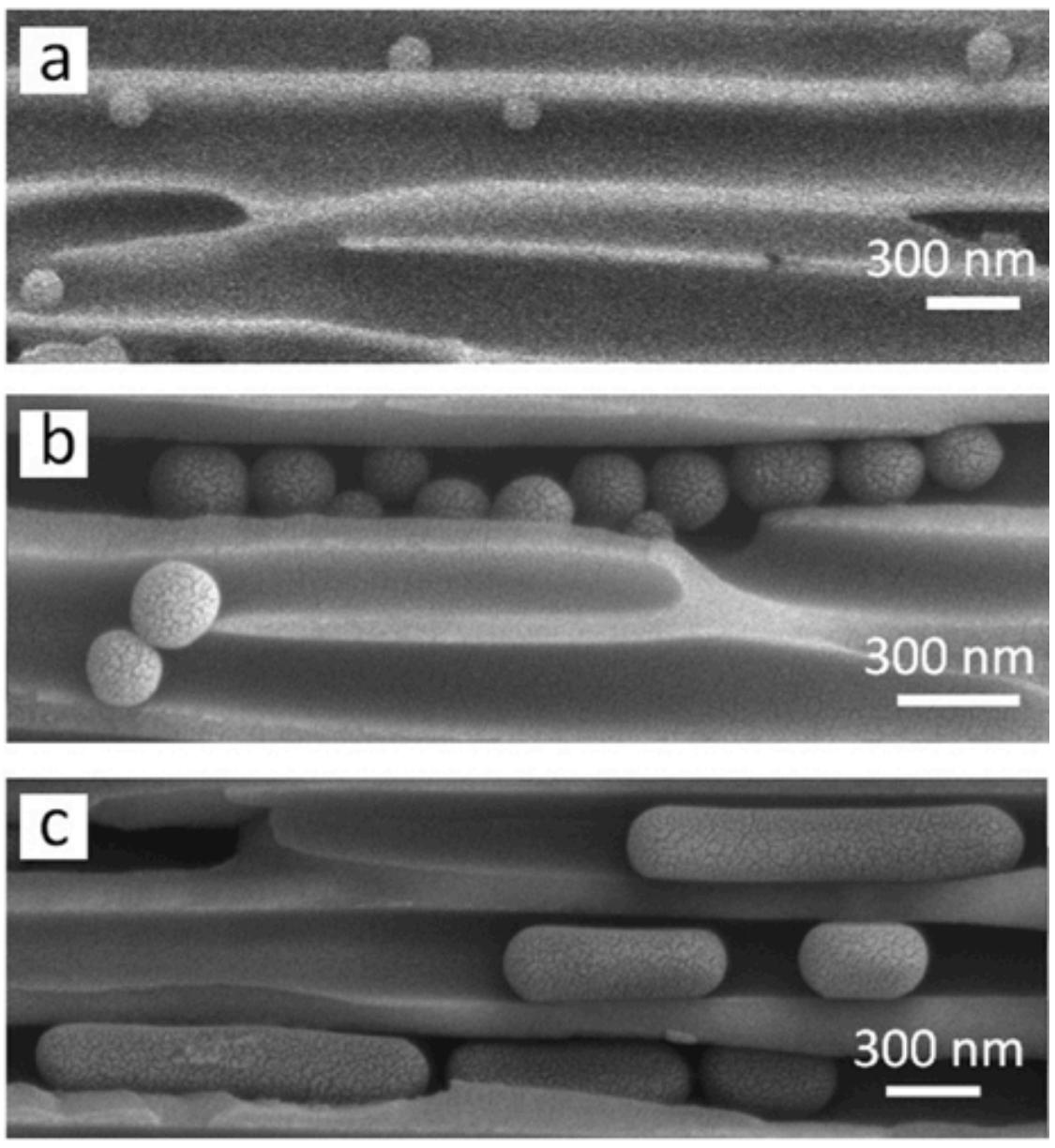

**Figure 10.** SEM micrographs of PS nanostructures embedded in the cleaved alumina nanopores. (a) Nanospheres from a 10 mg/mL PS/THF solution. (b) Nanospheres from a 20 mg/mL PS/THF solution.(c) Nanocapsules and nanorods from a 40 mg/mL PS/THF solution. Reproduced with permission [139] Copyright 2009: American Chemical Society

Many ordered, smaller, multicomponent and more complex polymer geometries can be achieved by combining dewetting processes, physical confinement and polymers and nanoparticles [158, 159]. Very recently, Chen et al [137] reported the formation of nanopeapod-like polymer structures from polystyrene (PS) and poly (methyl methacrylate) (PMMA) using a novel double-solution wetting method in the nanopores of anodic aluminum oxide (AAO) templates and selection of appropriate solvents, dimethylformamide (DMF) for PS and acetic acid for PMMA. The method is based on the stronger interaction between acetic acid and aluminum oxide than that between DMF and aluminum oxide. As a consequence, the PMMA solution preferentially wets the pore walls of the templates and the PS solution is isolated in the center of the nanopores. If the solvent is evaporated, the formation of nanopeapod-like PS/PMMA nanostructures takes places, as observed in Figure 11, where the shell and the core are composed of PMMA and PS, respectively. The authors explained that the mechanism involved in the formation of the nanostructures is related to the Rayleigh-instability.

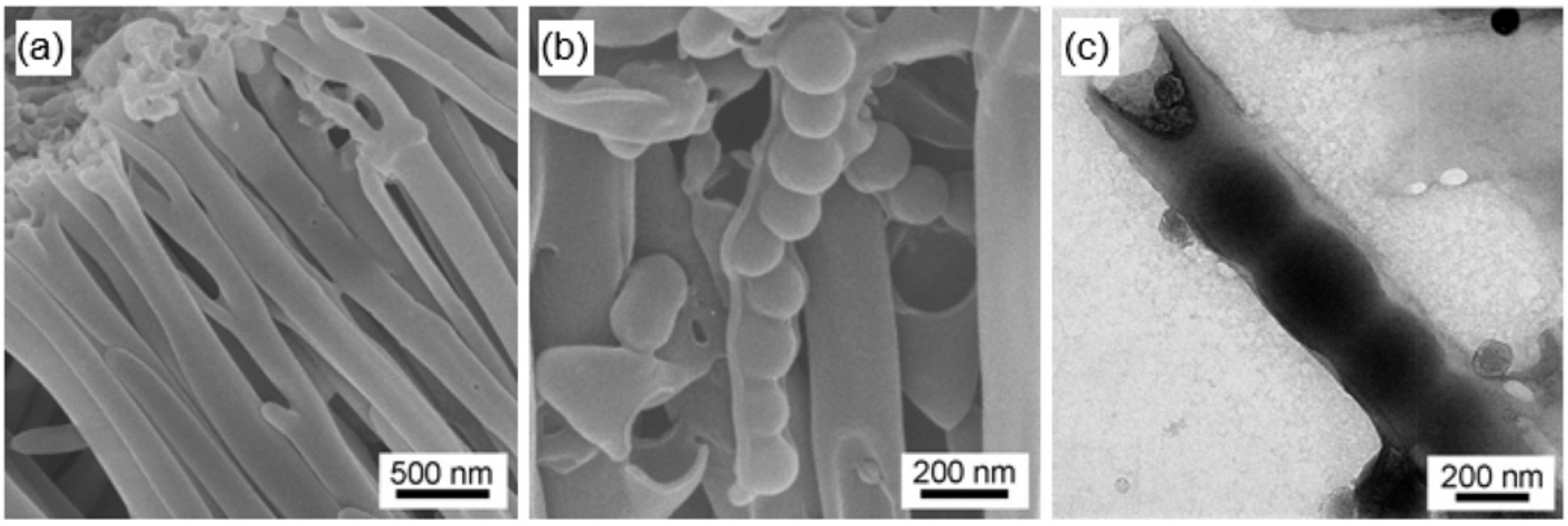

**Figure 11**. SEM (a and b) and TEM (c) images of peapod-like PS (Mw: 35 kg/mol) /PMMA (Mw: 97 kg/mol) nanostructures. The nanostructures are prepared by dipping AAO templates in PS solutions in DMF, followed by immersing the samples into 5 wt % PMMA solutions in acetic acid. The PS concentrations for the SEM (a and b) and TEM (c) samples are 1 and 5 wt %, respectively. In part b, some PS nanospheres are only covered partially by PMMA nanotubes, and the embedded PS nanospheres can be observed. Reproduced with permission [137] Copyright 2014: American Chemical Society

*3.1.1. Classification and selection of templates synthesis method*

Depending on the process involved and conditions, polymer infiltration methods described above can be classified into two main groups: methods based on applied forces and methods based on wetting phenomena which are, by far, the most common. This group of infiltration methods is based on the wetting properties of liquids onto solid surfaces and they have the peculiarity of being spontaneous. Depending on the process involved in the infiltration, three different methods can be considered: ***melt wetting***, classified at the same type in complete wetting regime, known as precursor film method and partial wetting regime, known as capillary method, and solution wetting and other sophisticated methods. Figure 12 summarizes all the infiltration methods of polymers within AAO membranes.

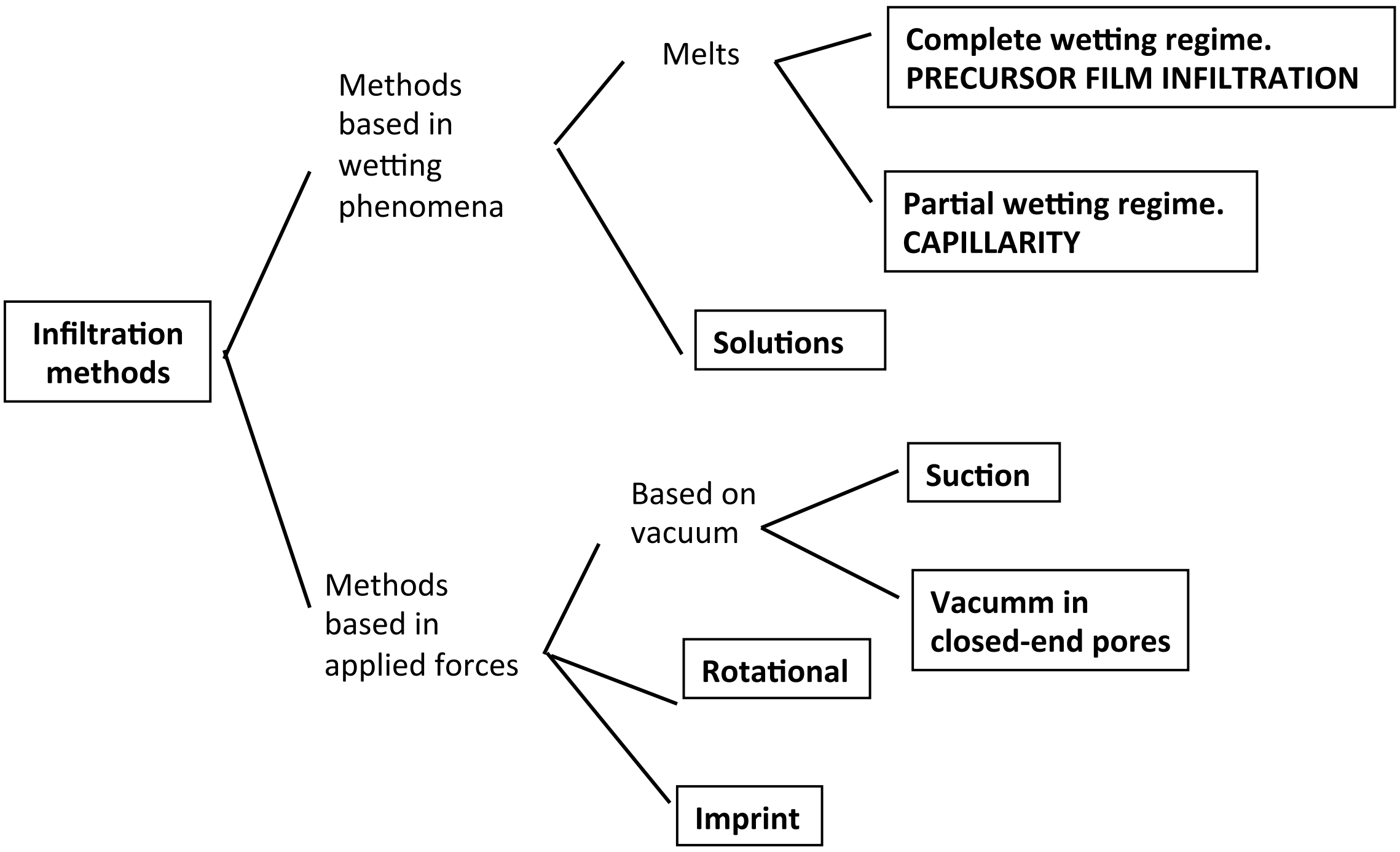

**Figure 12.** Summary of different infiltration methods of polymers within AAO templates. Reproduced with permission [100] Copyright 2012: Elsevier

It is important to note that other methods of infiltration based on vacuum, rotation or a combination of different complex processes have also been reported to prepare polymer nanostructures, based on AAO template synthesis methods[93].

In order to select the most suitable infiltration process to obtain polymer nanostructures using AAO templates some general guidelines can be suggested: when possible, the use of melts instead of solutions is generally recommended to obtain both nanofibers and nanotubes (in both wetting regimes: capillary rise and precursor wetting infiltration). Spontaneous melt infiltrations are usually more reproducible than solution since solvent related problems such as evaporation time, incomplete evaporation, lack of material in nanotubes or nanofibers, etc., are avoided. However, in solution wetting infiltration there is a set of parameters that can be adjusted (concentration, quality of the solvent, etc.), and this fact could lead to adjust a higher number of characteristic parameters in the nanostructures, such as, wall thickness or periodic modulations in nanotubes. This suggestion is also applicable when polymer-based composites are needed.

Although for the fabrication of polymer nanotubes, precursor wetting is recommended in the case of possible degradation, solution wetting should be used. If polymer-based composite nanotubes are required, precursor wetting is also recommended, when possible, for the same reasons as for pure polymeric nanotubes.

For the preparation of solid polymer nanofibers of high aspect ratio, the diameter of the nanostructure required must be taken into account. In the case of low diameters (approx. <70-100 nm), precursor film wetting infiltration is recommended. However, for high diameter nanofibers (approx. >70-100 nm), melt capillary infiltration is preferred even though it is a time consuming process. Solution wetting is an alternative employed for the fabrication of any diameter solid nanofibers if the polymer can be degraded. For solid polymer added nanoparticles composite nanofibers, the same procedure described or vacuum-induced infiltration through a membrane, always with concentrated solutions, can be followed.

Additionally, one can also obtain spherical nanoparticles using AAO templates by several procedures: (i) controlled evaporation of polymer solutions within pores, (ii) by promoting instabilities of infiltrated solutions and subsequent swelling of the formed nanoblocks, (iii) or by means of dewetting of curved polymer films in contact to a non-solvent liquid.

*3.1.2 Polymer nanostructures from polymer precursor (Variety of polymer and polymer composite nanostructures)*

In this section, we summarize on Table 1 many examples of different polymer nanostructures: nanofibers, nanotubes and nanospheres, obtained from a great variety of different polymers. Some of the samples contain embedded nanoparticles into the polymer matrices: $Fe_3O_4$; $Lu_3O_4$, etc. Moreover, we also report the infiltration method and preparation conditions together with the corresponding references

**Table 1. Polymers, infiltration conditions and references**

| ***POLYMER*** | ***MORPHOLOGY*** | ***AAO Diameter/ Length*** | ***INFILTRATION METHOD*** | ***MELT/SOL*** | ***CONDITIONS Tª/Time*** | ***REF.*** |
|---|---|---|---|---|---|---|
| PS | NR/NT | 35-360nm/ 100-50µm | Wetting Precursor Film | Melt | 200ºC/ 30min | [147] |
| PS | NT | 360nm/ 50µm | Solution wetting | Solution 0.5 wt% in toluene | RT/ 2days | [147] |
| PS | NS | 330nm/ 50µm | Solution wetting | Solution 2wt% in Toluene | RT/ 1days | [134] |
| PS | NR/NT | 200nm/ 60µm | Wetting Precursor Film | Melt | 130ºC/ 2h 205ºC/ 2h | [131] |
| PS | NR | 200nm/ 0.6-1.6µm | Wetting Precursor Film | Melt | 200ºC/ 5-20min | [117] |
| PS-b-PMMA | NR | 25-400nm | Wetting Precursor Film | Melt | 180ºC/ 24h | [58] |
| PS-b-PBD | Cylindrically NR | 80nm/ 80µm | Capillarity | Melt | 125ºC/ 4days | [133] |
| PS-b-P2VP | NR | 330nm/ 1.5µm | Capillarity | Melt | 220ºC/ 2days | [128] |
| PS-b-P4VP | NR | 25-330nm | Capillarity | Melt | 195ºC/ 2days | [156] |
| PS-b-P4VP_Au | NR | 25-330nm | Capillarity | Melt | 195ºC/ 2days | [156] |
| PS-co-MOP | NF | 28-35nm | Wetting Precursor Film | Melt | 210-230ºC/ 225-360min | [153] |
| PS/FePt | NT/NR | 360nm/ 50µm | Wetting Precursor Film | Melt | 200ºC/ 30min | [160] |
| PS/FePt | NF | 360nm/ 50µm | Solution wetting | Solution 10% in THF | RT/ 2days | [160] |
| PS/$La_xSr_{(1-x)}MnO_3$ | NT | 360nm/ 100µm | Wetting Precursor Film | Melt | 200ºC/ 30min | [100] |
| PS/$La_xSr_{(1-x)}MnO_3$ | NF | 360nm/ 100µm | Solution wetting | Solution 10% in THF | RT/ 2days | [100] |
| PS | NS | 200nm/ 60µm | Solution wetting | Solution 1wt% ethylenglycol | 130ºC/ 10min | [134] |
| PMMA | NR | 50-65 nm | Precursor film | Solution 20 wt% in toluene | 110ºC/ 3h | [145] |
| PMMA | NR-NT | 100-400 nm | Solution wetting | Solution 5 wt% chloroform | 60ºC/ 12h | [132] |
| PMMA | NT | 150-400 nm/ 60µm | Solution wetting | Solution 5% in THF | RT | [45] |
| PMMA | NR | 28, 35, 65 nm | Capillarity | Melt | 220ºC/ 12 h | [71] |
| PMMA | NR-NT | 170 nm | Capillarity | Melt | 120ºC/ 40 h | [147] |
| PMMA+$Lu_2O_3$ | NR | 330 nm | Capillarity | Melt | 130ºC/ 1 week | [155] |
| PMMA+$Lu_2O_3$ | NT | 330 nm | Precursor film | Melt | 190ºC/1 h | [155] |
| PVDF | NF | 20,35,60 nm/ 100µm | Precursor film | Melt | 240ºC/ 45min | [161] |
| PVDF | NS/NR | 330 nm/ | Solution wetting | Solution 2 wt% , 15% in | RT/ 1day | [100] |
| | | 100μm | | DMF | | |
| PVDF_SWCNTs | NR/NT | 35,330 nm/ 100μm | Solution wetting | Melt | 240ºC/ 6 h | [100] |
| P(VDF-TrFE) | NF NT | 35 nm/ 100μm 400 nm/ 150 μm | Precursor film | Melt | 250ºC/ 50min | [100] |
| PA | NR | 120 nm | | Photo crosslink PO 77F resin | | [161] |
| PFA | NR/NF | 200-350nm/0.7 μm 200-350nm/ 50μm | Solution | Polymerization | 120ºC/ 8h | [162] |
| PTFE | NR | 25-400nm | Precursor film | Melt | 400ºC/ 40min | [163] |
| P3HT | NF | 35nm | Precursor film | Melt | 200ºC/ 30 h | [100] |
| PTFE | NR | 120 nm | Precursor film | Melt | 400 ºC / 40min | [164] |
| PEEK | NR | 120 nm | Precursor film | Melt | 390 ºC / 25min | [101] |
| PEO | NR NT | 25-35nm/ 100μm 400nm/ 100μm | Precursor film | Melt | 140ºC/ 1h 110ºC/ 1h | [165] |
| PEO_PCL | NF | 25-400nm/ 100μm | Precursor film | Melt | 100ºC/ 12h | [130] |
| PP | NF | 25-380nm/ 80-100μm | Precursor film | Melt | 200ºC/ 20h | [125] |
| PE_Au | NR | 60 nm | Solution wetting | Solution10wt% in toluene, 1.5wt% dodecanethiol | capped-AuNPs in Toluene | [166] |
| PVC_$Fe_3O_4$ | NF | 70nm/ 2μm | Solution wetting | Solution 10wt% in THF | 60ºC/ overnight | [146] |
| PI | NF | 25-400nm/ 100μm | Precursor film | Melt | 20ºC/ 24h | [129] |
| PDMS | NF | 26-60nm/ 100μm | Capillarity | Melt | RT/ 3 days | [152] |
| PCL | NF | 25-200nm/ 100μm | Precursor film | Melt | 100ºC/ 12h | [167] |

## 3.2. Polymerization in AAO templates

### *3.2.1. Introduction*

The "in-situ" polymerization of a monomer in the AAO template (nanoreactor) is an alternative to prepare polymer nanostructures, other than by the infiltration of polymers (nanomolding). In fact, is a complementary way to obtain tailored thermoplastic polymer nanostructures when the polymer infiltration process must be carried out at high temperature and/or for a relatively long time, from hours to days, and the only way to obtain nanostructures from thermoset polymers, for which the polymer infiltration by melting is impossible [162, 168, 169]. It also offers important advantages in the study of the polymerization kinetics and reaction modelling in confinement, in comparison to other nanoconfined systems since i) each AAO nanocavity is a watertight compartment and, therefore, can be considered as a nanoreactor, ii) the dimensions of AAO nanopores are not only well defined but also easily tailored; and iii) the polymer can also be extracted[170-174]. Moreover, one important fact is the feasibility to directly produce polymer nanostructures for applications.

The direct polymerization of monomers in the reduced space of porous aluminum oxide templates started with the initial work of Martin et al, on the synthesis of conductive polymer nanostructures[116]. Since then, some interesting works have been also reported. Among them, it is worth mentioning, a recent review by Jani et al. that summarizes the experiments carried out to modify the top or inside of AAO surfaces by polymer grafting and plasma polymerization of different monomers and polymers via atom transfer radical polymerization (ATRP), reversible addition-fragmentation chain transfer polymerization (RAFT) or plasma-induced graft polymerization[32]. For instance, Cui et al reported that through surface-initiated atom transfer radical polymerization using a commercial porous anodic aluminum oxide membrane, poly(N-

isopropylacrylamide)-co-(N,N'-methylenebisacrylamide) (PNIPAM-co-MBAA) nanotubes were prepared with different composite ratios and it was found that the tubular wall strongly depends on the monomer concentration [175, 176] (Figure 13).

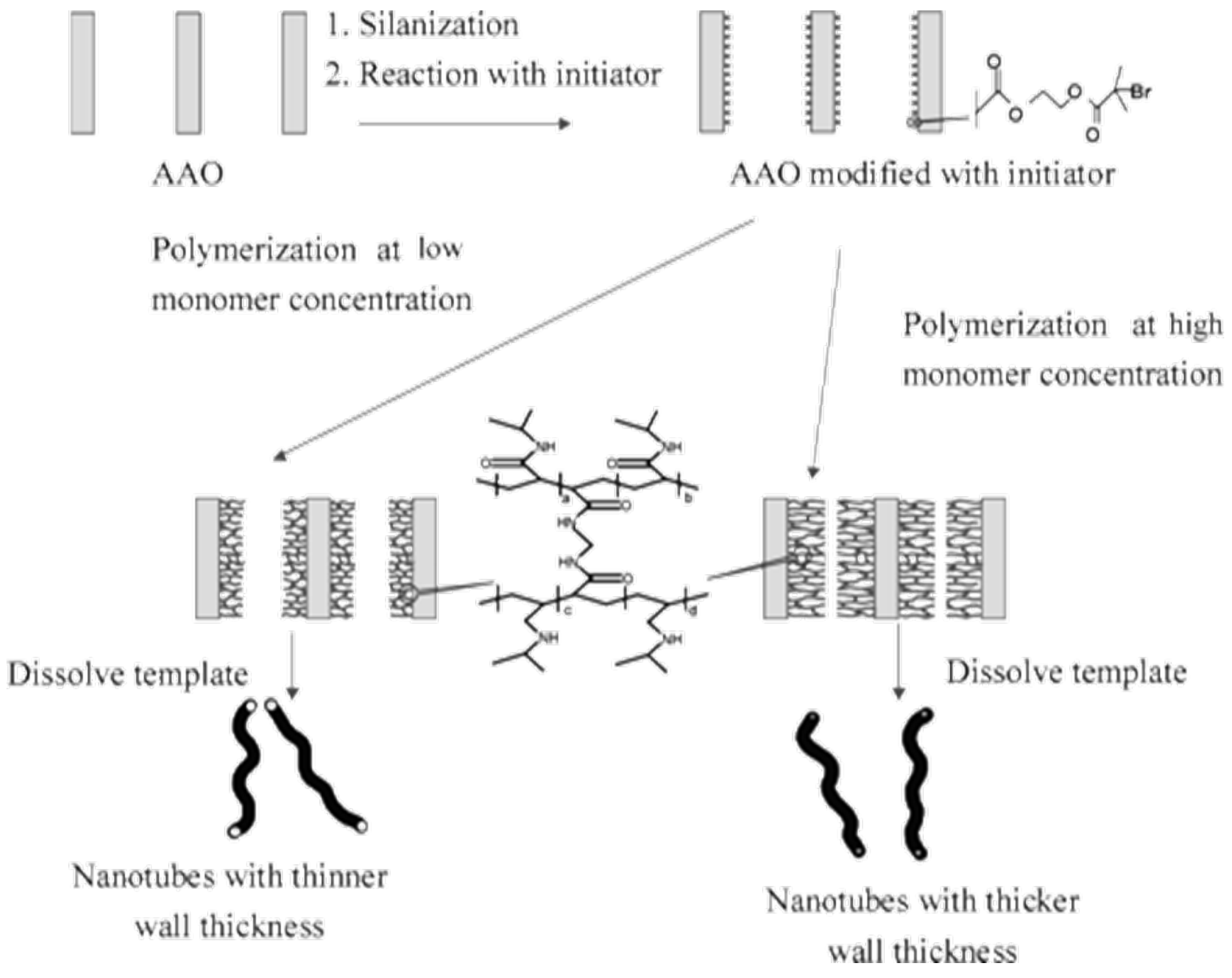

**Figure 13.** Schematic Illustration in Fabricating PNIPAM-*co*-MBAA Nanotubes with Different Wall Thicknesses in Porous AAO Membrane. Reproduced with permission [176] Copyright 2006: American Chemical Society

Some years later, much effort has been carried by Gorman et al [177] to study the polymerization of poly(methyl methacrylate) (PMMA) on porous anodically etched aluminum oxide and silicon substrates via surface-initiated atom transfer radical polymerization . They reported than using hydrogen fluoride, the chains could be cleaved from the substrates, as was evidenced by infrared spectroscopy. The molecular weights and molecular weight distributions of PMMA analyzed directly on these substrates, using direct ionization mass spectrometry and matrix-assisted laser desorption ionization mass spectrometry, lead to the conclusion that under the same polymerization conditions, PMMA grown on both p-Si and AAO substrates had a much lower molecular weight and a broader molecular weight distribution than that grown in solution. Confinement effects imposed by the pores during the polymerization are proposed as the likely mechanism for the reduced growth rates and more polydisperse chains.

Other attempts to explore the direct synthesis of polymers in AAO nanocavities have been done by J W Back and coworkers who studied the fabrication of conducting PEDOT nanotubes by vapor deposition polymerization using AAO templates [178]; Nair et al [179] reported the Ziegler-Natta polymerization of ethylene inside nanochannels of AAO template; functional acidic pyrrole containing oxidizable monomers have been template-polymerized using a hard AAO template in liquid phase polymerization conditions [180]; Polybenzyl glutamate has been surface grafted within nanoporous AAO templates and optically characterized [181, 182] and Grimm et al [161] reported the oligoetheracrylate photopolymerization with the aid of an 4 wt % of a free-radical photoinitiator into the AAO hard templates at room temperature and the formation of nanofibers consisting of cross-linked polyacrylate inside the nanopores that were connected with a cross-linked polyacrylate film on top of the AAO hard template (figure 14).

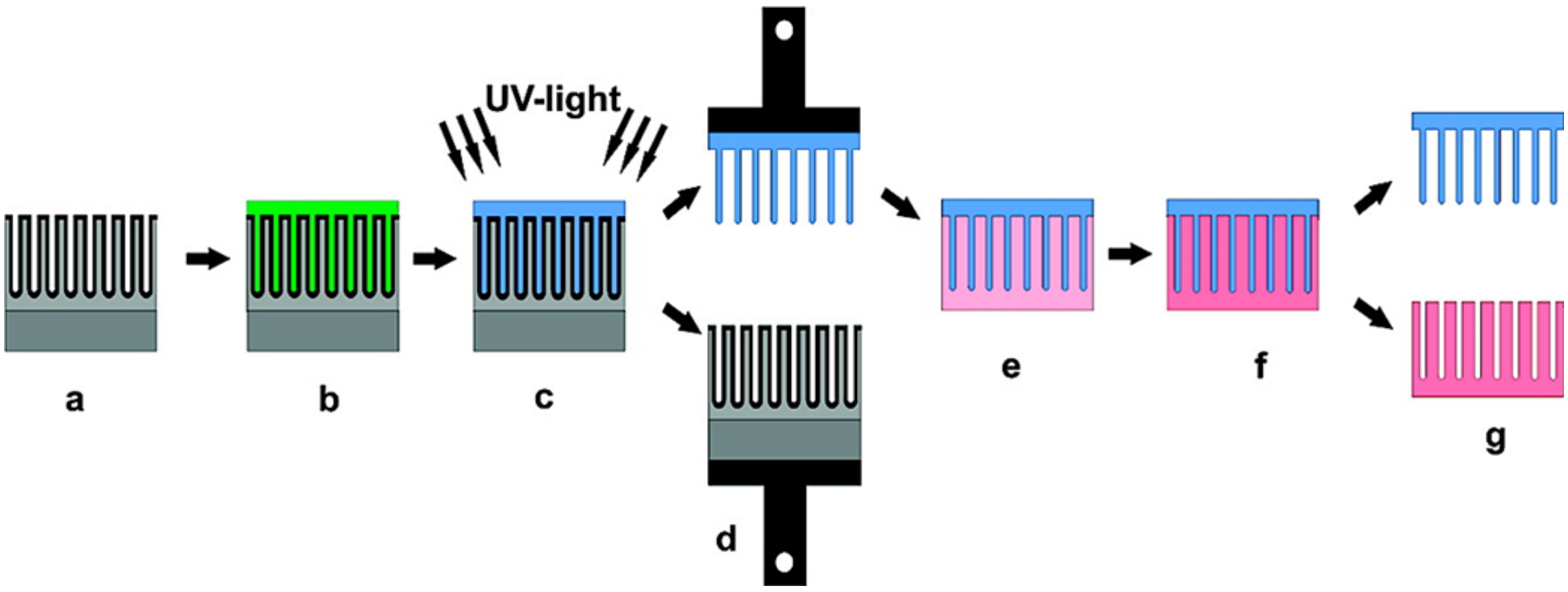

**Figure 14.** Schematic Diagram of the Nondestructive Replication of Self-Ordered AAO Based on the Use of Cross-Linked Polyacrylate Nanofiber Arrays As Secondary Molds. Reproduced with permission [161] Copyright 2008: American Chemical Society

Recently, a functional monomer was thermally polymerized inside the anodized aluminum oxide (AAO) channel into nanotubes, which were isolated and characterized to be semiconductive and blue fluorescent, and used as nano-containers of Fe3O4 nanoparticles to form magnetic nanocomposites[183]. Previously, polypyrrole (PPy) nanotubes with highly uniform surface and tunable wall thickness were fabricated by one-step vapor deposition polymerization using anodic aluminium oxide (AAO) template membranes and transformed into carbon nanotubes through a carbonization process[184].

Moreover, it was reported a simple method of utilizing anodized aluminum oxide as a reproducible template for fabricating high-aspect-ratio uniformly bent polymeric nanopillars that can be used as a physical adhesive[185]. In this paper, it is shown how to achieve straight high-aspect-ratio nanopillars with concepts of the work of adhesion and lateral collapse between polymer pillars without serious damage to the master template and is demonstrated, with the support of manufacturing polymeric nanopillars from the reusable AAO, a simple route to asymmetric dry adhesive nanopillars bent by residual stresses. However, with a few exceptions described below, in all the above papers the kinetics of the vinyl polymerization or how the kinetics is affected by the confinement have not been studied and neither if the chemical structure of synthesized polymers have been modified compared to bulk polymerization.

*3.2.2. Effect on polymerization kinetics and chemical structure*

In a recent work [168], the polymerization of styrene in the AAO nanotemplates has been carried out as an example of vinyl polymerization by free radicals mechanism. A porous anodic aluminum oxide (AAO) device of 35 nm of diameter was used as a nanoreactor both to produce in one step PS nanostructures and to study the radical polymerization kinetics of styrene (St) in confinement at 60 C and the results were compared to those of polymerization in bulk.

In order to estimate the monomer conversion at different times of polymerization, a Confocal Raman Microscopy (CRM) study was undertaken to monitor the consumption of styrene monomer inside the AAO nanocavities, from the surface up to the bottom of the template, since CRM spectroscopy method enables to chemically analyze the polymer or monomer along the nanopores by identifying specific bands of the monomer. The signal at 1630 $cm^{-1}$ corresponds to the C=C stretching band, which is present only in the monomer, while the signal at 1600 $cm^{-1}$ corresponds to C=C aromatic stretching band, present in both monomer and polymer, and thus represents 100% of the sample. Thus, by studying monomer conversion and the formation of the polymer through the AAO cavities as a function of time has been possible to establish the polymerization kinetics in confinement and the results compared to bulk polymerization. The plot of polymerization rate ($R_p$) in confinement versus time shows a sharp increase up to 2.5 h reaction time followed by a slight decrease. Under the same polymerization conditions, the reaction carried out in the AAO nanoreactor exhibits a value of Rp almost three times higher than the reaction rate carried out in bulk. The observed behaviors suggest a catalytic effect of the AAO template on the rate of polymerization and provides good evidence of the confinement effect (Figure 15). Recently, this same methodology was generalized and applied to obtain superhydrophobic polymer morphologies nanostructures by choosing the appropriate monomer and by tailoring the dimension of AAO cylindrical nanocavities to obtain the desired morphology[162].

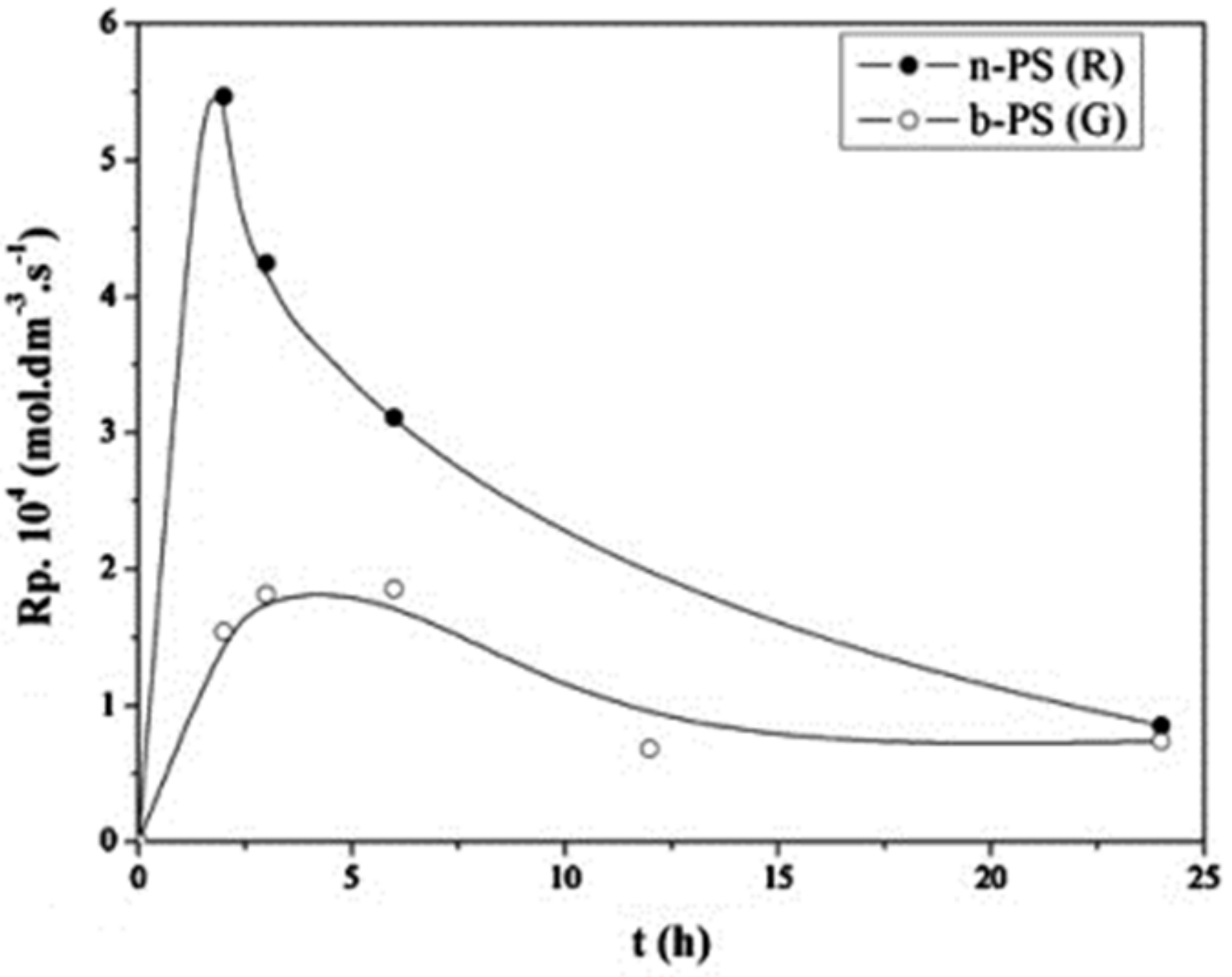

**Figure 15.** Rate of polymerization of St under confinement by Raman (R) determination and bulk by gravimetric (G) determination. Reaction conditions: [AIBN] ¼ 0.47% w/v; T : 70C . Reproduced with permission [168] Copyright 2013: Elsevier

To the best of our knowledge, no other key examples have been found in the literature regarding kinetic studies of free radical vinyl polymerization processes within AAO nanoreactors. Nevertheles, as polymer synthesis is expected to lead to functionalized polymer nanostructures with unique properties, it seems obvious that investigations in progress will soon appear in the literature. In regard to polycondensation processes, only one or two attempts have been reported to prepare polymer nanostructures by this method [186, 187]. In the first one, high-aspect-ratio polymeric nanopillars, based on UV curable polyurethane were obtained from AAO templates on a large scale without lateral collapse and the sticking problem between the mold and the nanopillars, and asymmetric uniform bending was imparted to the nanopillars by the evaporation of metals on the nanopillars with tilt. The UV curable polyurethane consisted of a functionalized prepolymer with an acrylate group for cross-linking, a photoinitiator, and some additives (Figure 16)

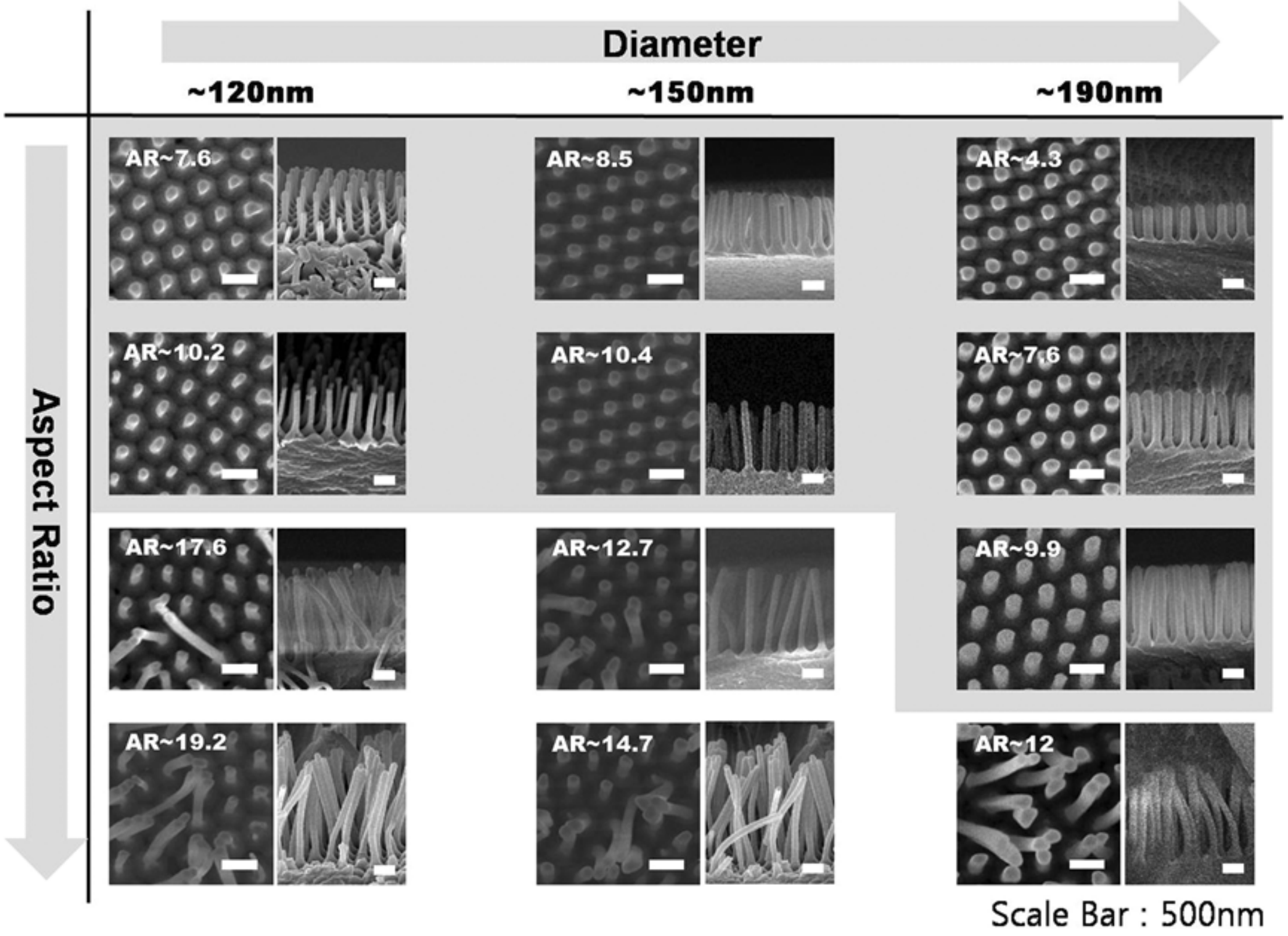

**Figure 16** SEM images of the top and side views of the polymer nanopillars of different diameters and heights with a fixed center-to-center distance (~500 nm). Images on the left and right sides of each panel are top and cross-sectional views of the nanopillars, respectively. The regime in the gray background indicates the optimal dimensional conditions for the manufacture of straight, uniform nanopillar arrays. Reproduced with permission [185]Copyright 2006: American Chemical Society

Although no reported as yet, a second attempt to prepare polyurethane nanofibers has been succesfully developed by a reaction of condensation of an isocyanate and a dietylen glycol in AAO templates (figure 17) [186].

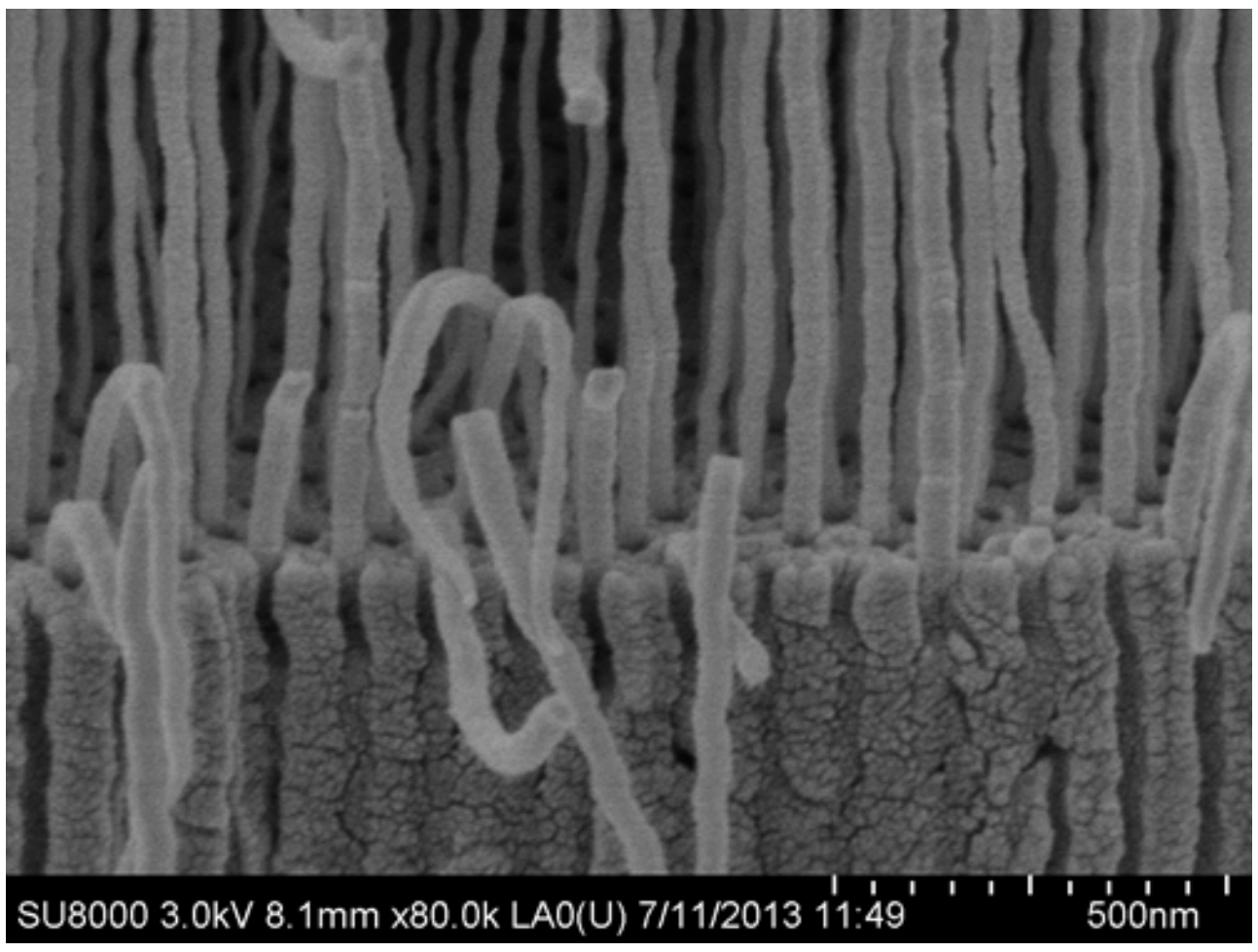

**Figure 17.** SEM image of the side view of polyurethane nanofibers[186].

Concerning the final chemical structure of polymer nanostructures obtained by infiltration in the AAO templates, it is obvious that the stereoregularity, molecular weight, polydispersity, etc are the same as those of the polymer precursor. In contrast, the chemical characteristics of the same polymer nanostructures synthesized from the monomer in AAO templates could be different from those corresponding to polymers obtained in bulk due to confinement effects on the polymerization reaction. To the best of our knowledge, there are no systematic studies in the literature about the effect of the degree of confinement on the chemical characteristics of polymers synthesized in AAO templates. In this case, it is generally accepted that nanoconfinement of free radical polymerization can impact the molecular weight ($M_w$) and molecular weight distribution (DPI) of the polymer produced, with results in the literature generally indicating an increase in molecular weight and a concomitant decrease in polydispersity index [173, 187].

In the case of free radical polymerization of styrene in AAO nanocavities, the polystyrene nanostructures extracted from the AAO templates have been characterized by SEC and NMR spectroscopy, as a function of the percentage of conversion. Up to 50% conversion, no significant differences in the values of $M_w$ or PDI for bulk and confinement polymerization processes were found, which means that the growth radical termination mechanisms is the same in both cases [168]. At high conversion (>65%), it was observed that $M_w$ and PDI values are higher for bulk polymerization than nanopolymerization and the difference is more pronounced as the reaction conversion progresses. This behavior was attributed to the better control of the temperature inside of the nanomold, in comparison to the bulk polymerization, where the heat transfer will occur in a very much higher volume. Concerning the steoregularity of the nanostructured polymer, NMR results showed that nanopolymerized PS has a higher degree of syndiotacticity than bulk polymerized. Although the authors cannot attribute this difference to any particular confinement effect, they speculate if the difference in the diffusion of growing chains in confinement with regard to the bulk and/or nanocavities wall obstruction, could favor a kind of regiorigidity of the chain that would allow a rather stereoregular growing. From these results, it is thus evident the influence of nanoconfinement both on the styrene polymerization and the polymer microstructure, therefore, these changes will be accompanied by the variation in the physical properties of the polymer. Nonetheless, considering that confinement and surface effects may not be necessarily identical to all the monomer and polymerization processes, it seems obvious the need of systematic studies of polymerization in confinement, polymer characterization and modelling.

### *3.2.3 Modeling of polymerization in nanoconfinement*

Very recently, Begum et al. [173, 188] studied the modeling of methyl methacrylate free radical polymerization in borosilicate nanopores finding that the confinement accelerates the reaction rate as a function of the nanopores size. In contrast, a recent study has found that free radical polymerization of perfluorodecyl acrylate (PFA) in AAO nanopores, showed a significant reduction in the rate of polymerization with respect to the reaction carried out in bulk [162]. The authors considered that the situation observed in this case is similar in nature to dispersed phase polymerizations such as miniemulsion and emulsion polymerization, where the concentration of radicals in the dispersed phase varies depending on the volume of monomer droplets/polymer particles dispersed in water [189, 190]. From this assumption a kinetics polymerization model was proposed in which the reaction was confined to a cylindrical space defined by the dimensions of the AAO template used, rather than being confined to droplets within a continuous medium figure 18a. The authors developed a mathematical model to account for the compartmentalization effects in such a case. The model was constructed assuming that the radicals were confined to short cylindrical sections of an infinite cylinder. The confinement of radicals to a volume defined by the radius of the compartment and a given length is a result of the lower effective volume experienced by the radicals in a cylindrical space compared to an infinite 3 dimensional space as in bulk polymerization. The results of the model together with the experimental data and the predicted rate for a bulk polymerization are shown in Figure 18b.

A good agreement between the experimental data and the compartmentalized model was obtained while the corresponding bulk free radical polymerization would proceed at a much faster rate. Thus, the current model accurately predicts that as the nanocompartment diameter is decreased the probability of termination is increased due to a higher local radical concentration resulting in a slower rate of polymerization.

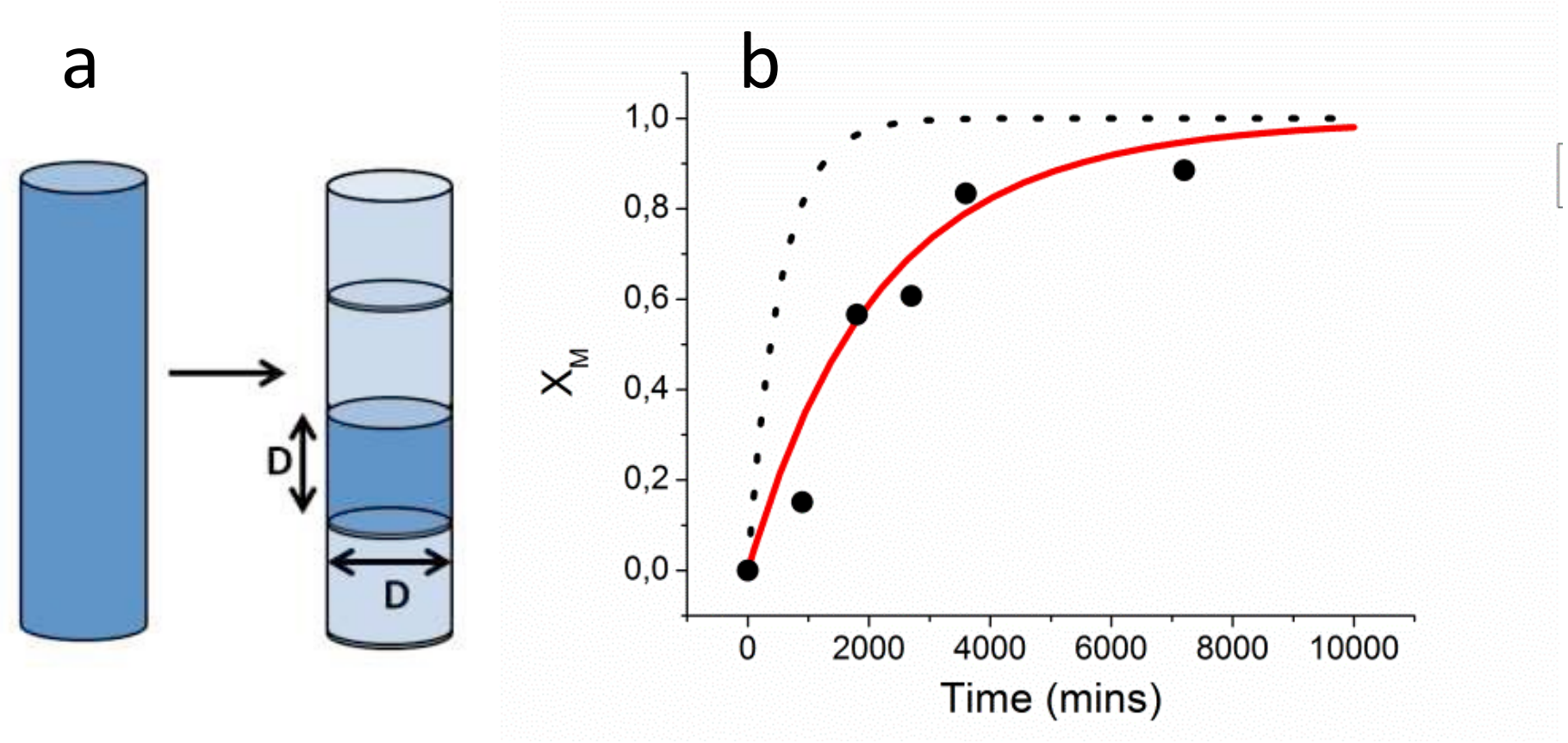

**Figure 18.** Schematic showing the volume of each cylindrical section according to the model (a) and Graph of conversion versus time for polymerization of PFA in AAO template. Symbols represent experimental points while the dotted and solid lines represent the model prediction for bulk and nanoconfined polymerization respectively (b). The rate constants used are reported in Reference[162]

# 4. Confinement effects on the physical processes of polymers

The physical behavior and the properties of nanosized materials often differ from those of their bulk counterparts. Moreover, the size of the material itself becomes one of the variables that determine its physical behavior and, thus, can be employed to tune its properties. The rationale for this is that the size of the material comes into conflict with the characteristic lengthscales associated with a certain physical process. Thus, the physical process takes place in a different manner, leading, eventually, to a new physical behavior of the material.

Such kinds of effects associated to low dimensionality are highly noticeable in polymer materials. Since the polymer molecule itself has nanoscale dimensions, many of physical processes in which the polymer chain is involved present length scales in the range of nm (or tens of nm), hence, are likely to be sensitive to the confinement at the nanoscale. That is the case of structural features such as crystallization process, the conformation of the chain, enthalpy driven molecular self-assembly, liquid-crystal organization, and the like. Likewise, a number of dynamical processes will also be affected by the reduced size, namely segmental dynamics, Rouse dynamics, or chain reptation. Hence, all the materials properties depending on any of the physical processes mentioned will be sensible to nanoscale confinement as well, i.e. optical properties, electrical properties, mechanical properties, etc. Thereby, polymer materials can experience low dimensionaly effects at scales notably larger that those where confinement effect are shown in inorganic materials, as quantum-confinement is not required. This might be an advantage against them when the time comes to design materials with novel properties and functionalities.

## 4.1 Confinement effects on structural features

### *4.1.1 Confinement effects on crystallization*

The crystallization of polymers is frequently altered when the material is confined to nanoscale pores, as both nucleation of crystals and growth of crystals are among processes having lengths scales in the rage of nanometers or tens of nanometers [191]. Several reviews addressing the crystallization of thermoplastics inside the nanopores of AAO templates have been published so far [192-194], including the article by Michell and Müller in this volume. Therefore, in order to avoid any subject overlap we opted for not including such section in our article. Hence, we highly encourage the reading of those articles by Steinhart [194] and Müller [192, 193] (as well as the paper in this volume) for a detailed bibliographic information on this field.

### *4.1.2 Structure of amorphous polymers*

Disordered polymer chains, like those in the molten state or below $T_g$ in amorphous polymers, might be also considered nanosized entities, as typical values of the radius of gyrations, $R_g$, of Gaussian chain conformations range typically from 3 to 30 nm. Therefore, the conformation of disordered chains can also be affected when they are confined into nanoscopic cavities, such as the nanopores of AAO templates. Moreover, the nature of interaction with the confining surface -attractive, neutral or repulsive- may lead also to anisotropic conformations of chains located at the interfaces. Furthermore, these deviations from the isotropic statistic chain as well as polymer-pore wall contact points are expected to influence the entanglement network [149, 195, 196], with subsequent consequences in chain diffusion [197, 198] and rheological properties [199-202], as will be shown below.

In this context AAO templates stand as ideal platform for the study of confinement effects on chain conformation and the entanglement network as: (i) The degree of confinement can be easily modulated from values significantly below $2R_g$ up to values much larger than $2R_g$; (ii) Well defined cylindrical confinement allows for the study of anisotropies; (iii) the nature of polymer-wall interaction can be tuned by chemical functionalization of the hydroxilated pore walls. Nonetheless, these studies are hitherto scarce in the literature [149, 199, 203, 204].

Noirez et al. reported the more detailed study so far on the chain conformation of an amorphous polymer, PS, confined into nanopores of AAO templates [204]. The analysis of the polymer form factor obtained by small angle neutron scattering revealed that the chain extension was similar along the transverse and longitudinal direction to the AAO pore axes (Figure 19). Thus, they concluded that the PS chain remains isotropic in both configurations inside the nanopores. Likewise, Martín and coworkers [149] as well as Lagrené et al. [203] did not detect any influence of confinement on molten PEO chains inside AAO templates.

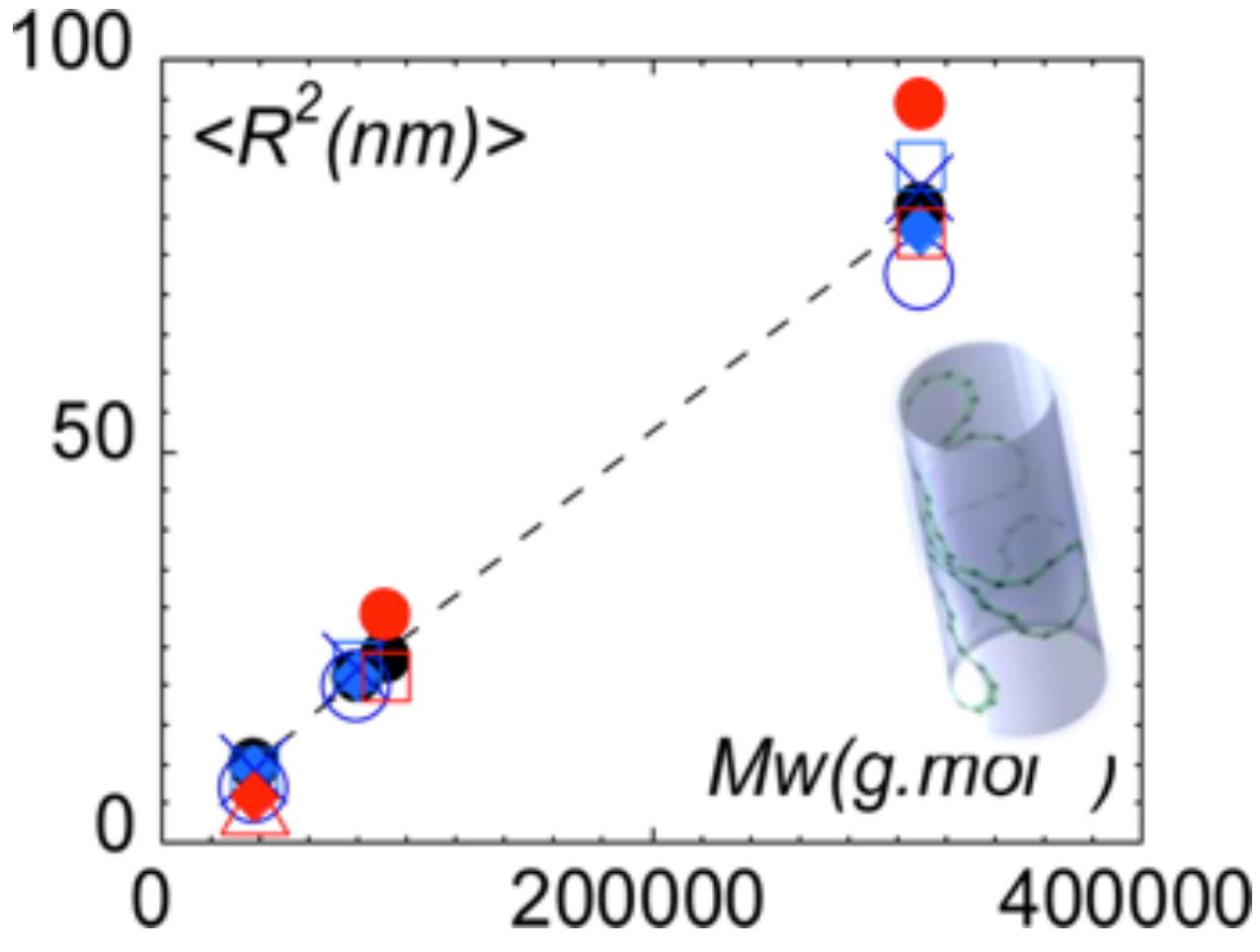

**Figure 19.** Chain dimensions versus AAO pore diameter. The black symbols correspond to bulk to the component of the radius of gyration of bulk polymer, blue symbols correspond to hat in in 180 nm diameter nanotubes and the red ones to components in 35 nm diameter. Reprinted with permission of [204] Copyright (2013) American Chemical Society.

The entanglement network of molten high molecular weight polymers is likely to be more affected by the confinement than chain conformation [149, 199, 205]. Shin et al. reported reduction of the viscosity of PS chains during capillary flow along AAO nanopores with dimensions smaller than that of the chain in bulk [199]. Also, they observed a weak molecular weight dependence of flow. These results were interpreted in terms of a strong reduction of the density of entanglements. Similar result was obtained by Martín et al. by neutron spin echo spectroscopy when analysed PEO melts within strong confinement (dimensions of the chain were 2.4 times larger than lateral dimensions of the pores) [149]. Specifically, a 15 % dilution of the entanglement network was observed. In contrast, Lagrené et al. did not detect any variation by performing similar experiments [203].

### *4.1.3 Supramolecular structures*

It is well known that 2D confinement and the influence of pore walls influence the self-assembly of block copolymers having immiscible blocks and, thus, their microphase structure. The same must be true for liquid crystalline polymers. However, in this review none of these topics will be addressed. In the case of block-copolymers, because Steinhart has recently reported a very complete review article, which includes the description of a number of theoretical and experimental works, as well as general conclusions on the organization of block copolymers within nanopores [194]. Furthermore, reviews on theory of block copolymer self-assembly in confined geometries have been also reported [206]. On the other hand, self-organization of liquid crystalline polymers confined into AAO templates is not reviewed here due to the few reports on the topic, which prevents drawing general conclusions [207].

## 4.2. Confinement effects on dynamical processes

As stated before, finite size effects are considered to originate directly from the hindering of a certain physical process. These effects are due to the fact that the associated length scale of the physical property in question comes into conflict with the dimensions of the material itself. However, the real influence of the size limitation over such processes is often hard to identify and thus has become a matter of controversy among the scientific community [208, 209]. One of the main reasons behind that might be the fact that finite size effects are frequently masked by other effects, which are also connected to low dimensionality but cannot be directly ascribed to the hindering of the physical processes due to reduced space. Among such effects are interfacial interactions between polymer molecules and the confining surface, i.e. substrates, pore walls, etc. Furthermore, confinement-induced changes on the microstructure and packing of polymers may also play a role in its dynamical behavior, as structure and dynamics are strongly correlated [155].

Although size effects have been observed in local dynamical processes [71, 210, 211], they are more noticeable at larger scales, i.e., in the segmental dynamics and in the chain dynamics.

### *4.2.1 Tg, segmental dynamics*

The typical finite size effects on segmental relaxation of polymers have been proposed to be *i)* the broadening of relaxation time distribution function, *ii)* the acceleration or deceleration of dynamics and *iii)* a different dependence of the relaxation time on the temperature. Nevertheless, dimensions of pores in AAO templates are likely to be far from the length-scales associated with the segmental dynamics [212]. Thus these kinds of pure finite size effects are hard to be observed in the AAO-polymer system. Instead, the adsorption of polymer chains on the hydroxilated pore walls of AAO templates seems to play mayor roll on the segmental dynamics –and thus on the $T_g$- of polymers embedded in AAO templates. The dimensions of the AAO nanopores, however, come into conflict with the length scale of crystalline structures, which hence are altered in these pores. Consequently, indirect effects of pore confinement on segmental dynamics through structural changes might be a general feature of this system.

The first investigation on segmental dynamics of polymers confined in well defined, ordered AAO templates was reported by Duran et al [213]. They synthesized the polypeptide poly (γ-bencil-L-glutamate) (PBLG) in AAOs with pore diameters ranging from 25 to 400 nm and the dynamical process was analyzed by dielectric spectroscopy. The segmental dynamics of PBLG located in 200 and 400 nm in diameter pores corresponded to that of bulk PBLG, which is characterized by a "fragile" temperature dependence of the relaxation times. In contrast, they reported a change from "fragile" to "strong" dynamic behavior in PBLG confined in templates with pore diameters equal or smaller than 65 nm. Moreover, they observed a reduction of the effective glass temperature of 50 K. Both were discussed in terms of newly formed hydrogen bonds between the silanol group used for the synthesis and the peptide backbone (Figure 20).

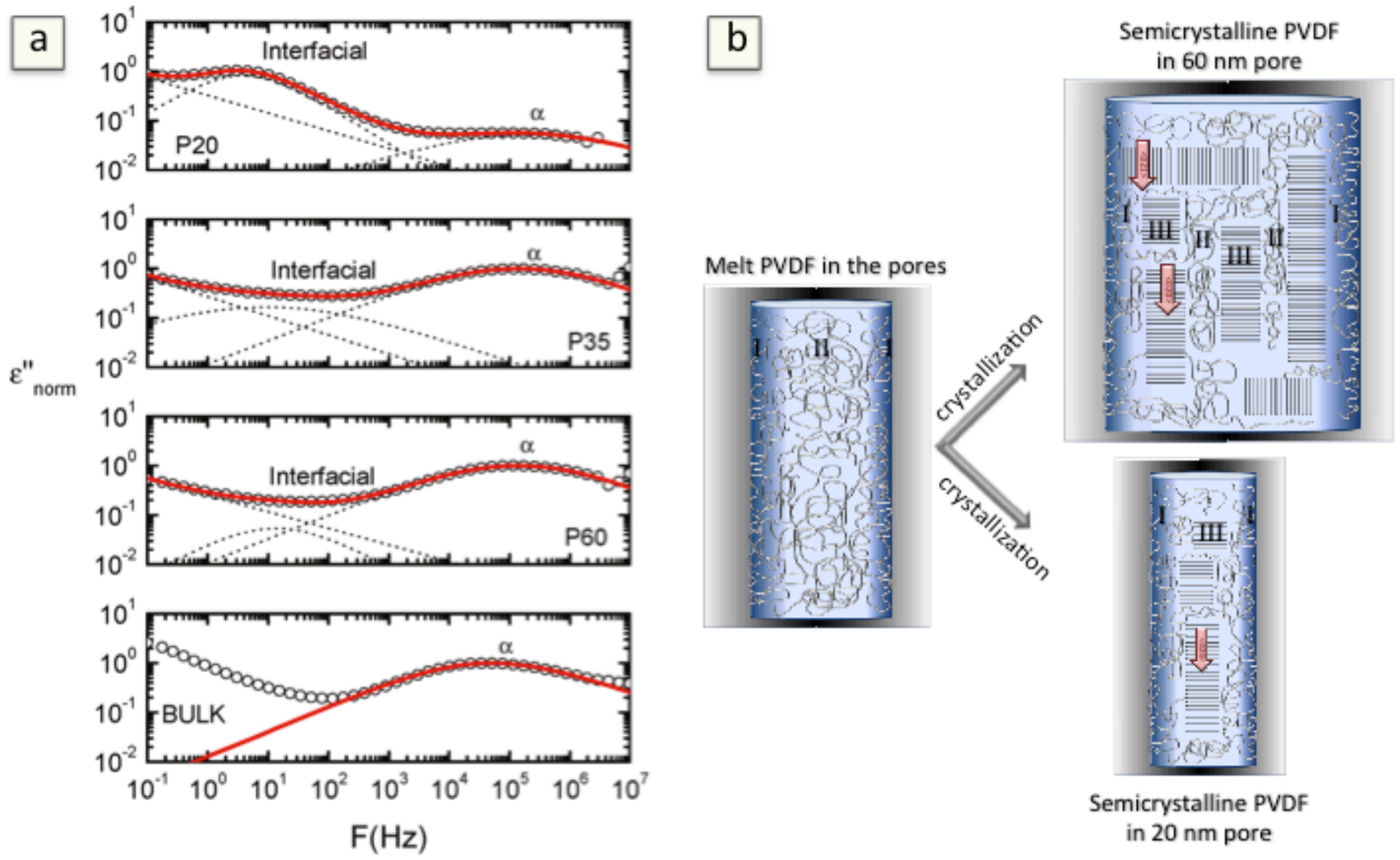

**Figure 20.** (a) ε'' values as a function of frequency for the PVDF confined into AAO templates of different pore diameters. The dotted curves in the P20 (20 nm pore), P35 (35 nm pore), and P60 (60 nm pore) samples indicate separate contributions from the α and interfacial relaxation, respectively. In the bulk samples it is possible to observe also the high-frequency tail of the $\alpha_c$ relaxation [148]. (b) Scheme of the proposed model for layers with different mobility. During infiltration from the melt, an adsorbed layer (I) in contact with the walls is formed. Upon cooling, and depending on the pore size, one may find two situations: in large diameter pores, i.e. 60 nm, there is volume enough to accommodate more than a parallel lamella (III). Therefore, there is an amorphous interlamellar region, which relaxes similar to that of the bulk. Moreover, the amorphous region adsorbed to pore wall relaxes in a particular way, as compared to the bulk. On the contrary, the 20 nm pores have volume to accommodate a single lamella oriented flat on to pore walls (III), the amorphous phase is mainly included in the adsorbed layer (I), and therefore, the main relaxation in this sample is the highly constrained one [100]. Reproduced with permission [148],copyright 2009 American Chemical Society; [100], copyright 2012.

Almost at the same time, Martín and coworkers reported on the complex connection between confinement, microstructure and segmental dynamics in semicrystalline polymers, employing poly(vinylidene fluoride) (PVDF) as model system [148]. The polymer was infiltrated into pores having diameters of 20, 35, and 65 nm and its α relaxation was studied by dielectric spectroscopy. The dielectric spectrum of PVDF confined into 35 and 65 nm pores was similar to that of the bulk PVDF, although the α relaxation was faster at low temperature in the nonoconfined PVDF. This enhancement of the dynamics was claimed to be consequence of the lower crystallinity of the nanoconfined PVDF as compared to that of the bulk. It is well know that the presence of crystals frequently produces effects similar to those of the confinement, namely an increase of the α relaxation time and a broadening of the relaxation distribution function. A most striking effect was observed in PVDF confined into 20 nm in diameter pores, in which the bulk-like α relaxation was replaced by a much slower dynamical process. This highly constrained relaxation was associated with the polymer-alumina interfacial layer and showed a nearly Arrhenius type dependence on temperature, which suggests the loss of cooperativity of the motion.

Serghei et al. employed dielectric spectroscopy to investigate the glassy dynamics of poly-2-vinylpyridine (P2VP) during capillary flow through the nanopores of AAO templates [214]. The used pores presented diameters of 200, 40 and 18 nm. No significant changes on the dynamics of P2VP were found, which suggests the absence of a low mobile interfacial layers close to the AAO pore wall during flow as well as alteration of segmental dynamics due to changes in the chain conformation and packing due to the unidirectional flow.

Suzuki et al. observed the broadening of the α relaxation peak – also that of the β process- for polyethyelene oxide (PEO) confined into AAO templates with pores ranging from 400 to 25 nm [211]. The broadening of relaxation times is indicative of the presence of different environments and interactions of the PEO dipoles at the time-scales of the segmental -and secondary- process. Hence, they speculated about the possibility of slower and faster dynamics due to density modulations in PEO layers adsorbed to the pore walls or due to a distribution of crystalline domains. Similar broadening was observed by the same group for poly(ε-polycaprolactone) (PCL) confined into nanopores with diameters ranging from 60 nm to 25 nm [215]. Furthermore, they observed the slowing down of the segmental mobility in this system. Conversely, in a subsequent report, the same group reported a speed-up of the segmental dynamics of the PCL component of a Poly(ethylene oxide-b-ε-caprolactone) (PEO-b-PCL) block copolymer infiltrated in AAO templates, as compared to bulk PEO-b-PCL [130]. Maiz et al. infiltrated the phase separating polystyrene-block-poly(4-vinylpyridine) (PS-b-P4VP) block copolymer into AAO templates and studied the segmental dynamics of the minority component P4VP [156]. Within 25 and 35 nm in diameter pores, P4VP presented a faster dynamics, more likely related to the relevance of surface effects. Moreover, they showed that the presence of gold nanoparticles in the system enhanced the effect. On the contrary, they measured a slowing down of the dynamics of P4VP block 60 and 330 nm pores, which was attributed to interactions with the pore walls. A similar non-uniform dependence of the relaxation time on the degree of confinement has been reported by Blaszczyk-Lezak et al [71]. Whereas the segmental dynamics of PMMA in pores with diameters of 65 and 35 nm was slower than that of the bulk, the process was faster in 25 nm in diameter pores. They claimed that two opposite effects influence the dynamics of PMMA in confinement: on one hand, attractive interfacial interactions with walls of AAO templates which slow down the chain motions; on the other hand, the confinement effect that speeds-up the dynamics of the PMMA chains. The surface effects would dominate at low pore diameters, while confinement would rule at higher dimensions. The PMMA-AAO system was also selected by Li et al. Interestingly, their calorimetric experiments showed a double glass transition behavior in PMMA confined in AAO nanopores of 80 nm in diameter [216].

One of the $T_g$ was lower than the bulk value, while the other was much higher. Hence, they proposed a two-layer model: polymer chains near the pore walls show the increased $T_g$ due to the strong interfacial interactions, while chains located in the core volume would experience a reduction of the packing density, which would decrease the $T_g$. (Figure 21)

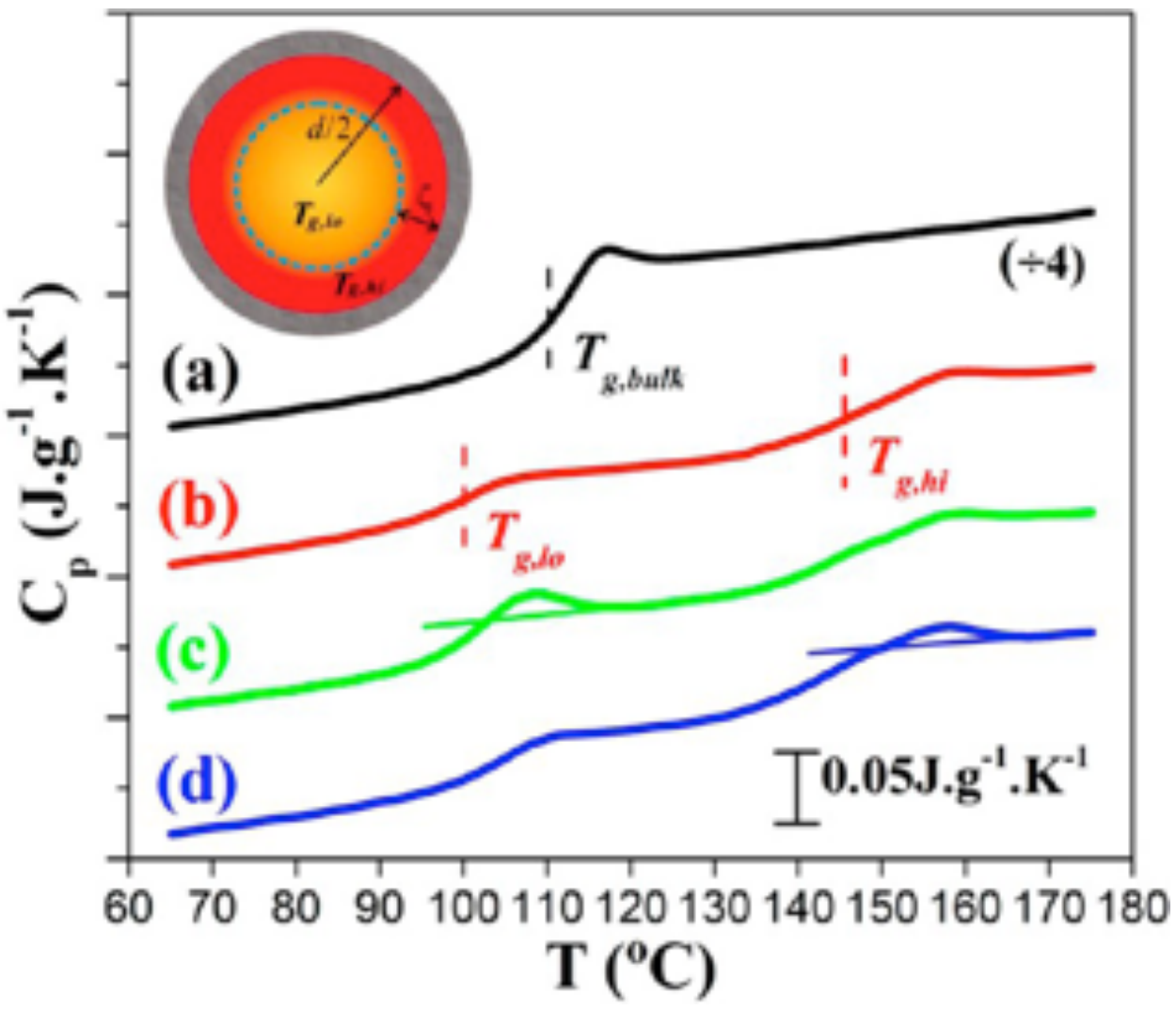

**Figure 21.** Normalized DSC traces of (a) bulk PMMA, (b) PMMA confined in 80 nm AAO template, and PMMA filled AAO samples annealed at temperatures 10 K below the low temperature $T_g$ (c) and 10 K below the high temperature $T_g$ (d) for 2 h. Both, the cooling and the heating rate equal 10 K/min. The dashed lines indicate the $T_g$ values and the solid lines are guides to the eye to identify the changes of the enthalpy relaxation peaks after sub- $T_g$ annealing. The inset illustrates the possible $T_g$ distribution of PMMA confined in AAO nanopores. Reproduced with permission [216], copyright 2013: American Chemical Society.

Lastly, the segmental dynamics of unentangled polyisoprene (PI) confined in nanopores ranging from 25 to 400 nm was studied by Alexandris et al. by means of dielectric spectroscopy [129]. The glass temperature of PI was unaffected by confinement, but a broadening of the distribution of the relaxation times was observed, which was associated with the adsorption of PI chains to the pore wall. Mapesa et al. have reported the bulk-like structural relaxation of nanoconfined PI [217].

### *4.2.2 Chain Dynamics*

Lateral dimensions of nanopores in AAO templates can be comparable to dimensions of polymer chains. Thus, pure finite size effects can take place on dynamical processes at this scale due to changes in the chain conformation, packed or interact either mutally or with the confining cavity. Chain dynamics can be generally described in terms of Rouse and Reptation models. In the Rouse model the chain motion is determined by the balance of entropic and viscous forces as chain relaxation modes (Rouse modes), which are governed by a single relaxation time. Moreover, high molecular weight polymers mutually interpenetrate restricting their motions at large time scales. These topological interactions, named entanglements, are characterized by their distances along the chain. In the Reptation model, the effect of entanglements is described as a tube of diameter along the chain profile, which restricts the chain dynamics.

First investigation on the large-scale dynamics of polymer chains confined in self-ordered AAO templates was performed by Shin et al [199]. Investigating the capillary flow of polystyrene (PS) melts along pores with diameters below 30 nm, they observed a significant enhancement of the chain mobility. Moreover, they observed a weak molecular-weight-dependence of the dynamics in the nanoconfined PS. These results were explained, as stated before, in terms of a reduction of intermolecular entanglements due to confinement.

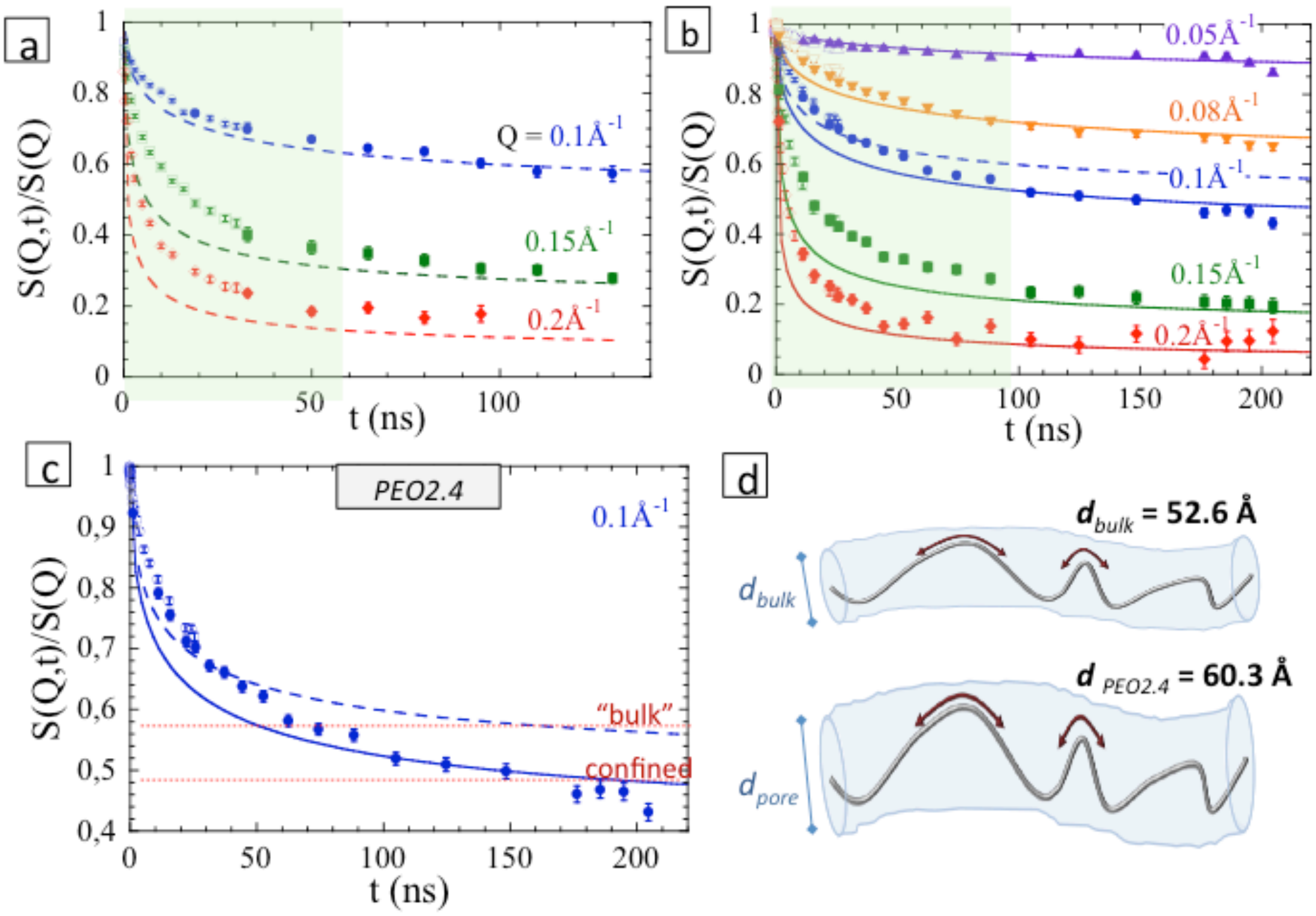

**Figure 22.** Dynamic structure factors of samples slightly confined (a) and strongly confined (b) PEO chains within pores. (c) Normalized dynamic structure factor of the highly confined chains at the $Q$ = 0.1 Å $^{-1}$. Solid line: fits of results of confined PEO with the bulk Rouse frequency value delivering a tube diameter value of 60.3 Å. Dashed lines: de Gennes curves fitting the bulk PEO behavior. (d) Scheme where the enlargement of the tube is represented. Reproduced with permission [100], Copyright 2012

Martín, Krutyeva, et al. came to similar conclusions through a series of studies, in which the dynamics of PEO chains confined in AAO templates with pore diameters of 20 and 35 nm was studied [149, 218]. In their first analysis, carried out by quasielastic neutron scattering [218], they found that the dynamics of the confined polymer is indistinguishable from that in the bulk in the $Q$-range approx. 0.2 Å$^{-1}$< $Q$ < 1 Å$^{-1}$. In this range the incoherent scattering function can be well described by the Rouse theory, and showed no indication of confined dynamics. At the high $Q$-limit, around $Q$ = 1.4 Å$^{-1}$ an anisotropic slowing down of the dynamics was observed, being more pronounced in the direction perpendicular to the pore axis. This effect was attributed to the interactions between the pore walls and polymer segments within a ~1 nm layer. Neutron Spin Echo (NSE) Spectroscopy, however, reveled a clear slowing down of the dynamics at intermediate timescales compared to the bulk behavior for both slightly and strongly confined PEO chains (Figure 22a and 22b) [149]. These experiments suggested a higher impact of confinement on the longer wavelength Rouse modes than on the local scale, in agreement with quasielastic neutron scattering data. One possible explanation for the observed reduction of the intermediate scale dynamics could be some adsorption effect on the surface of the pore walls. But more interestingly, NSE experiments revealed that the confined dynamic structure factor decayed to significantly lower values, which indicates a less restricted dynamics (larger effective tube diameter reflecting a less dense entanglement network) (Figure 22c and 22d). A value of effective tube diameter 60.3 Å for the strongly confined system was thereby measured, which represents an increase of about 15% in the tube diameter compared to bulk PEO and, thus, the microscopic evidence that in the presence of hard confinement the lateral tube size marginally increases. Similar experiments of NSE were carried out by Lagrené et al. on PEO confined in 18 nm in diameter pores [203, 219]. However, their experiments did not show evidences that confinement affects neither the dynamical properties of the polymer nor the diameter of the reputation tube.

NSE spectroscopy was also used by Krutyeva et al. to assess unentangled PDMS melt within 25 nm in diameter pores [152] (Figure 23a and 23b). They found that the dynamic behavior of PDMS was strongly affected by the confinement and followed a two-phase model: one phase consisted of free bulk-like chains and the other phase was conformed by confined polymers. The confined phase is characterized by a vanishing center-of-mass diffusion and by a suppression of long wavelength Rouse modes. It is, hence, ascribed to a multiple anchored, topologically confined chains. According to their model, 75% of the chains are integrated in the confined phase and 25 % are bulk-like

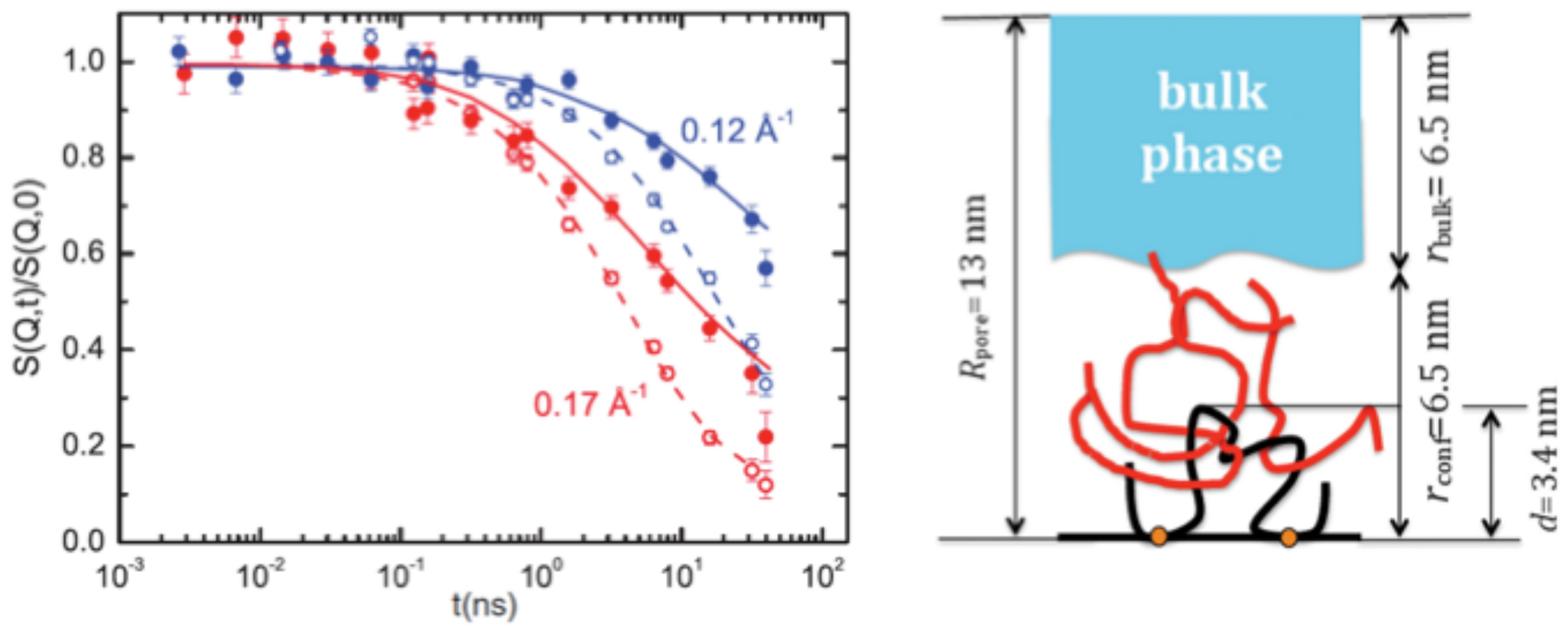

**Figure 23.** (a) SNS-NSE data for bulk PDMS (open symbols) and PDMS confined in AAO nanopores (filled symbols). Dashed lines present the Rouse curves. Solid lines represent the final result (see text). (b) Schematic representation of the artificial surface-induced entanglements in the confined polymer melt. The black line represents the chain adsorbed on the surface of an AAO nanopore, and the red lines show entangled chains in the confined phase. Reproduced with permission [152], Copyright 2013: American Physical Society. APS Journals.

By means of nuclear magnetic resonance (NMR) techniques, Ok et al. concluded that geometric confinement leads to significantly more anisotropic chain fluctuations than predicted by the tube model on time scales beyond the entanglement time. Employing 60 and 20 nm pores, they were able to detect a 3 nm polybutadiene (PB) layer with reduced dynamics. Interestingly, they highlighted the potential drawbacks inherent to the use of disordered AAO membranes in investigations of confined polymer dynamics due to the heterogeneity of the pores in these filters. Likewise, Hoffmann and coworkers used NMR techniques to assess the dynamical behavior of entangled PB chains within 60 20 nm in diameter pores. Their results suggested that whereas glassy dynamics in PB did not show evidences of confinement effects, at chain level, Rouse dynamics slowed down. Thus these results are in agreement with those by Krutyeva, Martín et al [149, 218].

Chain dynamics can also be studied through the analysis of the normal model by dielectric spectroscopy in A-type polymers, i.e. polymers with nonzero dipole moment parallel to the chain contour. Thereby, the normal mode refers to the relaxation of the global chain. Alaxandris et al. studied PI melts with different $M_w$ confined into AAO templates [129]. A remarkable broadening of the distribution of relaxation times of the chain modes revealed that global chain relaxation was severely retarded for nanoconfined PI. This was interpreted in terms of chain adsorptions to the pore wall.

### 4.3. Confinement effects on mechanical properties

Confined dynamics and structures influence mechanical properties of polymer nanostructures. For instance, rigidity of polyanurate one-dimensional nanostructures has shown to be affected by confinement-induced molecular orientation, as demonstrated Duran et al [220]. They observed that precursors of polycyanurate thermosets showed interface-induced anisotropic ordering, characterized by uniform uniaxial orientation with respect to the AAO pore axes, prior to be cross-linked. That uniaxial molecular orientation resulted in a Young's modulus one order of magnitude higher than that reported for bulk polycyanurate, as revealed by atomic force microscopy. Martín et al studied the mechanical response of a semi-free-standing thin film of poly[[9-(1-octylnonyl)-9H-carbazole-2,7-diyl]-2,5-thiophenediyl-2,1,3-benzothiadiazole-4,7-diyl-2,5-thiophenediyl] (PCDTBT) fabricated from AAO templates [221]. Thanks to the free-standing nature of some regions in the film, they manage to characterize the Young modulus of the PCDTBT thin film without the influence of any substrate. This turned to be 30 % lower than that of the bulk polymer and 45 % lower than that of a thin film of the same thickness deposited onto glass. That evidences the fact that real influence of size limitations on the properties of polymer materials is frequently masked by artefacts and interfacial interactions between chains and the confining surface, the substrate in this case. An enhancement of dynamics at the free interfaces due to the free-standing nature of the film, together with the reduction of the crystallinity in the confined materials were proposed to be at the origin of this softening. Furthermore, deep indentation experiments –up to the rupture of the film- enabled them to determine important mechanical features of the nanostructure, such as the bending rigidity (κ = 10−15 J), yield strain (75%) and rupture strain (550%), which were only plausible due to the free-standing configuration (Figure 24).

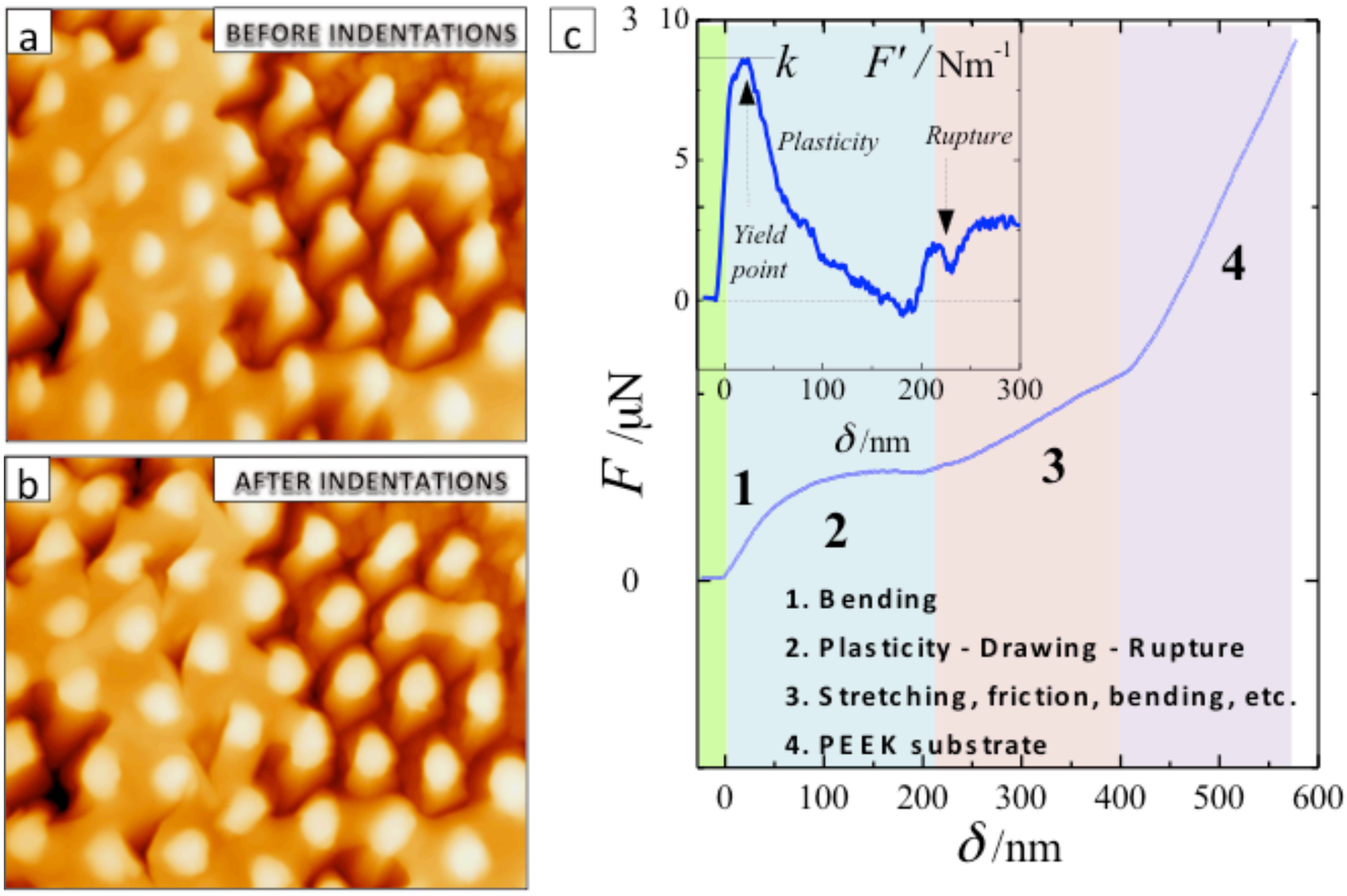

**Figure 24** (a) and (b) AFM images of the same area of the PCDTBT semi-free-standing thin film before and after indentations, respectively, where the damage cause by the indentations in the film can be seen. (c) Representative force indentation curve for the deep indentation experiments. The different membrane deformation mechanisms have been indicated. The inset shows the derivative of the force indentation curve where different relevant points / regimes can be identified. Reproduced with permission [221], Copyright 2013: American Chemical Society.

### 4.4. Confinement effects on thermal properties

Thermal conductivity of polymer of one-dimensional nanostructures has been demonstrated to depend strongly on their structural features [222-224]. All papers addressing the thermal conductivity of polymers confined in AAO templates have hitherto dealt with semicrystalline polymers. It is well known that crystals show intrinsically higher conductivity than amorphous regions, in such a way that thermal conductivity of semicrystalline polymers is usually higher than that of amorphous polymers. Thermal conductivity of semicrystalline polymers is known to depend on both the degree of crystallinity and the orientation of their structural elements, i.e. molecules, aggregates, crystals, etc. [225] Such crystal orientation, moreover, leads to a large anisotropy in the thermal transport of these materials, which can be commonly understood considering that phonons can propagate more efficiently along certain crystallographic directions that offer lower thermal resistance.

Cao et al. reported on the thermal conductivity of PE nanowires of 100 and 200 nm in nominal diameters measured by the laser flash technique [222, 224]. They estimated thermal conductivity values for single PE nanowires of 26.5 W/mK, which means an increase of 2 orders of magnitude with respect to bulk PE. They claimed that the increase is due to chain alignment in the nanowires. However, the measured samples consisted of collapsed bundles of nanowires and thus, these measurement might be influenced by the different environments experienced by the nanowires at interior positions of the bunch, from those at external positions, free nanowires, etc.

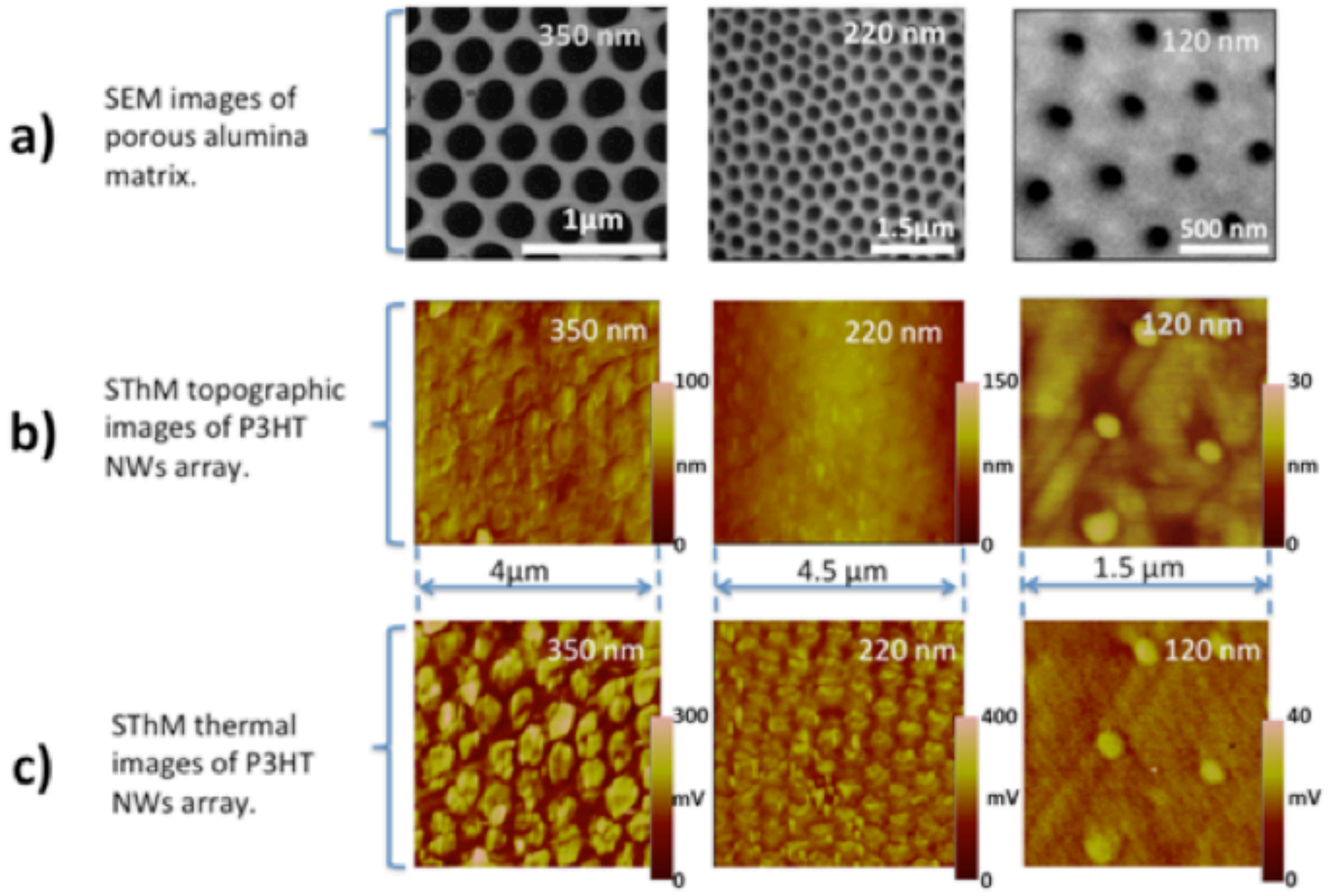

**Figure 25** (a) SEM pictures of the three different diameter size porous alumina matrices used to embed P3HT nanowires (b) topography of the filled templates and (c) (thermal images of P3HT nanowires taken using 3ω-SThM. Reproduced with permission [223], copyright 2014.

In order to be able to understand the thermal behavior of nanowires, thus, low dimensionality effects on polymer materials, Muñoz et al. measured the thermal transport in P3HT nanowires under well controlled boundary conditions, i.e., embedded into the AAO templates (Figure 25) [223]. By means of a complex scanning thermal microscopy operated in 3ω configuration, they were able to observe a reduction of the thermal conductivity, from 2.29±0.15W/K·m to 0.5±0.24W/K·m, when the diameter of nanowires is reduced from 350 nm to 120 nm. According to WAXS measurements, such reduction was proposed to be consequence of an increasing presence of crystals oriented laying the [100] direction parallel to nanowire long axis, as decreasing the diameter of nanopores. Their argument was based on the fact that both the extended polymer chain direction ($c$ axis) and the π-π stacking direction ($b$ axis) in the P3HT crystal are expected to present little thermal resistance, because strong conjugated covalent bonds along the chain direction ([001] direction) and the compact π-π stacking (along the [010] direction) would facilitate the phonon transport along those crystallographic directions. In contrast, the [100] crystallographic direction is the one along which the alternation of layers of thiophenic backbone chains and aliphatic side chains take place. Thereby, insulating aliphatic regions separate the more conductive thiophenic layers, which may introduce additional thermal boundary resistances in the crystal structure along that direction, decreasing, thus, the thermal conductivity.

**4.5. Confinement effects on rheological properties**

The flow behavior of nanoconfined polymers is expected to be sensitive to the entanglement network of the molten polymer as well as to the capability of the polymer chain to diffuse along the nanopores. High surface-energy cylindrical nanopores of AAO templates allows for the assessment of the flow of confined polymer melts during capillary rise. The time required to fill the nanopores a polymer melt by capillary action can be coarsely obtained by [158, 226]:

$$t = 2\eta z^2 / (R\, \gamma_{olymer/air} \cos\theta) \qquad (1)$$

Where $t$ is the time, η is the viscosity, $z$ is the height of the liquid column, $R$ is the hydraulic radius (the cross sectional area of a stream divided by the wetting perimeter, which for well-defined cylindrical pores is equal to the half of the pore radius), $\gamma_{polymer/air}$ is the surface tension of the polymer, and $\theta$ is the contact angle at the polymer-pore wall interface.

The equation above was employed by Xiang et al. to estimate the time required by a poly (styrene-b-butadiene) (PS-b-PB) copolymer melts to fill the nanopores [58, 227].

Using the viscosity value of bulk PS-b-PB, they demonstrated that the calculated time was in remarkably good agreement with the time used experimentally.

The same method was employed by Martín et al. to obtain the viscosity value of P3HT melt as flowing through 350 nm in diameter pores [215]. Although a notable low viscosity was reported for the polymer in the nanochannels in that communication, no influence of reduced dimensions was detected. Apparently, the low dimensionality effects on large-scale dynamics, such as diffusion of chains, are exclusively observed within nanocavities in the range the radius of gyration of the polymer, as reported by Shin et al. [228]. They investigated the capillary flow of PS melts along pores of 15 nm and observed a significant enhancement of the chain diffusion, as well as a weak molecular-weight-dependence of the dynamics. These results were explained in terms of a dilution of the entanglement network in highly confined polymer melts. Likewise, Hu et al. also found a decrease of the viscosity of PE in their flow through pores with relatively high distribution of diameters [216].

# 5. Modulated properties of nanostructured polymers from AAO templates: Examples of applications.

The low dimensionality effects on the structure of polymers confined in nanopores, i.e. preferential orientations of crystals, suppression of crystallization, dilution of the entanglement network, etc., as well as effects on dynamical processess, i.e. shifts of $T_g$, enhanced chain diffusion, etc., have noticeable impact on the properties of the materials they integrate. The properties of the polymer nanostructures are also very dependent on the morphology, for example, gecko lizards, bio-inspired, icephobic surfaces, nanomotors, microelectronics, optical devices, plasmonic antennas, nanoimprinting, all of them are terms related to nano or micro aspects of a polymer surface. Accordingly, research efforts have been focused on developing "patterned" surfaces by different approaches, depending on the application pursued. This kind of "patterned" surfaces can be provided by hierarchical one-dimensional, concave, tubular, spherical or branched nanostructures, etc, obtained by templating, microlithography, or a combination of them, and an appropriate selection of polymer material or composite. An obstacle to the wide-spread use of hierarchical nanostructures is the difficulty to control and determine the surface properties of the structures because of the lack of appropriate characterization techniques.

However, in the last few years, many examples of *nanostructures with modulated polymer properties for potential applications* have been reported in the literature. In the following a selection of different patterned nanostructures that stress the potential interest in a particular application is reported

## 5.1. Biorelated properties

### *5.1.1. Biosensors*

Nanoporous alumina membranes are employed in the construction of electrochemical biosensors that show a good operational stability [229, 230]. They provide an available relatively high surface area for the retention of enzymes or related bioactive compounds which is important in order to enhance the sensitivity of the biosensor and facilitate the diffusion of electroactive species towards the electrode surface allowing the development of biosensors that show a good operational stability.

Enzyme molecules are entrapped in the membrane nanopores, retaining their biocatalytic activity, and then the modified membranes are attached to the electrode surface. Good results have been obtained for the immobilization of glucose oxidase in a set of home-made alumina membranes of controlled pore size and thickness. An external coverage with a thin layer of the biopolymer chitosan avoided enzyme leaching thus enhancing enzyme stability. The enzyme-modified membranes have been attached to the surface of a platinum electrode for the biosensor construction (figure 26). The biosensor response is directly related to the membrane dimensions and, in addition, the enzymatic response is linear in a wide concentration range up to 20 mM as required for some clinical applications, i.e. determination of blood glucose levels[231]. Other clinical applications developed for enzyme biosensors immobilized in AAO templates include urea determination in urine sample[232]. Chemiluminescent biosensors have also been fabricated through physical immobilization of alkaline phosphatase via electrostatic layer-by-layer assembly to nylon nanorods synthesized by AAO membrane templating which significantly influences the catalytic activity of the enzyme [233].

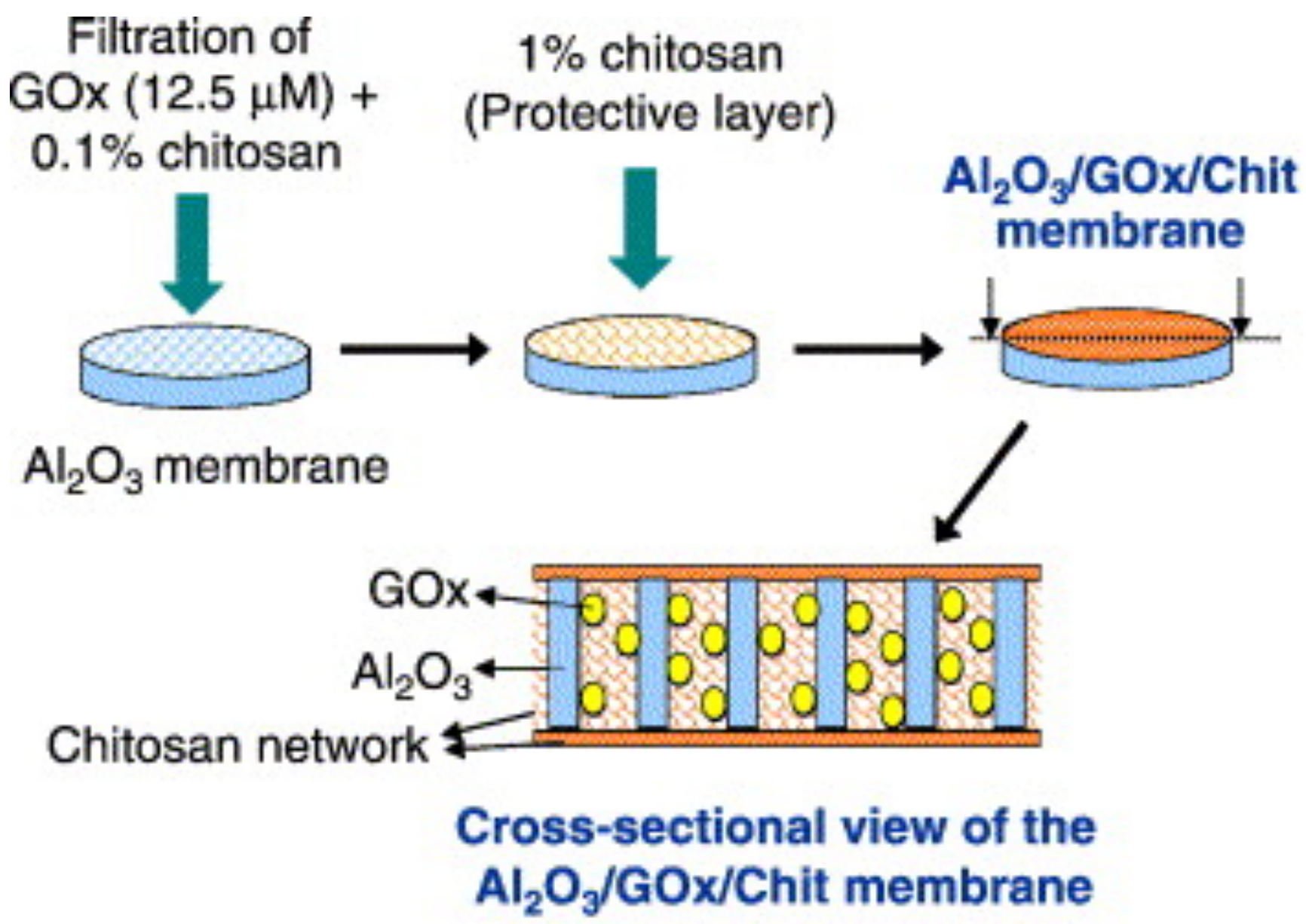

**Figure 26.** Schematic representation of the process of immobilization of glucose oxidase in nanoporous alumina membranes. Reproduced with permission[231]. Copyright 2006: Elsevier.

*5.1.2. Polymer scaffolds for cell adhesion*

In numerous studies, cells have been reported to be able to sense micro and nanoscale geometric cues from their environments through the phenomenon of contact guidance. The interactions between the cells and the topographically different substrates lead to specific cell responses and result in different cell functions[234, 235]. With the development of nano- and micro-engineering technologies, reconstructing 3D physical microenvironment *in vitro* with a spatiotemporal control becomes feasible[236]. In particular, the fabrication of polymer scaffolds by means of AAO templates combines advantages of top-down lithography (well-defined topography) and self-assembly (low costs, high throughput, feature size in the 100-nm range) and yield ordered arrays of polymer nanorods. These arrays of polymer nanorods act as three-dimensional scaffolds for cell growth, as obvious from the massive emission of filopodia contacting the tips of the rods, forming eventually dense cell layers. For example, poly-DL-lactide (PDLLA) rod arrays were fabricated in AAO templates with a pore diameter of 180 nm, a pore depth of 600 nm or 1 μm and a lattice constant of 500 nm. It was proven than depending on the array design, cells recognize the rod arrays either as surface providing rods as cues sensed by filopodia or as a quasi-three dimensional scaffold[237].

In another example, poly(ε-caprolactone) (PCL) hierarchical patterned nanowires with highly ordered nano-microscaled topography was developed with a double template method involving the employment of AAO membranes as a first template and an Al grid as the second template[238]. The results revealed that the hierarchical patterned PCL nanowires showed higher capability of protein adsorption and better cell growth than the PCL film with smooth surface.

### 5.2. Surface Properties. Applications as highly adhesive and hydrophobic surfaces

The fabrication of arrays of straight polymeric nanorods with a high aspect ratio constitutes a biomimetic approach for the preparation of highly adhesive surfaces [239-243]. This is based on the fact that these nanostructured surfaces can mimic gecko's foot-hair and provide a similar adhesion [244-246]. Gecko toes present arrays of millions of fine hairs with a high aspect ratio (setae), which split into submicron or nanoscale fine termini (spatulae), and these hairs enable geckos to make extensive contact with almost any surface by van der Waals and possibly capillary forces[247, 248]. Arrays of branched contact elements have been fabricated by replication of AAO with hierarchical pore morphologies in order to replicate as close as possible the topographic features found in geckos[249, 250]. The tip shape in hairy polymer arrays also modulates the adhesion properties of the material. By simple mechanical shaping, flat, foot-like, and pancake-like tip shapes roughly corresponding to three main types of biological contact shapes were accessible. The normal dry adhesion was enhanced in the polymer array with pancake-like tips as can be observed in figure 27.

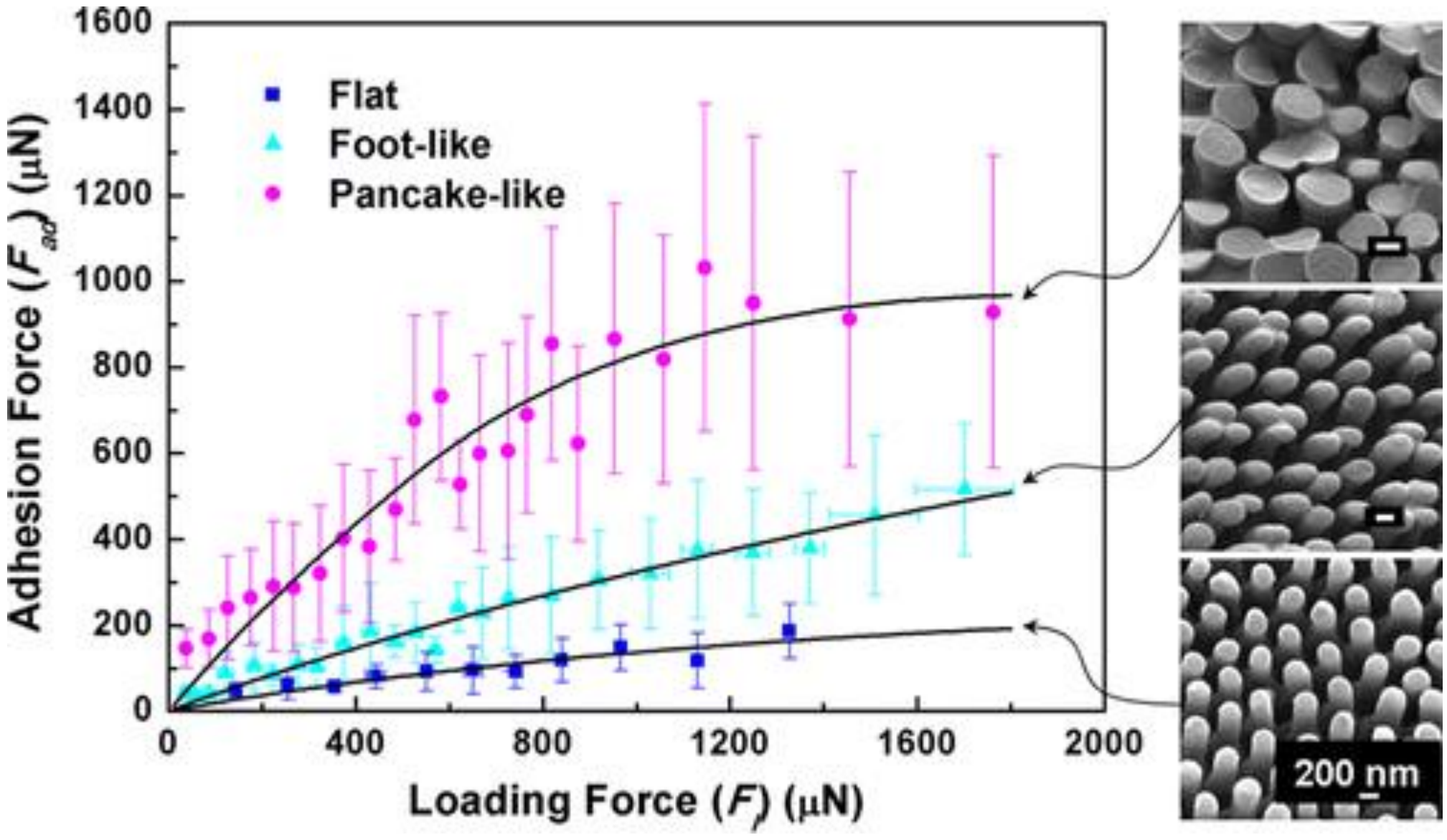

**Figure 27.** Dependence of $F_{ad}$ on $F_l$ for PS nanorod arrays with flat tips obtained by cold pressing, foot like tips obtained by shear pressing, and pancake-like tips obtained by hot pressing. Reproduced with permission [243] Copyright 2012: American Chemical Society.

The properties of adhesion can be modulated by different stimuli., i.e. chemical functionality and topography [251], pressure changes[252] or even humidity. As an example, adhesion on arrays of porous fibrillar adhesive pads consisting of the amphiphilic block copolymer PS-*b*-P2VP can be reversibly switched between low-adhesion states at low humidity and high-adhesion states at high humidity[128].

Polymers with the same chemical composition can provide different properties to the surface simply by altering the nanostructure [240, 245, 253]. In particular, water repellence or hydrophobicity of a surface is an important characteristic which depends not only on the chemical nature of the polymer material but also on the micro and nanoscale surface roughness[254]. In rough morphologies, air pockets can be trapped between the surface and water droplets and, therefore, highly hydrophobic morphologies can be achieved, being reflected directly from the contact angle measurements. The replication of the nanoporous of AAO templates with a hydrophobic polymer is a good strategy to induce the roughness on the polymer surfaces thus increasing the amount of air cubicles trapped between the surface and water droplet. With this idea in mind, Salsamendi et al [162] reported the preparation of superhydrophobic nanostructured polymer surfaces from a PFA infiltrated in the nanocavities of AAO templates of different porous diameter 150, 250 and 300nm and same length of 1.5 μm. They studied the effect of the AAO nanocavities pore size/length on the hydrophobicity of the PFA surfaces, obtained after removing the AAO templates, by contact angle measurements. While the contact angle for bulk PFA surface was 114º, for nanostructured PFA is 159 º, much higher than its analogous non-nanostructured PFA. Moreover, in the superhydrohobic nanostructured surface, the "lotus effect" was observed which has a low sliding angle of 8 º. So, results demonstrated the increase in hydrophobicity of the same starting material after nanostructuration by using AAO templates

### 5. 3. Optical properties. Applications in optical devices

Optical properties of polymer nanowires and nanotubes are highly influenced by the molecular arrangement in the nanowires or nanotubes. This is especially noticeable for conjugated polymers, as demonstrated by O'Carrol, Moynihan et al. in a series of papers [255-261]. They found that *c*-axis of PFO crystal (parallel to chains) was preferentially alingned parallel to nanopore long axis when confined to nanowires prepared from wetting disordered AAO membranes by PFO melts [259], which led to the anisotropic emission of the nanowires. Same authors managed to produce PFO nanowires and nanotubes in the β-phase by infiltrating solutions into the same nanopores. The β-phase exhibits the largest intra-chain correlation and conjugation length of all the phases of PFO [262], thus exhibiting enhanced optical properties, such as a reduction of the threshold for observation of amplified spontaneous emission [263]. Hence, they managed to use such confinement-induced polymorph and orientation for achieving nanostructured devices with notable electroluminescence (Figure 28a and 28b) [258, 261] and lasing properties [256]. Furthermore, Liu et al. investigated the photoluminescence of poly[3-(2-methoxyphenyl)tiophene] (PMP-Th) nanowires within AAO nanopores and observed a blue-shift in the spectra and a notable enhancement in the intensity [264]. Such enhancement was attributed to Föster energy transfer from oxigen vacancies in the AAO to PMP-Th.

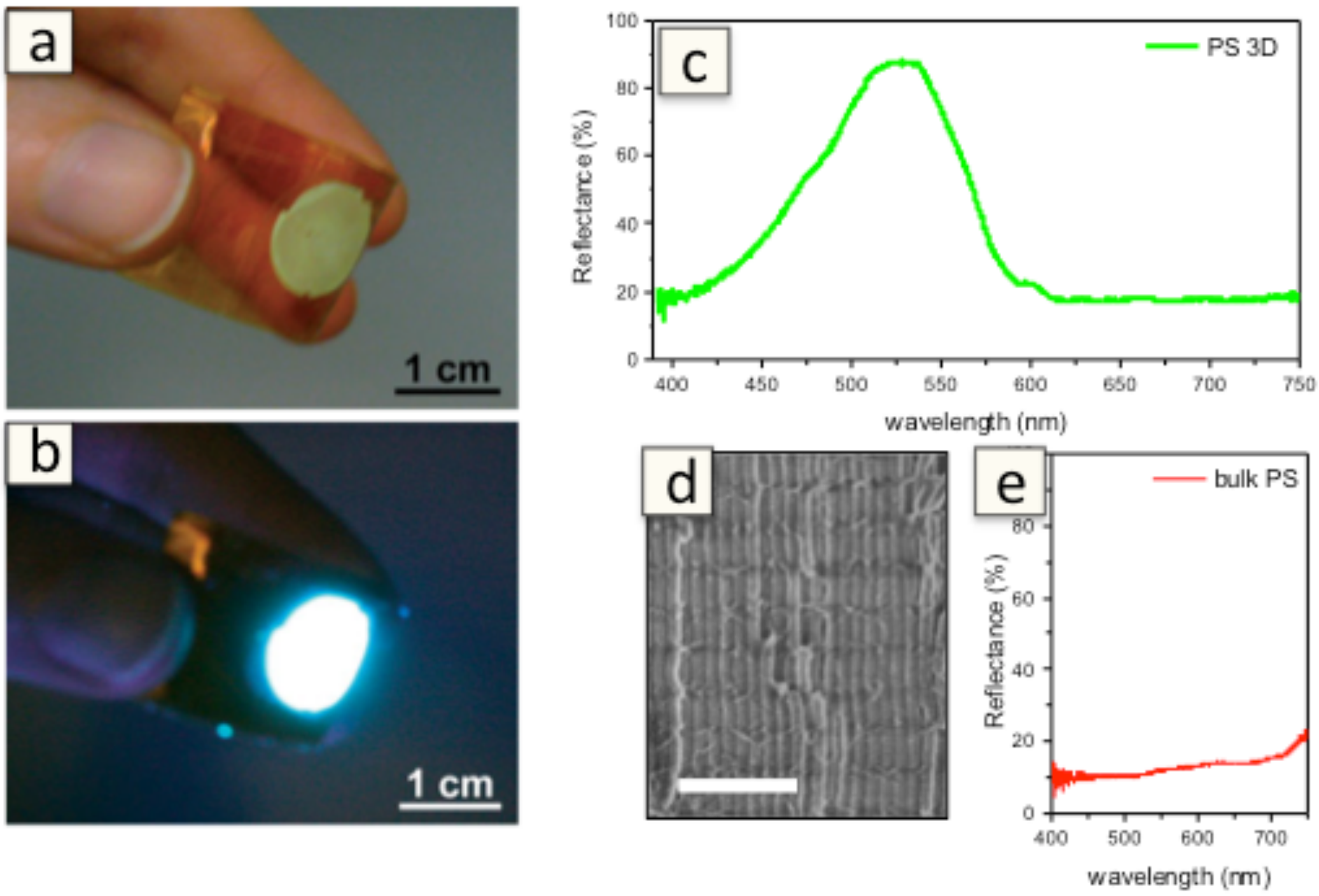

**Figure 28.** (a) and (b) Photographs of a PFO nanowire array mounted onto flexible tape under white light illumination and in darkness under UV excitation, respectively [258]. (c) Reflectance spectra of the 3D PS nanowire network shown in (d). (e) Reflectance spectrum of a non-porous PS (named as „bulk PS") [90]. Reproduced with permission [258] Copyright 2008: John Wiley & Sons; [90], Copyright 2014: Nature Publishing Group.

Martín et al. evaluated the optical properties of a 3D PS nanowire network by means of UVvis spectroscopy [90]. Such 3D PS nanostructure was prepared by the replication of a 3D AAO template (Figure 28d). Figure 28c shows the reflectance spectrum of the PS 3D nanowire network, where an intense optical bad gap at ≈ 525 nm can be observed as a consequence of the periodic alternation of layers with different refractive index. This reflection band at ≈ 525 nm was responsible of the strong green color of that PS piece - in contrast to bulk PS, which is colorless (Figure 28e) demonstrating the possibility of producing distributed Bragg reflectors (one-dimensional photonic crystals) from "simple" and cheap commodity plastics as PS.

### 5.4. Electrical functionalities for potential applications

Electrical conductivity of conjugated polymers also changes when the material is embedded into nanopores due to structural changes taking place in the nanoconfined material. Thus, chain orientation phenomena induced by interactions with AAO surfaces, or by the space reduction itself, may induce strong enhancement of electrical conductivity. This fact was first demonstrated by Martin et al, who managed to polymerize different conjugated polymers, such as polypyrrole, polyaniline, etc., into the nanopores of disordered AAO membranes [265, 266].

In the case of the more recent, soluble, semicrystalline conjugated polymers, electrons or holes are known to be transported differently along each crystallographic orientation. Thus, the conductivity of these materials is very sensible to crystalline features, i.e crystal orientation, size, crystallinity, and the like. In an analogous way to thermal transport, the chain direction and the π-π stacking direction in the crystals are expected to present little electrical resistance. This agrees with results reported by Byun et al. on P3HT nanotubes prepared by solution wetting [267]. Electrical conductivity of these nanotubes was measured by a two-point probe method and found to be about 10 times higher than that of a continuous P3HT film due to the fact that nanotubes showed uniaxial crystal orientation with the π-π stacking direction parallel to the longitudinal axis of the nanotubes. Moreover, Hu et al. prepared 80 nm in diameter P3HT nanowires doped with 2,3,5,6-tetrafluoro-7,7,8,8-tertacyanoquinodimethane (F4-TCNQ) [268]. They measured the conductivity of individual nanowires by four probe scanning tunneling microscopy and showed that they had superior conductivity than thin films (Figure 29). Huang et al. have recently reported the electrical characterization of P3HT nanotubes of around 200 nm diameters by melt wetting [269]. Their field effect transistor devices based on single P3HT nanotubes showed carrier mobility of 0.14 ± 0.02 $cm^2/V$.

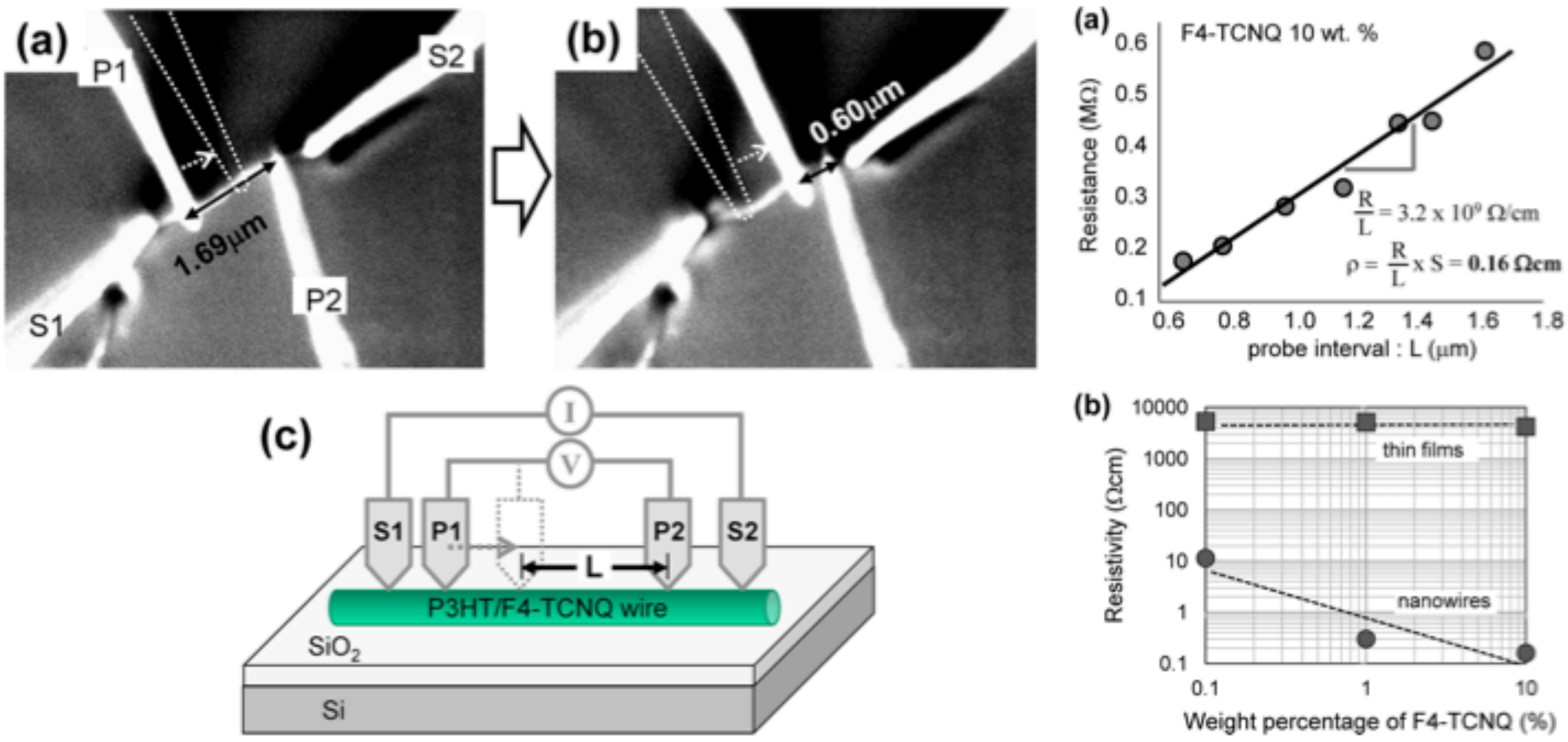

**Figure 29.** (a, b) SEM images of the four-probe STM measurements. The outer pair of probes (S1 and S2) is fixed on both sides of the wire to provide electrical current. The inner pair (P1 and P2) monitors the voltage drop at each interval, where the position of P1 is varied from 1.69 to 0.60 μm while P2 remains fixed at a certain position. (c) Schematic of the measurement procedure. (a) Interval−resistance line obtained from the P3HT/F4-TCNQ nanowire with 10 wt % F4-TCNQ. The slope of the line (R/ L) together with the cross-section of the nanowire (S ≈ 5 Å~ 103 nm2) provides a resistivity (ρ). (b) F4-TCNQ concentration dependence of resistivities for composite nanowires (circles) and thin films (squares). Reproduced with permission [268], Copyright 2013, Americal Chemical Society.

Joo et al. synthesized nanotubes and nanowires of polypirrole, poly(3,4-ethylenedioxythiophene) (PEDOT) and polyaniline by using AAO templates through electrochemical polymerization and observed some differences with respect they bulk counterparts [270]. Shirai et al. also synthesized PEDOT nanowires using AAO membranes as templates and reported that their conductivity and carrier mobility was dependent on the a nanowire size [271]. They found that lower diameter nanowires, having 50 nm in diameter, showed a carrier mobility of 2.0 $cm^2/V$ and proposed to be consequence of a higher conjugation length of the polymer in the nanowires as compared to bulk PEDOT. Moreover, PEDOT nanotubes were synthesized by vapor deposition polymerization inside nanopores of AAO templates [178], which presented conductivity values of 2,000 S/cm, as measured by conductive scanning probe microscopy.

*5.4.1. Ferroelectric properties*

Ferroelectric (and piezoelectric) properties of polymers are expected to be also influenced by confinement as they have their origin in the dipoles of the crystal. The most important ferroelectric polymers is poly (vynilidene-co-trifluoroethylene) (P(VDF-TrFE)). It has multiple polymorphs, being the β-phase the most polar phase, as chains of all-trans (*tttt*) conformation pack with dipoles parallel to a common axis in a pseudohexagonal configuration. The highly ordered and aligned dipoles in the β-phase give rise to strong ferroelectric and piezoelectric behavior.

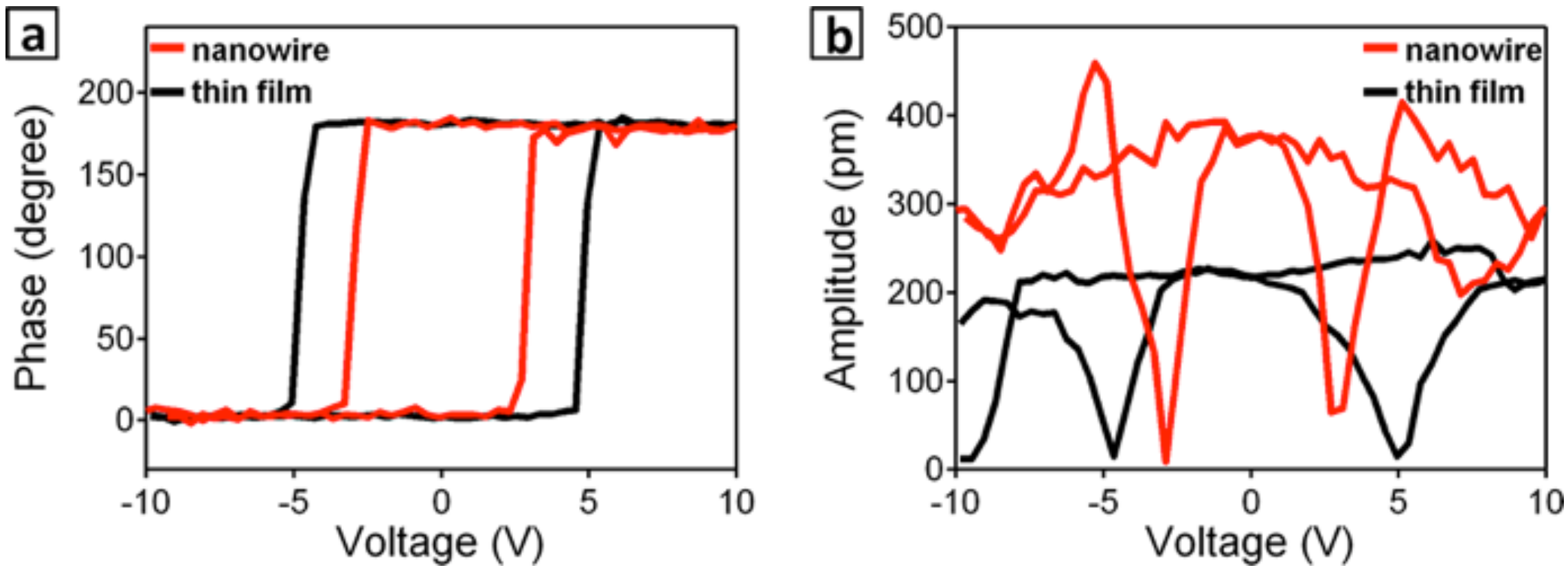

**Figure 30.** Piezoresponse force microscopy (a) phase-voltage and (b) amplitude-voltage hysteresis loops in the "off" state of a P(VDF-TrFE) nanowire of 60 nm diameter, compared to the ones of a uniform thin film of 60 nm thickness. Reproduced with permission [272], Copyright 2013 American Chemical Society.

Lutkenhaus et al. studied P(VDF-TrFE) confined in AAO templates of different diameter pores [273]. They observed that crystallization of the polymer in the β-phase was enhanced and the non-ferroelectric phase was suppressed. The Curie transition, in contrast, was only slightly affected. Upon heating, the Curie transition coincided with melting; upon cooling, the Curie transition region narrowed. Moreover, a second broad Curie transition near room temperature was observed for P(VDF-TrFE) confined within 15 nm pores. Shigne et al. also observed that ferroelectricity was retained in P(DF-TrFE) confined in nanopores and that the paraelectric to ferroelectric transition was not much affected by confinement [127]. Choi et al. studied how the ferroelectric loop was affected by the transition from solid nanorods to hollow nanotubes as increasing the pore diameter diameter in the AAO templates [274]. They observed that the polarization increased rapidly as increasing the nanotube diameter from 38 to 65 nm, as crystal size and crystallinity in nanorods increased with the diameter, which resulted in an enhancement of the ferroelectric properties. With a further increase to 72 nm, the polarization value decreased abruptly and maintained a relatively low value for all samples with diameters greater than 72 nm. This was associated with the formation of P(VDF-TrFE) hollow nanotubes of reduced crystallinity or crystal quality, compared to the nanorods. Furthermore, Wu et al. prepared P(VDF-TrFE) nanowires by solution wetting and observed that preferential orientation of β-phase crystals translated into improved ferroelectric and piezoelectric properties, such as lower coercive field and increased piezoelectric coefficient (figure 30) [272].

## 6. Conclusions

This review summarizes the hundreds of works reported in the literature to fabricate nanostructured materials from polymers and polymer-based composites, prepared by template synthesis method using anodic aluminum oxide templates (AAO) as nanomold and nanoreactor. From them, the following conclusions can be extracted:

The design of AAO templates with suitable adjusted porous dimensions can be easily achieved, by a two anodisation process, starting from ultrahigh aluminum laminates. AAO templates present great homogeneity and offer high flexibility regarding pore diameters and lengths. Moreover, more complex geometries are also feasible

By an adequate selection of the polymer infiltration process in the AAO template, it is possible to prepare a huge variety of polymers and polymer-based composites of well-defined shape and dimensions, from nanorods to nanotubes, nanospheres, nanocapsules or more complex nanostructures as peapod-like morphologies. The "in-situ" polymerization of monomers allows the fabrication of nanostructures from thermoplastic to thermoset polymers in one step

The study of relevant physical properties of different polymers confined in the nanocavities of AAO templates reveal, as a general conclusion, that local and segmental dynamics are affected in a great number of polymers, the same conclusions can be inferred from the study of thermal and mechanical properties.

The fabrication strategy followed through AAO templates is also highly appropriate for the preparation of different types of polymer nanostructures and morphologies to be employed as scaffolds for cell adhesion, in surface properties related applications, optical and electrical applications.

From the present review it can also be concluded that there are still new challenges in the field of nanotechnology, through the development of polymer miniaturized devices, nanostructured morphologies and "functional" polymer hybrid nanostructures, etc . Moreover, the study of "in-situ" polymerization in AAO templates has already started and, in a near future, many other thermoset polymer nanostructures will be fabricated.

## ACKNOWLEDGMENTS

Authors are gratefully acknowledged to the Ministerio de Economía y Competitividad for their financial support (MAT2011-24797). Rebeca Hernández thanks MEC for a Ramon y Cajal contract.

## REFERENCES

[1] Tran HD, Li D, Kaner RB. One-Dimensional Conducting Polymer Nanostructures: Bulk Synthesis and Applications. Advanced Materials. 2009;21:1487-99.

[2] Sheng Zhao Y, Yao J. Organic one-dimensional nanostructures: construction and optoelectronic properties. New Jersey: Wiley and sons; 2013.

[3] Sun Y, Dong W, Wang H, Huang Y, Gu H, Yang VC, et al. Template synthesis of PMAA@chitosan hollow nanorods for docetaxel delivery. Polymer Chemistry. 2013;4:2489-95.

[4] Lee Y, Park SH, Kim KB, Lee JK. Fabrication of Hierarchical Structures on a Polymer Surface to Mimic Natural Superhydrophobic Surfaces. Advanced Materials. 2007;19:2330-5.

[5] Xia Y, Yang P, Sun Y, Wu Y, Mayers B, Gates B, et al. One-Dimensional Nanostructures: Synthesis, Characterization, and Applications. Advanced Materials. 2003;15:353-89.

[6] Kriha O, Becker M, Lehmann M, Kriha D, Krieglstein J, Yosef M, et al. Connection of Hippocampal Neurons by Magnetically Controlled Movement of Short Electrospun Polymer Fibers—A Route to Magnetic Micromanipulators. Advanced Materials. 2007;19:2483-5.

[7] Stoiljkovic A, Agarwal S. Short Electrospun Fibers by UV Cutting Method. Macromolecular Materials and Engineering. 2008;293:895-9.

[8] Lu X, Wang C, Wei Y. One-Dimensional Composite Nanomaterials: Synthesis by Electrospinning and Their Applications. Small. 2009;5:2349-70.

[9] Frenot A, Chronakis IS. Polymer nanofibers assembled by electrospinning. Current Opinion in Colloid & Interface Science. 2003;8:64-75.

[10] Li D, Xia Y. Electrospinning of Nanofibers: Reinventing the Wheel? Advanced Materials. 2004;16:1151-70.

[11] Sun Z, Zussman E, Yarin AL, Wendorff JH, Greiner A. Compound Core–Shell Polymer Nanofibers by Co-Electrospinning. Advanced Materials. 2003;15:1929-32.

[12] Li D, Xia Y. Direct Fabrication of Composite and Ceramic Hollow Nanofibers by Electrospinning. Nano Letters. 2004;4:933-8.

[13] Dror Y, Salalha W, Avrahami R, Zussman E, Yarin AL, Dersch R, et al. One-Step Production of Polymeric Microtubes by Co-electrospinning. Small. 2007;3:1064-73.

[14] Xu X, Zhuang X, Chen X, Wang X, Yang L, Jing X. Preparation of Core-Sheath Composite Nanofibers by Emulsion Electrospinning. Macromolecular Rapid Communications. 2006;27:1637-42.

[15] Biswas A, Bayer IS, Biris AS, Wang T, Dervishi E, Faupel F. Advances in top–down and bottom–up surface nanofabrication: Techniques, applications & future prospects. Advances in Colloid and Interface Science. 2012;170:2-27.

[16] Barth JV, Costantini G, Kern K. Engineering atomic and molecular nanostructures at surfaces. Nature. 2005;437:671-9.

[17] Whitesides GM, Boncheva M. Beyond molecules: Self-assembly of mesoscopic and macroscopic components. Proceedings of the National Academy of Sciences. 2002;99:4769-74.

[18] Palmer LC, Stupp SI. Molecular Self-Assembly into One-Dimensional Nanostructures. Accounts of Chemical Research. 2008;41:1674-84.

[19] Raez J, Barjovanu R, Massey JA, Winnik MA, Manners I. Self-Assembled Organometallic Block Copolymer Nanotubes. Angewandte Chemie International Edition. 2000;39:3862-5.

[20] Liu G, Qiao L, Guo A. Diblock Copolymer Nanofibers. Macromolecules. 1996;29:5508-10.

[21] Liu G. Nanofibers. Advanced Materials. 1997;9:437-9.

[22] Liu G, Yan X, Qiu X, Li Z. Fractionation and Solution Properties of PS-b-PCEMA-b-PtBA Nanofibers. Macromolecules. 2002;35:7742-7.

[23] Choi IS, Li XH, Simanek EE, Akaba R, Whitesides GM. Self-assembly of hydrogen-bonded polymeric rods based on the cyanuric acid center dot melamine lattice. Chemistry of Materials. 1999;11:684-90.

[24] Menard E, Meitl MA, Sun Y, Park J-U, Shir DJ-L, Nam Y-S, et al. Micro- and Nanopatterning Techniques for Organic Electronic and Optoelectronic Systems. Chemical Reviews. 2007;107:1117-60.
[25] DJ Lipomi, RV Martinez, L Cademartiri, Whitesides aG. Soft Lithographic Approaches to Nanofabrication in Polymer Science: A Comprehensive Reference: Elsevier; 2012.
[26] Pokroy B, Epstein AK, Persson-Gulda MCM, Aizenberg J. Fabrication of Bioinspired Actuated Nanostructures with Arbitrary Geometry and Stiffness. Advanced Materials. 2009;21:463-9.
[27] Qin D, Xia Y, Whitesides GM. Soft lithography for micro- and nanoscale patterning. Nat Protocols. 2010;5:491-502.
[28] Zheng Z, Azzaroni O, Zhou F, Huck WTS. Topography Printing to Locally Control Wettability. Journal of the American Chemical Society. 2006;128:7730-1.
[29] Courbin L, Denieul E, Dressaire E, Roper M, Ajdari A, Stone HA. Imbibition by polygonal spreading on microdecorated surfaces. Nat Mater. 2007;6:661-4.
[30] Tan JL, Tien J, Pirone DM, Gray DS, Bhadriraju K, Chen CS. Cells lying on a bed of microneedles: An approach to isolate mechanical force. Proceedings of the National Academy of Sciences. 2003;100:1484-9.
[31] Roca-Cusachs P, Rico F, Martínez E, Toset J, Farré R, Navajas D. Stability of Microfabricated High Aspect Ratio Structures in Poly(dimethylsiloxane). Langmuir. 2005;21:5542-8.
[32] Md Jani AM, Losic D, Voelcker NH. Nanoporous anodic aluminium oxide: Advances in surface engineering and emerging applications. Progress in Materials Science. 2013;58:636-704.
[33] Wehrspohn RB. Ordered Porous Nanostructures and Applications: Springer; 2005.
[34] Wang J-G, Wei B, Kang F. Facile synthesis of hierarchical conducting polypyrrole nanostructures via a reactive template of MnO2 and their application in supercapacitors. Rsc Advances. 2014;4:199-202.
[35] Wei Z, Zhang Z, Wan M. Formation Mechanism of Self-Assembled Polyaniline Micro/Nanotubes. Langmuir. 2002;18:917-21.
[36] Jang J, Yoon H. Formation Mechanism of Conducting Polypyrrole Nanotubes in Reverse Micelle Systems. Langmuir. 2005;21:11484-9.
[37] Jang J, Yoon H. Facile fabrication of polypyrrole nanotubes using reverse microemulsion polymerization. Chemical Communications. 2003:720-1.
[38] Tingyang D, Xiaoming Y, Yun L. Controlled growth of polypyrrole nanotubule/wire in the presence of a cationic surfactant. Nanotechnology. 2006;17:3028.
[39] Lee JI, Cho SH, Park S-M, Kim JK, Kim JK, Yu J-W, et al. Highly aligned ultrahigh density arrays of conducting polymer nanorods using block copolymer templates. Nano Letters. 2008;8:2315-20.
[40] Chen K, Tang L, Xia Y, Wang Y. Silver(I)-Coordinated Organogel-Templated Fabrication of 3D Networks of Polymer Nanotubes. Langmuir. 2008;24:13838-41.
[41] Starbird R, Garcia-Gonzalez CA, Smirnova I, Krautschneider WH, Bauhofer W. Synthesis of an organic conductive porous material using starch aerogels as template for chronic invasive electrodes. Materials Science & Engineering C-Materials for Biological Applications. 2014;37:177-83.
[42] Tena-Zaera R, Elias J, Wang G, Lévy-Clément C. Role of Chloride Ions on Electrochemical Deposition of ZnO Nanowire Arrays from O2 Reduction. The Journal of Physical Chemistry C. 2007;111:16706-11.
[43] Kokotov M, Hodes G. Reliable chemical bath deposition of ZnO films with controllable morphology from ethanolamine-based solutions using KMnO4 substrate activation. Journal of Materials Chemistry. 2009;19:3847-54.
[44] Doebbelin M, Tena-Zaera R, Carrasco PM, Sarasua JR, Cabanero G, Mecerreyes D. Electrochemical Synthesis of Poly(3,4-ethylenedioxythiophene) Nanotube Arrays Using ZnO Templates. Journal of Polymer Science Part a-Polymer Chemistry. 2010;48:4648-53.

[45] Yuan W, Lu Z, Liu J, Wang H, Li CM. ZnO nanowire array-templated LbL self-assembled polyelectrolyte nanotube arrays and application for charged drug delivery. Nanotechnology. 2013;24.
[46] R.L. Fleischer, P.B. Price, Walker RM. Nuclear Tracks in Solids. Berkeley: University of California Press; 1975.
[47] Masuda H, Fukuda K. Ordered Metal Nanohole Arrays Made by a Two-Step Replication of Honeycomb Structures of Anodic Alumina. Science. 1995;268:1466-8.
[48] Lehmann V, Foell H. Formation mechanism and properties of electrochemically etched trenches in n-type silicon. Journal of The Electrochemical Society. 1990;137:653-9.
[49] Jessensky O, Müller F, Gösele U. Self-organized formation of hexagonal pore arrays in anodic alumina. Applied Physics Letters. 1998;72:1173-5.
[50] Lehmann V. Physics of macropore formation in low doped n-type silicon. Journal of The Electrochemical Society. 1993;140:2836-43.
[51] Martin CR. Nanomaterials: A membrane-based synthetic approach. Science. 1994;266:1961-6.
[52] Martin CR. Membrane-based synthesis of nanomaterials. Chemistry of Materials. 1996;8:1739-46.
[53] Steinhart M, Wehrspohn RB, Gösele U, Wendorff JH. Nanotubes by template wetting: A modular assembly system. Angewandte Chemie - International Edition. 2004;43:1334-44.
[54] Xiao R, Cho SI, Liu R, Lee SB. Controlled Electrochemical Synthesis of Conductive Polymer Nanotube Structures. Journal of the American Chemical Society. 2007;129:4483-9.
[55] Byun J, Lee JI, Kwon S, Jeon G, Kim JK. Highly Ordered Nanoporous Alumina on Conducting Substrates with Adhesion Enhanced by Surface Modification: Universal Templates for Ultrahigh-Density Arrays of Nanorods. Advanced Materials. 2010;22:2028-+.
[56] Tang C, Lennon EM, Fredrickson GH, Kramer EJ, Hawker CJ. Evolution of block copolymer lithography to highly ordered square arrays. Science. 2008;322:429-32.
[57] Chen J-T, Zhang M, Yang L, Collins M, Parks J, Avallone A, et al. Templated nanostructured PS-b-PEO nanotubes. Journal of Polymer Science Part B-Polymer Physics. 2007;45:2912-7.
[58] Xiang H, Shin K, Kim T, Moon SI, McCarthy TJ, Russell TP. Block Copolymers under Cylindrical Confinement. Macromolecules. 2004;37:5660-4.
[59] Shin K, Xiang H, Moon SI, Kim T, McCarthy TJ, Russell TP. Curving and Frustrating Flatland. Science. 2004;306:76.
[60] Sun Y, Steinhart M, Zschech D, Adhikari R, Michler GH, Gösele U. Diameter-Dependence of the Morphology of PS-b-PMMA Nanorods Confined Within Ordered Porous Alumina Templates. Macromolecular Rapid Communications. 2005;26:369-75.
[61] Chen JT, Shin K, Leiston-Belanger JM, Zhang M, Russell TP. Amorphous Carbon Nanotubes with Tunable Properties via Template Wetting. Advanced Functional Materials. 2006;16:1476-80.
[62] Rodriguez AT, Chen M, Chen Z, Brinker CJ, Fan H. Nanoporous Carbon Nanotubes Synthesized through Confined Hydrogen-Bonding Self-Assembly. Journal of the American Chemical Society. 2006;128:9276-7.
[63] Decher G. Fuzzy Nanoassemblies: Toward Layered Polymeric Multicomposites. Science. 1997;277:1232-7.
[64] Decher G, Hong JD, Schmitt J. Buildup of ultrathin multilayer films by a self-assembly process: III. Consecutively alternating adsorption of anionic and cationic polyelectrolytes on charged surfaces. Thin Solid Films. 1992;210–211, Part 2:831-5.
[65] Dubacheva GV, Dumy P, Auzely R, Schaaf P, Boulmedais F, Jierry L, et al. Unlimited growth of host-guest multilayer films based on functionalized neutral polymers. Soft Matter. 2010;6:3747-50.
[66] Zhang L, Vidyasagar A, Lutkenhaus JL. Fabrication and thermal analysis of layer-by-layer micro- and nanotubes. Current Opinion in Colloid & Interface Science. 2012;17:114-21.

[67] Cho Y, Lee W, Jhon YK, Genzer J, Char K. Polymer nanotubules obtained by layer-by-layer deposition within AAO-membrane templates with sub-100-nm pore diameters. Small. 2010;6:2683-9.
[68] Raoufi M, Tranchida D, Schönherr H. Pushing the Size Limits in the Replication of Nanopores in Anodized Aluminum Oxide via the Layer-by-Layer Deposition of Polyelectrolytes. Langmuir. 2012;28:10091-6.
[69] Wang Y, Angelatos AS, Caruso F. Template Synthesis of Nanostructured Materials via Layer-by-Layer Assembly†. Chemistry of Materials. 2007;20:848-58.
[70] Ai S, Lu G, He Q, Li J. Highly Flexible Polyelectrolyte Nanotubes. Journal of the American Chemical Society. 2003;125:11140-1.
[71] Blaszczyk-Lezak I, Hernández M, Mijangos C. One Dimensional PMMA Nanofibers from AAO Templates. Evidence of Confinement Effects by Dielectric and Raman Analysis. Macromolecules. 2013;46:4995-5002.
[72] Keller F, Hunter MS, Robinson DL. Structural Features of Oxide Coatings on Aluminum. Journal of The Electrochemical Society. 1953;100:411-9.
[73] Thompson GE, Furneaux RC, Wood GC, Richardson JA, Goode JS. Nucleation and growth of porous anodic films on aluminium. Nature. 1978;272:433-5.
[74] Thompson GE, Wood GC. Porous anodic film formation on aluminium. Nature. 1981;290:230-2.
[75] Masuda H, Satoh M. Fabrication of Gold Nanodot Array Using Anodic Porous Alumina as an Evaporation Mask. Japanese Journal of Applied Physics. 1996;35.
[76] Li AP, Muller F, Birner A, Nielsch K, Gosele U. Hexagonal pore arrays with a 50--420 nm interpore distance formed by self-organization in anodic alumina. Journal of Applied Physics. 1998;84:6023-6.
[77] Li A-P, Müller F, Birner A, Nielsch K, Gösele U. Fabrication and Microstructuring of Hexagonally Ordered Two-Dimensional Nanopore Arrays in Anodic Alumina. Advanced Materials. 1999;11:483-7.
[78] Lee W, Ji R, Gosele U, Nielsch K. Fast fabrication of long-range ordered porous alumina membranes by hard anodization. Nat Mater. 2006;5:741-7.
[79] Lee W, Schwirn K, Steinhart M, Pippel E, Scholz R, Gosele U. Structural engineering of nanoporous anodic aluminium oxide by pulse anodization of aluminium. Nature Nanotechnology. 2008;3:234-9.
[80] Masuda H, Hasegwa F, Ono S. Self-Ordering of Cell Arrangement of Anodic Porous Alumina Formed in Sulfuric Acid Solution. Journal of The Electrochemical Society. 1997;144:L127-L30.
[81] Masuda H, Yada K, Osaka A. Self-ordering of cell configuration of anodic porous alumina with large-size pores in phosphoric acid solution. Japanese Journal of Applied Physics. 1998;37.
[82] Masuda H, Asoh H, Watanabe M, Nishio K, Nakao M, Tamamura T. Square and Triangular Nanohole Array Architectures in Anodic Alumina. Advanced Materials. 2001;13:189-92.
[83] Masuda H, Yada K, Osaka A. Self-ordering of cell configuration of anodic porous alumina with large-pore-size in phosphoric acid solution Japanese Journal of Applied Physics. 1998;37.
[84] Nishinaga O, Kikuchi T, Natsui S, Suzuki RO. Rapid fabrication of self-ordered porous alumina with 10-/sub-10-nm-scale nanostructures by selenic acid anodizing. Sci Rep. 2013;3.
[85] Choi J, Nielsch K, Reiche M, Wehrspohn RB, Gosele U. Fabrication of monodomain alumina pore arrays with an interpore distance smaller than the lattice constant of the imprint stamp. Journal of Vacuum Science & Technology B: Microelectronics and Nanometer Structures. 2003;21:763-6.
[86] Schwirn K, Lee W, Hillebrand R, Steinhart M, Nielsch K, Gösele U. Self-Ordered Anodic Aluminum Oxide Formed by $H_2SO_4$ Hard Anodization. ACS Nano. 2008;2:302-10.
[87] Manzano CV, Martín J, Martín-González MS. Ultra-narrow 12 nm pore diameter self-ordered anodic alumina templates. Microporous and Mesoporous Materials. 2014;184:177-83.

[88] Matsui Y, Nishio K, Masuda H. Highly Ordered Anodic Porous Alumina with 13-nm Hole Intervals Using a 2D Array of Monodisperse Nanoparticles As a Template. Small. 2006;2:522-5.
[89] Losic D, Lillo M. Porous Alumina with Shaped Pore Geometries and Complex Pore Architectures Fabricated by Cyclic Anodization. Small. 2009;5:1392-7.
[90] Martín J, Martín-González M, Francisco Fernández J, Caballero-Calero O. Ordered three-dimensional interconnected nanoarchitectures in anodic porous alumina. Nat Commun. 2014;5.
[91] Lee W, Park S-J. Porous Anodic Aluminum Oxide: Anodization and Templated Synthesis of Functional Nanostructures. Chemical Reviews. 2014;114:7487-556.
[92] Sulka GD. Highly Ordered Anodic Porous Alumina Formation by Self-Organized Anodizing: Wiley-VCH Verlag GmbH & Co. KGaA; 2008.
[93] Martín J, Mijangos C. Tailored Polymer-Based Nanofibers and Nanotubes by Means of Different Infiltration Methods into Alumina Nanopores. Langmuir. 2009;25:1181-7.
[94] O'Sullivan JP, Wood GC. The Morphology and Mechanism of Formation of Porous Anodic Films on Aluminium. Proceedings of the Royal Society of London A Mathematical and Physical Sciences. 1970;317:511-43.
[95] Martín J, Manzano CV, Martín-González M. In-depth study of self-ordered porous alumina in the 140–400 nm pore diameter range. Microporous and Mesoporous Materials. 2012;151:311-6.
[96] Sun C, Luo J, Wu L, Zhang J. Self-Ordered Anodic Alumina with Continuously Tunable Pore Intervals from 410 to 530 nm. ACS Applied Materials & Interfaces. 2010;2:1299-302.
[97] Martín J, Nogales A, Martín-González M. The Smectic–Isotropic Transition of P3HT Determines the Formation of Nanowires or Nanotubes into Porous Templates. Macromolecules. 2013.
[98] Jessensky O, Muller F, Gosele U. Self-organized formation of hexagonal pore arrays in anodic alumina. Applied Physics Letters. 1998;72:1173-5.
[99] Masuda H, Takenaka K, Ishii T, Nishio K. Long-range-ordered anodica porous alumina with less-than-30 nm hole interval. Japan Journal of Applied Physics. 2006;45.
[100] Martín J, Maiz J, Sacristan J, Mijangos C. Tailored polymer-based nanorods and nanotubes by "template synthesis": From preparation to applications. Polymer. 2012;53:1149-66.
[101] Martin J, Martin-Gonzalez M. The use of PEEK nanorod arrays for the fabrication of nanoporous surfaces under high temperature: SiNx example. Nanoscale. 2012;4:5608-13.
[102] Masuda H, Nagae M, Morikawa T, Nishio K. Long-range-ordered anodic porous alumina with reduced hole interval formed in highly concentrated sulfuric acid solution. The Japan journal of Applied Physics. 2006;45.
[103] Chen W, Wu J-S, Xia X-H. Porous Anodic Alumina with Continuously Manipulated Pore/Cell Size. ACS Nano. 2008;2:959-65.
[104] Krishnan R, Thompson CV. Monodomain High-Aspect-Ratio 2D and 3D Ordered Porous Alumina Structures with Independently Controlled Pore Spacing and Diameter. Advanced Materials. 2007;19:988-92.
[105] Lee W, Scholz R, Goÿsele U. A Continuous Process for Structurally Well-Defined Al2O3 Nanotubes Based on Pulse Anodization of Aluminum. Nano Letters. 2008;8:2155-60.
[106] Lee W, Kim J-C. Highly ordered porous alumina with tailor-made pore structures fabricated by pulse anodization. Nanotechnology. 2010;21:485304.
[107] Sulka GD, Hnida K. Distributed Bragg reflector based on porous anodic alumina fabricated by pulse anodization. Nanotechnology. 2012;23:075303.
[108] Sulka GD, Brzózka A, Liu L. Fabrication of diameter-modulated and ultrathin porous nanowires in anodic aluminum oxide templates. Electrochimica Acta. 2011;56:4972-9.
[109] Losic D. Preparation of Porous Anodic Alumina with Periodically Perforated Pores. Langmuir. 2009;25:5426-31.

[110] Lee W, Kim J-C, Gösele U. Spontaneous Current Oscillations during Hard Anodization of Aluminum under Potentiostatic Conditions. Advanced Functional Materials. 2010;20:21-7.
[111] Li J, Papadopoulos C, Xu J. Nanoelectronics: Growing Y-junction carbon nanotubes. Nature. 1999;402:253-4.
[112] Ho AYY, Gao H, Lam YC, Rodríguez I. Controlled Fabrication of Multitiered Three-Dimensional Nanostructures in Porous Alumina. Advanced Functional Materials. 2008;18:2057-63.
[113] Meng G, Jung YJ, Cao A, Vajtai R, Ajayan PM. Controlled fabrication of hierarchically branched nanopores, nanotubes, and nanowires. Proceedings of the National Academy of Sciences of the United States of America. 2005;102:7074-8.
[114] Chen B, Xu Q, Zhao X, Zhu X, Kong M, Meng G. Branched Silicon Nanotubes and Metal Nanowires via AAO-Template-Assistant Approach. Advanced Functional Materials. 2010;20:3791-6.
[115] Cheng W, Steinhart M, Gosele U, Wehrspohn RB. Tree-like alumina nanopores generated in a non-steady-state anodization. Journal of Materials Chemistry. 2007;17.
[116] Penner RM, Martin CR. Controlling the Morphology of Electronically Conductive Polymers. Journal of The Electrochemical Society. 1986;133:2206-7.
[117] Steinhart M, Senz S, Wehrspohn RB, Gösele U, Wendorff JH. Curvature-Directed Crystallization of Poly(vinylidene difluoride) in Nanotube Walls. Macromolecules. 2003;36:3646-51.
[118] Steinhart M, Wendorff JH, Greiner A, Wehrspohn RB, Nielsch K, Schilling J, et al. Polymer Nanotubes by Wetting of Ordered Porous Templates. Science. 2002;296:1997.
[119] Kriha O, Zhao L, Pippel E, Gösele U, Wehrspohn RB, Wendorff JH, et al. Organic Tube/Rod Hybrid Nanofibers with Adjustable Segment Lengths by Bidirectional Template Wetting. Advanced Functional Materials. 2007;17:1327-32.
[120] Schlitt S, Greiner A, Wendorff JH. Cylindrical Polymer Nanostructures by Solution Template Wetting. Macromolecules. 2008;41:3228-34.
[121] Steinhart M, Zimmermann S, Göring P, Schaper AK, Gösele U, Weder C, et al. Liquid Crystalline Nanowires in Porous Alumina: Geometric Confinement versus Influence of Pore Walls. Nano Letters. 2004;5:429-34.
[122] Wang Y, Gösele U, Steinhart M. Mesoporous Polymer Nanofibers by Infiltration of Block Copolymers with Sacrificial Domains into Porous Alumina. Chemistry of Materials. 2007;20:379-81.
[123] Grimm S, Martin J, Rodriguez G, Fernandez-Gutierrez M, Mathwig K, Wehrspohn RB, et al. Cellular interactions of biodegradable nanorod arrays prepared by nondestructive extraction from nanoporous alumina. Journal of Materials Chemistry. 2010;20:3171-7.
[124] Ok S, Steinhart M, Şerbescu A, Franz C, Vaca Chávez F, Saalwächter K. Confinement Effects on Chain Dynamics and Local Chain Order in Entangled Polymer Melts. Macromolecules. 2010;43:4429-34.
[125] Duran H, Steinhart M, Butt H-Jr, Floudas G. From Heterogeneous to Homogeneous Nucleation of Isotactic Poly(propylene) Confined to Nanoporous Alumina. Nano Letters. 2011;11:1671-5.
[126] Hofmann M, Hofmann A, Herrmann S, Ok C, Franz D, Kruk K, et al. Polymer Dynamics of Polybutadiene in Nanoscopic Confinement As Revealed by Field Cycling1H NMR. Macromolecules. 2011;44:4017-21.
[127] Shingne N, Geuss M, Hartmann-Azanza B, Steinhart M, Thurn-Albrecht T. Formation, morphology and internal structure of one-dimensional nanostructures of the ferroelectric polymer P(VDF-TrFE). Polymer. 2013;54:2737-44.
[128] Xue L, Kovalev A, Dening K, Eichler-Volf A, Eickmeier H, Haase M, et al. Reversible Adhesion Switching of Porous Fibrillar Adhesive Pads by Humidity. Nano Letters. 2013;13:5541-8.

[129] Alexandris S, Sakellariou G, Steinhart M, Floudas G. Dynamics of Unentangled cis-1,4-Polyisoprene Confined to Nanoporous Alumina. Macromolecules. 2014;47:3895-900.
[130] Suzuki Y, Duran H, Steinhart M, Butt H-J, Floudas G. Suppression of Poly(ethylene oxide) Crystallization in Diblock Copolymers of Poly(ethylene oxide)-b-poly(ε-caprolactone) Confined to Nanoporous Alumina. Macromolecules. 2014;47:1793-800.
[131] Zhang M, Dobriyal P, Chen J-T, Russell TP, Olmo J, Merry A. Wetting Transition in Cylindrical Alumina Nanopores with Polymer Melts. Nano Letters. 2006;6:1075-9.
[132] Chen J-T, Zhang M, Russell TP. Instabilities in Nanoporous Media. Nano Letters. 2007;7:183-7.
[133] Dobriyal P, Xiang H, Kazuyuki M, Chen J-T, Jinnai H, Russell TP. Cylindrically Confined Diblock Copolymers. Macromolecules. 2009;42:9082-8.
[134] Chen D, Zhao W, Wei D, Russell TP. Dewetting on Curved Interfaces: A Simple Route to Polymer Nanostructures. Macromolecules. 2011;44:8020-7.
[135] Xiang H, Shin K, Kim T, Moon SI, McCarthy TJ, Russell TP. From Cylinders to Helices upon Confinement. Macromolecules. 2005;38:1055-6.
[136] Chen J-T, Chen D, Russell TP. Fabrication of Hierarchical Structures by Wetting Porous Templates with Polymer Microspheres. Langmuir. 2009;25:4331-5.
[137] Chen J-T, Wei T-H, Chang C-W, Ko H-W, Chu C-W, Chi M-H, et al. Fabrication of Polymer Nanopeapods in the Nanopores of Anodic Aluminum Oxide Templates Using a Double-Solution Wetting Method. Macromolecules. 2014;47:5227-35.
[138] Tsai C-C, Chen J-T. Rayleigh Instability in Polymer Thin Films Coated in the Nanopores of Anodic Aluminum Oxide Templates. Langmuir. 2013;30:387-93.
[139] Feng X, Jin Z. Spontaneous Formation of Nanoscale Polymer Spheres, Capsules, or Rods by Evaporation of Polymer Solutions in Cylindrical Alumina Nanopores. Macromolecules. 2009;42:569-72.
[140] Mei S, Feng X, Jin Z. Fabrication of Polymer Nanospheres Based on Rayleigh Instability in Capillary Channels. Macromolecules. 2011;44:1615-20.
[141] Feng X, Mei S, Jin Z. Wettability Transition Induced Transformation and Entrapment of Polymer Nanostructures in Cylindrical Nanopores. Langmuir. 2011;27:14240-7.
[142] Hou P, Fan H, Jin Z. Spiral and Mesoporous Block Polymer Nanofibers Generated in Confined Nanochannels. Macromolecules. 2014;48:272-8.
[143] Lee MK, Lee J. A nano-frost array technique to prepare nanoporous PVDF membranes. Nanoscale. 2014;6:8642-8.
[144] Li J, Sattayasamitsathit S, Dong R, Gao W, Tam R, Feng X, et al. Template electrosynthesis of tailored-made helical nanoswimmers. Nanoscale. 2014;6:9415-20.
[145] Palacios R, Formentín P, Ferré-Borrull J, Pallarés J, Marsal LF. Polymer nanopillars using self-ordered nanoporous alumina templates. physica status solidi (c). 2009;6:1584-6.
[146] Martín J, Vázquez M, Hernández-Vélez M, Mijangos C. One-dimensional magnetopolymeric nanostructures with tailored sizes. Nanotechnology. 2008;19:175304.
[147] Martín J, Mijangos C. Tailored Polymer-Based Nanofibers and Nanotubes by Means of Different Infiltration Methods into Alumina Nanopores. Langmuir. 2008;25:1181-7.
[148] Martín J, Mijangos C, Sanz A, Ezquerra TA, Nogales A. Segmental Dynamics of Semicrystalline Poly(vinylidene fluoride) Nanorods. Macromolecules. 2009;42:5395-401.
[149] Martín J, Krutyeva M, Monkenbusch M, Arbe A, Allgaier J, Radulescu A, et al. Direct Observation of Confined Single Chain Dynamics by Neutron Scattering. Physical Review Letters. 2010;104:197801.
[150] Michell RM, Lorenzo AT, Müller AJ, Lin M-C, Chen H-L, Blaszczyk-Lezak I, et al. The Crystallization of Confined Polymers and Block Copolymers Infiltrated Within Alumina Nanotube Templates. Macromolecules. 2012;45:1517-28.
[151] Blaszczyk-Lezak I, Maiz J, Sacristán J, Mijangos C. Monitoring the Thermal Elimination of Infiltrated Polymer from AAO Templates: An Exhaustive Characterization after Polymer Extraction. Industrial & Engineering Chemistry Research. 2011;50:10883-8.

[152] Krutyeva M, Wischnewski A, Monkenbusch M, Willner L, Maiz J, Mijangos C, et al. Effect of Nanoconfinement on Polymer Dynamics: Surface Layers and Interphases. Physical Review Letters. 2013;110:108303.
[153] Giussi JM, Blaszczyk-Lezak I, Allegretti PE, Cortizo MS, Mijangos C. Tautomerizable styrenic copolymers confined in AAO templates. Polymer. 2013;54:5050-7.
[154] Martín J, Nogales A, Mijangos C. Directional Crystallization of 20 nm Width Polymer Nanorods by the Inducement of Heterogeneous Nuclei at Their Tips. Macromolecules. 2013;46:7415-22.
[155] Han X, Maiz J, Mijangos C, Zaldo C. Nanopatterned PMMA-Yb:Er/Tm:Lu 2 O 3 composites with visible upconversion emissions. Nanotechnology. 2014;25:205302.
[156] Maiz J, Zhao W, Gu Y, Lawrence J, Arbe A, Alegría A, et al. Dynamic study of polystyrene-block-poly(4-vinylpyridine) copolymer in bulk and confined in cylindrical nanopores. Polymer. 2014;55:4057-66.
[157] Lee JA, McCarthy TJ. Polymer Surface Modification: Topography Effects Leading to Extreme Wettability Behavior. Macromolecules. 2007;40:3965-9.
[158] Kim E, Xia Y, Whitesides GM. Polymer microstructures formed by moulding in capillaries. Nature. 1995;376:581-4.
[159] Xue L, Han Y. Pattern formation by dewetting of polymer thin film. Progress in Polymer Science. 2011;36:269-93.
[160] Martín J, Hernández-Vélez M, De Abril O, Luna C, Munoz-Martin A, Vázquez M, et al. Fabrication and characterization of polymer-based magnetic composite nanotubes and nanorods. European Polymer Journal. 2012;48:712-9.
[161] Grimm S, Giesa R, Sklarek K, Langner A, Gösele U, Schmidt H-W, et al. Nondestructive Replication of Self-Ordered Nanoporous Alumina Membranes via Cross-Linked Polyacrylate Nanofiber Arrays. Nano Letters. 2008;8:1954-9.
[162] Salsamendi M, Ballard N, Sanz B, Asúa JM, Mijangos C. Polymerization kinetics of a fluorinated monomer under confinement in AAO nanocavities. RSC Advances. 2015;accepted, ID. RA-ART-12-2014-016728.
[163] Martín J, Martín-González M, Del Campo A, Reinosa JJ, Fernández JF. Ordered arrays of polymeric nanopores by using inverse nanostructured PTFE surfaces. Nanotechnology. 2012;23.
[164] Jaime M, Marisol M-G, Adolfo del C, Julián JR, José Francisco F. Ordered arrays of polymeric nanopores by using inverse nanostructured PTFE surfaces. Nanotechnology. 2012;23:385305.
[165] Krutyeva M, Martin J, Arbe A, Colmenero J, Mijangos C, Schneider GJ, et al. Neutron scattering study of the dynamics of a polymer melt under nanoscopic confinement. Journal of Chemical Physics. 2009;131.
[166] Maiz J, Schäfer H, Trichy Rengarajan G, Hartmann-Azanza B, Eickmeier H, Haase M, et al. How gold nanoparticles influence crystallization of polyethylene in rigid cylindrical nanopores. Macromolecules. 2013;46:403-12.
[167] Suzuki Y, Duran H, Akram W, Steinhart M, Floudas G, Butt HJ. Multiple nucleation events and local dynamics of poly(ε-caprolactone) (PCL) confined to nanoporous alumina. Soft Matter. 2013;9:9189-98.
[168] Giussi JM, Blaszczyk-Lezak I, Cortizo MS, Mijangos C. In-situ polymerization of styrene in AAO nanocavities. Polymer. 2013;54:6886-93.
[169] Kim K, Jin J-I. Preparation of PPV Nanotubes and Nanorods and Carbonized Products Derived Therefrom. Nano Letters. 2001;1:631-6.
[170] Uemura T, Ono Y, Kitagawa K, Kitagawa S. Radical Polymerization of Vinyl Monomers in Porous Coordination Polymers: Nanochannel Size Effects on Reactivity, Molecular Weight, and Stereostructure. Macromolecules. 2007;41:87-94.
[171] Reddy CS, Arinstein A, Zussman E. Polymerization kinetics under confinement. Polymer Chemistry. 2011;2:835-9.

[172] Zhao H, Simon SL. Methyl methacrylate polymerization in nanoporous confinement. Polymer. 2011;52:4093-8.
[173] Begum F, Zhao H, Simon SL. Modeling methyl methacrylate free radical polymerization: Reaction in hydrophobic nanopores. Polymer. 2012;53:3261-8.
[174] Ramirez-Caballero GE, Mathkari A, Balbuena PB. Confinement-Induced Polymerization of Ethylene. The Journal of Physical Chemistry C. 2011;115:2134-9.
[175] Cui Y, Tao C, Zheng S, He Q, Ai S, Li J. Synthesis of Thermosensitive PNIPAM-co-MBAA Nanotubes by Atom Transfer Radical Polymerization within a Porous Membrane. Macromolecular Rapid Communications. 2005;26:1552-6.
[176] Cui Y, Tao C, Tian Y, He Q, Li J. Synthesis of PNIPAM-co-MBAA Copolymer Nanotubes with Composite Control. Langmuir. 2006;22:8205-8.
[177] Gorman CB, Petrie RJ, Genzer J. Effect of Substrate Geometry on Polymer Molecular Weight and Polydispersity during Surface-Initiated Polymerization. Macromolecules. 2008;41:4856-65.
[178] Back J-W, Lee S, Hwang C-R, Chi C-S, Kim J-Y. Fabrication of conducting PEDOT nanotubes using vapor deposition polymerization. Macromol Res. 2011;19:33-7.
[179] Nair S, Naredi P, Kim SH. Formation of High-Stress Phase and Extrusion of Polyethylene due to Nanoconfinements during Ziegler−Natta Polymerization Inside Nanochannels. The Journal of Physical Chemistry B. 2005;109:12491-7.
[180] Esman N, Lellouche J-P. Fabrication of functional polypyrrole (PolyPyr)-nanotubes using anodized aluminium oxide (AAO) template membranes. Compromising between effectiveness and mildness of template dissolution conditions for a safe release of PolyPyr-nanotubes. Polymer Chemistry. 2010;1:158-60.
[181] Lau KHA, Duran H, Knoll W. In situ Characterization of N-Carboxy Anhydride Polymerization in Nanoporous Anodic Alumina. The Journal of Physical Chemistry B. 2009;113:3179-89.
[182] Duran H, Gitsas A, Floudas G, Mondeshki M, Steinhart M, Knoll W. Poly(γ-benzyl-l-glutamate) Peptides Confined to Nanoporous Alumina: Pore Diameter Dependence of Self-Assembly and Segmental Dynamics. Macromolecules. 2009;42:2881-5.
[183] Lee L-C, Han H, Tsai Y-T, Fan G-L, Liu H-F, Wu C-C, et al. Template-assisted in situ polymerization for forming blue organic light-emitting nanotubes. Chemical Communications. 2014;50:8208-10.
[184] Jang J, Oh JH. A facile synthesis of polypyrrole nanotubes using a template-mediated vapor deposition polymerization and the conversion to carbon nanotubes. Chemical Communications. 2004:882-3.
[185] Choi MK, Yoon H, Lee K, Shin K. Simple Fabrication of Asymmetric High-Aspect-Ratio Polymer Nanopillars by Reusable AAO Templates. Langmuir. 2011;27:2132-7.
[186] Sanz B, Marcos A, Mijangos C. Polyurethane nanofibers prepared by polycondensation reaction in a AAO nanoreactor. *to be submitted*.
[187] Malvaldi M, Bruzzone S, Picchioni F. Confinement Effect in Diffusion-Controlled Stepwise Polymerization by Monte Carlo Simulation. The Journal of Physical Chemistry B. 2006;110:12281-8.
[188] Begum F, Simon SL. Modeling methyl methacrylate free radical polymerization in nanoporous confinement. Polymer. 2011;52:1539-45.
[189] Alduncin JA, Forcada J, Barandiaran MJ, Asua JM. On the main locus of radical formation in emulsion polymerization initiated by oil-soluble initiators. Journal of Polymer Science Part A: Polymer Chemistry. 1991;29:1265-70.
[190] Autran C, de la Cal JC, Asua JM. (Mini)emulsion Polymerization Kinetics Using Oil-Soluble Initiators. Macromolecules. 2007;40:6233-8.
[191] Strobl G. The Physics of Polymers: conceps for understanding their structures and behaviour. 3 ed. Berlin Heidelberg New York: Springer; 2007.

[192] Michell RM, Blaszczyk-Lezak I, Mijangos C, Müller AJ. Confinement effects on polymer crystallization: From droplets to alumina nanopores. Polymer. 2013;54:4059-77.
[193] Michell RM, Blaszczyk-Lezak I, Mijangos C, Müller AJ. Confined crystallization of polymers within anodic aluminum oxide templates. Journal of Polymer Science Part B: Polymer Physics. 2014;52:1179-94.
[194] Steinhart M. Supramolecular Organization of Polymeric Materials in Nanoporous Hard Templates. Advances in Polymer Science. 2008:123-87.
[195] Bae SC, Granick S. Molecular Motion at Soft and Hard Interfaces: From Phospholipid Bilayers to Polymers and Lubricants. Annual Review of Physical Chemistry. 2007;58:353-74.
[196] Sussman DM, Tung W-S, Winey KI, Schweizer KS, Riggleman RA. Entanglement Reduction and Anisotropic Chain and Primitive Path Conformations in Polymer Melts under Thin Film and Cylindrical Confinement. Macromolecules. 2014;47:6462-72.
[197] Zheng X, Rafailovich MH, Sokolov J, Strzhemechny Y, Schwarz SA, Sauer BB, et al. Long-Range Effects on Polymer Diffusion Induced by a Bounding Interface. Physical Review Letters. 1997;79.
[198] Lin EK, Wu W-l, Satija SK. Polymer Interdiffusion near an Attractive Solid Substrate. Macromolecules. 1997;30:7224-31.
[199] Shin K, Obukhov S, Chen J-T, Huh J, Hwang Y, Mok S, et al. Enhanced mobility of confined polymers. Nat Mater. 2007;6:961-5.
[200] Si L, Massa MV, Dalnoki-Veress K, Brown HR, Jones RAL. Chain Entanglement in Thin Freestanding Polymer Films. Physical Review Letters. 2005;94:127801.
[201] Rowland HD, King WP, Pethica JB, Cross GLW. Molecular Confinement Accelerates Deformation of Entangled Polymers During Squeeze Flow. Science. 2008;322:720-4.
[202] Barbero DR, Steiner U. Nonequilibrium Polymer Rheology in Spin-Cast Films. Physical Review Letters. 2009;102.
[203] Lagrené K, Zanotti J-M, Daoud M, Farago B, Judeinstein P, . Large-scale dynamics of a single polymer chain under severe confinement. Physical Review E. 2010;81:060801.
[204] Noirez L, Stillings C, Bardeau JF, Steinhart M, Schlitt S, Wendorff JH, et al. What Happens to Polymer Chains Confined in Rigid Cylindrical Inorganic (AAO) Nanopores. Macromolecules. 2013;46:4932-6.
[205] Hofmann M, Herrmann A, Ok S, Franz C, Kruk D, Saalwächter K, et al. Polymer Dynamics of Polybutadiene in Nanoscopic Confinement As Revealed by Field Cycling 1H NMR. Macromolecules. 2011;44:4017-21.
[206] Shi A-C, Li B. Self-assembly of diblock copolymers under confinement. Soft Matter. 2013;9:1398-413.
[207] Park K, Choi K, Lee JH, Park SH, Lee SC, Lee HS. Curvature-Driven Rigid Nanowire Orientation inside Nanotube Walls. ACS Macro Letters. 2011;1:110-4.
[208] McKenna GB. Ten (or more) years of dynamics in confinement: Perspectives for 2010. Eur Phys J Spec Top. 2010;189:285-302.
[209] Mataz A, Gregory BM. Effects of confinement on material behaviour at the nanometre size scale. Journal of Physics: Condensed Matter. 2005;17:R461.
[210] Kalogeras IM. Contradicting perturbations of the segmental and secondary relaxation dynamics of polymer strands constrained in nanopores. Acta Materialia. 2005;53:1621-30.
[211] Suzuki Y, Duran H, Steinhart M, Butt H-J, Floudas G. Homogeneous crystallization and local dynamics of poly(ethylene oxide) (PEO) confined to nanoporous alumina. Soft Matter. 2013.
[212] Cangialosi D, Alegría A, Colmenero J. Route to calculate the length scale for the glass transition in polymers. Physical Review E. 2007;76:011514.
[213] Duran H, Gitsas A, Floudas G, Mondeshki M, Steinhart M, Knoll W. Poly(Î³-benzyl-l-glutamate) Peptides Confined to Nanoporous Alumina: Pore Diameter Dependence of Self-Assembly and Segmental Dynamics. Macromolecules. 2009;42:2881-5.

[214] Serghei A, Chen D, Lee DH, Russell TP. Segmetal dynamics of polymers during capillary flow into nanopores. Soft Matter. 2010;6:1111-3.
[215] Suzuki Y, Duran H, Akram W, Steinhart M, Floudas G, Butt H-J. Multiple nucleation events and local dynamics of poly([varepsilon]-caprolactone) (PCL) confined to nanoporous alumina. Soft Matter. 2013;9:9189-98.
[216] Li L, Zhou D, Huang D, Xue G. Double Glass Transition Temperatures of Poly(methyl methacrylate) Confined in Alumina Nanotube Templates. Macromolecules. 2013;47:297-303.
[217] Mapesa EU, Popp L, Kiprop WK, Tress M, Kremer F. Molecular dynamics in 1- and 2-D confinement as studied for the case of poly(cis-1,4-isoprene). Soft Materials. 2014:null-null.
[218] Krutyeva M, Martin J, Arbe A, Colmenero J, Mijangos C, Schneider GJ, et al. Neutron scattering study of the dynamics of a polymer melt under nanoscopic confinement. The Journal of Chemical Physics. 2009;131:174901.
[219] Lagrené K, Zanotti JM, Daoud M, Farago B, Judeinstein P. Dynamical behavior of a single polymer chain under nanometric confinement. Eur Phys J Spec Top. 2010;189:231-7.
[220] Duran H, Yameen B, Geuss M, Kappl M, Steinhart M, Knoll W. Enhanced interfacial rigidity of 1D thermoset nanostructures by interface-induced liquid crystallinity. Journal of Materials Chemistry C. 2013;1:7758-65.
[221] Martín J, Muñoz M, Encinar M, Calleja M, Martín-González M. Fabrication and Mechanical Characterization of Semi-Free-Standing (Conjugated) Polymer Thin Films. Langmuir. 2013.
[222] Cao B-Y, Li Y-W, Kong J, Chen H, Xu Y, Yung K-L, et al. High thermal conductivity of polyethylene nanowire arrays fabricated by an improved nanoporous template wetting technique. Polymer. 2011;52:1711-5.
[223] Muñoz Rojo M, Martin J, grauby s, Borca-Tasciuc T, dilhaire s, Martin-Gonzalez MS. Decrease in Thermal Conductivity in Polymeric P3HT Nanowires by Size-Reduction induced by Crystal Orientation: New Approaches towards Organic Thermal Transport Engineering. Nanoscale. 2014.
[224] Cao B-Y, Kong J, Xu Y, Yung K-L, Cai A. Polymer Nanowire Arrays With High Thermal Conductivity and Superhydrophobicity Fabricated by a Nano-Molding Technique. Heat Transfer Engineering. 2012;34:131-9.
[225] Choy CL. Thermal conductivity of polymers. Polymer. 1977;18:984-1004.
[226] Suh KY, Kim YS, Lee HH. Capillary Force Lithography. Advanced Materials. 2001;13:1386-9.
[227] Xiang H, Shin K, Kim T, Moon S, Mccarthy TJ, Russell TP. The influence of confinement and curvature on the morphology of block copolymers. Journal of Polymer Science Part B: Polymer Physics. 2005;43:3377-83.
[228] Hu G, Cao B. Flows of Polymer Melts through Nanopores: Experiments and Modelling. Journal of Thermal Science and Technology. 2013;8:363-9.
[229] Ballarin B, Brumlik CJ, Lawson DR, Liang W, Van Dyke LS, Martin CR. Chemical sensors based on ultrathin-film composite membranes - A new concept in sensor design. Analytical Chemistry. 1992;64:2647-51.
[230] Myler S, Collyer SD, Bridge KA, Higson SPJ. Ultra-thin-polysiloxane-film-composite membranes for the optimisation of amperometric oxidase enzyme electrodes. Biosensors and Bioelectronics. 2002;17:35-43.
[231] Darder M, Aranda P, Hernández-Vélez M, Manova E, Ruiz-Hitzky E. Encapsulation of enzymes in alumina membranes of controlled pore size. Thin Solid Films. 2006;495:321-6.
[232] Yang Z, Zhang C. Single-enzyme nanoparticles based urea biosensor. Sensors and Actuators B: Chemical. 2013;188:313-7.
[233] Mark SS, Stolper SI, Baratti C, Park JY, Kricka LJ. Biofunctionalization of aqueous dispersed, alumina membrane-templated polymer nanorods for use in enzymatic chemiluminescence assays. Colloids and Surfaces B: Biointerfaces. 2008;65:230-8.

[234] Guilak F, Cohen DM, Estes BT, Gimble JM, Liedtke W, Chen CS. Control of Stem Cell Fate by Physical Interactions with the Extracellular Matrix. Cell Stem Cell. 2009;5:17-26.
[235] Martin P. Wound Healing--Aiming for Perfect Skin Regeneration. Science. 1997;276:75-81.
[236] Han YL, Wang S, Zhang X, Li Y, Huang G, Qi H, et al. Engineering physical microenvironment for stem cell based regenerative medicine. Drug Discovery Today. 2014;19:763-73.
[237] Gultepe E, Nagesha D, Sridhar S, Amiji M. Nanoporous inorganic membranes or coatings for sustained drug delivery in implantable devices. Advanced Drug Delivery Reviews. 2010;62:305-15.
[238] Du K, Gan Z. Cellular Interactions on Hierarchical Poly(ε-caprolactone) Nanowire Micropatterns. ACS Applied Materials & Interfaces. 2012;4:4643-50.
[239] Northen MT, Turner KL. A batch fabricated biomimetic dry adhesive. Nanotechnology. 2005;16:1159-66.
[240] del Campo A, Arzt E. Design Parameters and Current Fabrication Approaches for Developing Bioinspired Dry Adhesives. Macromolecular Bioscience. 2007;7:118-27.
[241] Northen MT, Turner KL. Batch fabrication and characterization of nanostructures for enhanced adhesion. Current Applied Physics. 2006;6:379-83.
[242] Northen MT, Greiner C, Arzt E, Turner KL. A Gecko-Inspired Reversible Adhesive. Advanced Materials. 2008;20:3905-+.
[243] Xue L, Kovalev A, Thoele F, Rengarajan GT, Steinhart M, Gorb SN. Tailoring Normal Adhesion of Arrays of Thermoplastic, Spring-like Polymer Nanorods by Shaping Nanorod Tips. Langmuir. 2012;28:10781-8.
[244] Geim AK, Dubonos SV, Grigorieva IV, Novoselov KS, Zhukov AA, Shapoval SY. Microfabricated adhesive mimicking gecko foot-hair. Nat Mater. 2003;2:461-3.
[245] del Campo A, Fernández-Blázquez JP. Bio-Inspired Reversible Adhesives for Dry and Wet Conditions. Biomimetic Approaches for Biomaterials Development: Wiley-VCH Verlag GmbH & Co. KGaA; 2012. p. 259-72.
[246] Northen MT, Turner KL. Meso-scale adhesion testing of integrated micro- and nano-scale structures. Sensors and Actuators a-Physical. 2006;130:583-7.
[247] Autumn K, Sitti M, Liang YA, Peattie AM, Hansen WR, Sponberg S, et al. Evidence for van der Waals adhesion in gecko setae. Proceedings of the National Academy of Sciences. 2002;99:12252-6.
[248] Huber G, Mantz H, Spolenak R, Mecke K, Jacobs K, Gorb SN, et al. Evidence for capillarity contributions to gecko adhesion from single spatula nanomechanical measurements. Proceedings of the National Academy of Sciences of the United States of America. 2005;102:16293-6.
[249] Ho AYY, Yeo LP, Lam YC, Rodríguez I. Fabrication and Analysis of Gecko-Inspired Hierarchical Polymer Nanosetae. ACS Nano. 2011;5:1897-906.
[250] Lee D, Lee S, Cho K. Hierarchical gecko-inspired nanohairs with a high aspect ratio induced by nanoyielding. Soft Matter. 2012;8:4905.
[251] Kamperman M, Synytska A. Switchable adhesion by chemical functionality and topography. Journal of Materials Chemistry. 2012;22:19390-401.
[252] Nadermann N, Ning J, Jagota A, Hui CY. Active Switching of Adhesion in a Film-Terminated Fibrillar Structure. Langmuir. 2010;26:15464-71.
[253] Rodríguez-Hernández J. Wrinkled interfaces: Taking advantage of surface instabilities to pattern polymer surfaces. Progress in Polymer Science.
[254] Law JBK, Ng AMH, He AY, Low HY. Bioinspired Ultrahigh Water Pinning Nanostructures. Langmuir. 2013;30:325-31.
[255] Moynihan S, Iacopino D, O'Carroll D, Lovera P, Redmond G. Template Synthesis of Highly Oriented Polyfluorene Nanotube Arrays†. Chemistry of Materials. 2007;20:996-1003.

[256] O'Carroll D, Lieberwirth I, Redmond G. Microcavity effects and optically pumped lasing in single conjugated polymer nanowires. Nat Nano. 2007;2:180-4.
[257] Moynihan S, Lovera P, O'Carroll D, Iacopino D, Redmond G. Alignment and Dynamic Manipulation of Conjugated Polymer Nanowires in Nematic Liquid Crystal Hosts. Advanced Materials. 2008;20:2497-502.
[258] O'Carroll D, Iacopino D, O'Riordan A, Lovera P, O'Connor É, O'Brien GA, et al. Poly(9,9-dioctylfluorene) Nanowires with Pronounced β-Phase Morphology: Synthesis, Characterization, and Optical Properties. Advanced Materials. 2008;20:42-8.
[259] O' Carroll D, Irwin J, Tanner DA, Redmond G. Polyfluorene nanowires with pronounced axial texturing prepared by melt-assisted template wetting. Materials Science and Engineering: B. 2008;147:298-302.
[260] O'Carroll D, Redmond G. Highly Anisotropic Luminescence from Poly(9,9-dioctylfluorene) Nanowires Doped with Orientationally Ordered β-Phase Polymer Chains. Chemistry of Materials. 2008;20:6501-8.
[261] O'Carroll D, Iacopino D, Redmond G. Luminescent Conjugated Polymer Nanowire Y-Junctions with On-Branch Molecular Anisotropy. Advanced Materials. 2009;21:1160-5.
[262] Grell M, Bradley DDC, Ungar G, Hill J, Whitehead KS. Interplay of Physical Structure and Photophysics for a Liquid Crystalline Polyfluorene. Macromolecules. 1999;32:5810-7.
[263] Shkunov MN, Österbacka R, Fujii A, Yoshino K, Vardeny ZV. Laser action in polydialkylfluorene films: Influence of low-temperature thermal treatment. Applied Physics Letters. 1999;74:1648-50.
[264] Liu X, Xu F, Li Z, Zhang W. Photoluminescence of poly(thiophene) nanowires confined in porous anodic alumina membrane. Polymer. 2008;49:2197-201.
[265] Martin CR. Template Synthesis of Electronically Conductive Polymer Nanostructures. Accounts of Chemical Research. 1995;28:61-8.
[266] Martin CR, Parthasarathy R, Menon V. Template synthesis of electronically conductive polymers - A new route for achieving higher electronic conductivities. Synthetic Metals. 1993;55:1165-70.
[267] Byun J, Kim Y, Jeon G, Kim JK. Ultrahigh Density Array of Free-Standing Poly(3-hexylthiophene) Nanotubes on Conducting Substrates via Solution Wetting. Macromolecules. 2011;44:8558-62.
[268] Hu J, Clark KW, Hayakawa R, Li A-P, Wakayama Y. Enhanced Electrical Conductivity in Poly(3-hexylthiophene)/Fluorinated Tetracyanoquinodimethane Nanowires Grown with a Porous Alumina Template. Langmuir. 2013;29:7266-70.
[269] Huang L-B, Xu Z-X, Chen X, Tian W, Han S-T, Zhou Y, et al. Poly(3-hexylthiophene) Nanotubes with Tunable Aspect Ratios and Charge Transport Properties. ACS Applied Materials & Interfaces. 2014;6:11874-81.
[270] Joo J, Park KT, Kim BH, Kim MS, Lee SY, Jeong CK, et al. Conducting Polymer Nanotube and Nanowire Synthesized by Using Nanoporous Template: Synthesis, Characteristics, and Applications. Synthetic Metals. 2003;135–136:7-9.
[271] Shirai Y, Takami S, Lasmono S, Iwai H, Chikyow T, Wakayama Y. Improvement in carrier mobility of poly(3,4-ethylenedioxythiophene) nanowires synthesized in porous alumina templates. Journal of Polymer Science Part B: Polymer Physics. 2011;49:1762-8.
[272] Wu Y, Gu Q, Ding G, Tong F, Hu Z, Jonas AM. Confinement Induced Preferential Orientation of Crystals and Enhancement of Properties in Ferroelectric Polymer Nanowires. ACS Macro Letters. 2013;2:535-8.
[273] Lutkenhaus JL, McEnnis K, Serghei A, Russell TP. Confinement Effects on Crystallization and Curie Transitions of Poly(vinylidene fluoride-co-trifluoroethylene). Macromolecules. 2010;43:3844-50.
[274] Choi K, Lee SC, Liang Y, Kim KJ, Lee HS. Transition from Nanorod to Nanotube of Poly(vinylidene trifluoroethylene) Ferroelectric Nanofiber. Macromolecules. 2013;46:3067-73.

## Table of abbreviations

| | |
|---|---|
| F4-TCNQ | 2,3,5,6-tetrafluoro-7,7,8,8-tertacyanoquinodimethane |
| 1DPN | One-dimensional polymeric nanostructures |
| AAO | anodic aluminum oxide |
| BCPs | block copolymers |
| CRM | Confocal Raman Microscopy |
| DMF | dimethylformamide |
| DPI | molecular weight distribution |
| HA | hard anodization |
| LbL | layer-by-layer |
| MA | mild anodization |
| $M_w$ | molecular weight |
| NF | nanofibers |
| NMR | nuclear magnetic resonance spectroscopy |
| NR | nanorods |
| NSE | Neutron Spin Echo |
| NT | nanotubes |
| P(VDF-TrFE) | poly (vynilidene-co-trifluoroethylene) |
| P2VP | poly-2-vinylpyridine |
| P3HT | poly (3-hexylthiophene) |
| PAH | polyallylamine hydrochloride |
| PB | polybutadiene |
| PBLG | poly (γ-bencil-L-glutamate) |
| PCDTBT | poly[[9-(1-octylnonyl)-9H-carbazole-2,7-diyl]-2,5-thiophenediyl-2,1,3-benzothiadiazole-4,7-diyl-2,5-thiophenediyl] |
| PCL | poly(ε-caprolactone) |
| PDLLA | poly-DL-lactide |
| PDMS | polydimethylsiloxane |
| PEDOT | poly(3,4-ethylenedioxythiophene) |
| PEEK | poly(ether-ether-ketone) |
| PEO | polyethyelene oxide |
| PEO-*b*-PCL | Poly(ethylene oxide-*b*-ε-caprolactone) |
| PFA | perfluorodecyl acrylate |
| PI | polyisoprene |
| PMMA | poly (methyl methacrylate) |
| PMP-Th | poly[3-(2-methoxyphenyl)tiophene] |
| PNIPAM-co-MBAA | poly(N-isopropylacrylamide)-co-(N,N'-methylenebisacrylamide) |
| PPy | polypyrrole |
| PS | polystyrene |
| PS/FePt | polystyrene with $Fe_{55}Pt_{45}$ nanoparticles |
| PS-*b*-P2VP | polystyrene-*b*-poly(2-vinyl pyridine) |
| PS-*b*-P4VP | polystyrene-*b*-poly(4-vinylpyridine) |
| PS-*b*-PAN | polystyrene-*b*-polyacrylonitrile |
| PS-*b*-PB | polystyrene-*b*-polybutadiene |
| PS-*b*-PMMA | polystyrene-*b*-poly(methylmethacrylate) |
| PSS | polystryrenesulfonic |
| PTFE | polytetrafluoroethylene |
| PVDF | poly(vinylidene fluoride) |
| SEC | size exclusion chromatography (SEC) |
| St-co-MOP | styrene-co-2-methyl-3-oxo-5-phenyl-4-pentenonitrile copolymer |